\documentclass[fleqn,usenatbib,useAMS]{mnras}

\usepackage{newtxtext,newtxmath}
\usepackage[T1]{fontenc}
\usepackage{ae,aecompl}

\usepackage{graphicx}
\usepackage{subfigure}
\usepackage{amsmath}
\usepackage{amssymb}	

\title[Exploding Core-Collapse Supernovae]{Revival of the Fittest: Exploding Core-Collapse Supernovae from 12 to 25 M$_{\odot}$}

\author[D. Vartanyan et. al]{David Vartanyan$^{1}$\thanks{E-mail: dvartany@princeton.edu},
Adam Burrows$^{1}$,
David Radice$^{1,2}$ ,
M. Aaron Skinner$^{3}$,
\newauthor
Joshua Dolence$^{4}$
\\
$^{1}$Department of Astrophysical Sciences, Princeton University, Princeton, NJ 08544\\
$^{2}$ Institute for Advanced Study, 1 Einstein Dr, Princeton NJ 08540\\
$^{3}$Livermore National Laboratory, 7000 East Ave., Livermore, CA 94550-923\\
$^{4}$CCS-2, Los Alamos National Laboratory, P.O. Box 1663 
Los Alamos, NM 87545\\
}

\date{Last updated 2018 January 1}

\pubyear{2018}

\begin{document}
\label{firstpage}
\pagerange{\pageref{firstpage}--\pageref{lastpage}}
\maketitle
\begin{abstract}
{We present results of 2D axisymmetric core-collapse supernova simulations, employing the F{\sc{ornax}} code, of nine progenitor models spanning 12 to 25 M$_{\odot}$. Four of the models explode with inelastic scattering off electrons and neutrons as well as the many-body correction to neutrino-nucleon scattering opacities. We show that these four models feature sharp Si-O interfaces in their density profiles, and that the corresponding dip in density reduces the accretion rate around the stalled shock and prompts explosion. The non-exploding models lack such a steep feature, highlighting the Si-O interface as one key to explosion. Furthermore, we show that all of the non-exploding models can be nudged to explosion with modest changes to macrophysical inputs, including moderate rotation and perturbations to infall velocities, as well as to microphysical inputs, including reasonable changes to neutrino-nucleon interaction rates, suggesting that all the models are perhaps close to criticality. Exploding models have energies of few 
$\times$10$^{50}$ ergs at the end of our simulation, and are rising, emphasizing the need to continue these simulations over larger grids and for longer times to reproduce the energies seen in Nature. Morphology of the explosion contributes to the explosion energy, with more isotropic ejecta producing larger explosion energies. We do not find evidence for the Lepton-number Emission Self-Sustained Asymmetry. Finally, we look at PNS properties and explore the role of dimension in our simulations. We find that convection in the proto-neutron star (PNS) produces larger PNS radii as well as greater ``$\nu_\mu$'' luminosities in 2D compared to 1D.}

\end{abstract}


\section{Introduction}
For over fifty years, since neutrinos were proposed by \cite{1966ApJ...143..626C} as critical to core-collapse supernovae, simulations have attempted, often unsuccessfully, to reproduce the robust explosions seen in Nature. Given recent detection of gravitational waves from compact mergers (\citealt{2016ApJ...818L..22A}; \citealt{2017PhRvL.119p1101A}), simultaneous detection of electromagnetic radiation, a neutrino signature (\citealt{2012PhRvD..86b4026O}), and gravitational waves from supernovae (\citealt{2009CQGra..26f3001O}; \citealt{2013ApJ...766...43M}; \citealt{2013ApJ...779L..18C}; \citealt{2013CRPhy..14..318K}; \citealt{2016ApJ...829L..14K}) represents a yet unbreached frontier and a probe of three of the four fundamental natural forces. Such observations will be tractable by second and third generator detectors (\citealt{2015PhRvD..92h4040Y}; \citealt{2017MNRAS.468.2032A}) and allow constraints on the explosion mechanism, progenitor mass, and equation of state (\citealt{2018arXiv180101914M}). Improvements over the years in understanding the multitude of microphysical interactions and in the capabilities of multi-dimensional simulations have combined together to improve our understanding of this central phenomenon.
  
Using the CHIMERA code, the Oak Ridge group (\citealt{2013ApJ...767L...6B}; \citealt{2016ApJ...818..123B}) found explosions for the 12-, 15-, 20-, and 25-M$_{\odot}$ progenitors from \citealt{2007PhR...442..269W} (henceforth, WH07), all roughly at the same post-bounce time and without the shock radius stalling. Employing PROMETHEUS-VERTEX, \cite{2016ApJ...825....6S} found later explosions over a spread of explosion times for the same four progenitors and 14 additional progenitors in the 11-28 M$_{\odot}$ mass range, from \cite{2002RvMP...74.1015W}. Both approaches use a ray-by-ray approach of multiple one-dimensional solutions to approximate multi-dimensional neutrino transport and include inelastic scattering of neutrinos off nucleons and electrons. However, earlier studies suggest that the ray-by-ray approach introduces axial artifacts and exaggerates anisotropies (\citealt{2006ApJS..164..130O}, \citealt{2016ApJ...831...81S}, \citealt{2015ApJ...800...10D}, \citealt{Burrows2018}) which may artificially promote explosion, at least in two dimensions. Recently updated results for the same progenitors by \cite{2018ApJ...854...63O} found explosions for all but the 12-M$_{\odot}$ progenitor using an M1 closure scheme for neutrino transport rather than the ray-by-ray approximation, but ignoring inelastic neutrino scattering processes. All their explosions were significantly delayed relative to \cite{2016ApJ...825....6S} (490 ms for the 15-M$_{\odot}$ model, 500 ms for the 20-M$_{\odot}$ model, and 270 ms for the 25-M$_{\odot}$ model). Though, \cite{2018ApJ...854...63O} found earlier explosions by 100-150 ms by correcting for inelastic neutrino scattering, this is insufficient to explain the delayed explosions using M1 transport rather than ray-by-ray. The lack of explosion for the 12-M$_{\odot}$ progenitor, and significantly delayed explosions for the remaining three models, buttress our argument that the ray-by-ray approach either allows an explosion where there would have been none, or accelerates the time of explosion,  at least in 2D. The potential artifacts of the ray-by-ray approach in three dimensions remain unclear.

\cite{2016MNRAS.461.3864A} (see also \citealt{2017arXiv171001282R}) found that late nuclear shell burning produces strong turbulent convection, which promotes supernova explosion. These results were iterated in 3D by various groups (see below). More recently, using the M1 closure for multi-dimensional neutrino transport, \cite{2018ApJ...852...28S} found 2D explosions abetted by using a general relativistic rather than Newtonian treatment of gravity. \cite{2017PhRvL.119x2702B} find that muon creation at the high temperatures in proto-neutron stars facilitates explosion in 2D. Thus, an interplay of turbulence, microphysics, and a proper treatment of gravity have been historically critical in producing supernovae explosions in two dimensions.

3D simulations have evolved in the decade since the early foray by \cite{2002ApJ...574L..65F} using a grey scheme for neutrino transport. Using PROMETHEUS-VERTEX, \cite{2015ApJ...801L..24M} found that the 9.6-M${\odot}$ progenitor explodes in 3D with faster shock expansion than in 2D. \cite{2015ApJ...808L..42M} found also explosion for a 20-M$_{\odot}$ progenitor, but only with a strangeness correction to the axial-vector coupling constant which may be too large to be physical (see \citealt{PhysRevLett.108.102001}, \citealt{2017PhRvD..95k4502G}).  Using ZEUS-MP and omitting heavy neutrinos, \cite{2012ApJ...749...98T} explode their 11.2-M$_{\odot}$ progenitor in 3D on a low-resolution grid with the IDSA scheme and the ray-by-ray approach to solve for multi-dimensional neutrino transport. Comparing to 2D, 3D resulted in increased neutrino dwell time in the gain region and more violent convection, but also increased neutrino cooling. Updating the IDSA scheme, including a leakage scheme for heavy neutrinos, and quadrupling the $\phi$ resolution, \cite{2014ApJ...786...83T} identified shock revival for all models, with delayed explosion at higher resolution and more robust explosions in 2D than 3D. More recently, the Garching group studied the 15-M$_{\odot}$ WH07 progenitor with various rotation models (\citealt{2018ApJ...852...28S}). They concluded that rapid rotation inhibits explosion in 2D but promotes it in 3D, citing the development of a SASI mode that compensates for reduced neutrino heating due to rotation. Notably, explosion set in shortly after the accretion of the Si/Si-O interface. Using FLASH, \cite{2013ApJ...778L...7C} produced explosions in 3D for their 15-M$_{\odot}$ progenitor when introducing perturbations to angular velocities. Such perturbations increased turbulent ram pressure (\citealt{2015ApJ...799....5C}), mediating explosion. \cite{2017MNRAS.472..491M} also presented the first simulations of the final minutes of iron core evolution in 3D, finding that asphericities in 3D progenitor structure enhance post-shock turbulence. Using COCONUT-FMT, \cite{2017MNRAS.472..491M} similarly found their 18-M$_{\odot}$ model to explode when the progenitor is allowed to evolve in 3D for the final five minutes of oxygen burning. More generally, some past multi-group 3D simulations either did not explode (\citealt{2013ApJ...770...66H}; \citealt{2014ApJ...792...96T}), or exploded later than 2D counterparts (\citealt{2013ApJ...775...35C}) more recent simulations suggest that 3D progenitors are only slightly less explosive (\citealt{2016ApJ...831...98R}; \citealt{2015ApJ...807L..31L}; see review by \citealt{2016PASA...33...48M}).

In \cite{Burrows2018}, we presented results of 2D simulations toggling a variety of physical processes, particularly inelastic neutrino scattering off electrons and nucleons and the many-body correction to neutrino-nucleon scattering opacities (\citealt{PhysRevC.95.025801}). We found that the results, particularly whether or not a model exploded, were sensitive to small changes in microphysics when near criticality for explosion. \cite{2017IAUS..331..107O} emphasized these results that explosion is sensitive to the many-body effect, with changes to the neutral-current scattering cross section at the 10-20\% level at densities above 10$^{12}$ g cm$^{-3}$ pushing all their models from 12-25 M$_{\odot}$ to explode.

Here, we present the comprehensive results of a series of 2D radiation/hydro simulations using F{\sc{ornax}} of a suite of nine progenitors spanning 12 to 25 M$_{\odot}$ performed on a grid extending out to 20,000 km. Our key findings are that four of our progenitors explode with the inclusion of inelastic scattering processes off electrons and nucleons as well as with the many-body correction to neutrino-nucleon scattering opacities. We show that the non-exploding models can also be nudged to explosion with the inclusion of additional physical inputs, such as modest rotation and perturbations to infall velocities. 

\begin{figure*}
\includegraphics[width=0.5\textwidth]{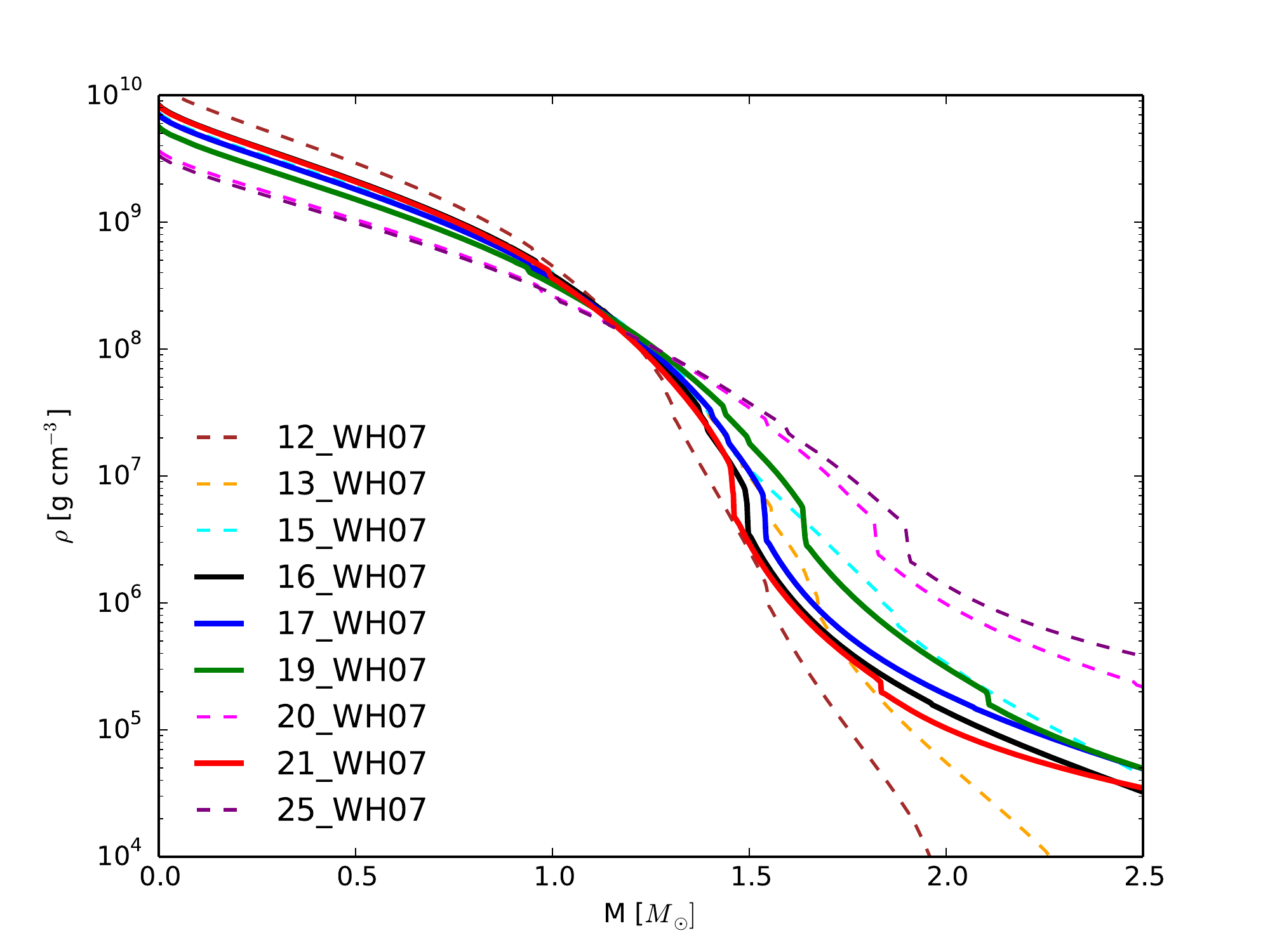}\hfill
\includegraphics[width=0.5\textwidth]{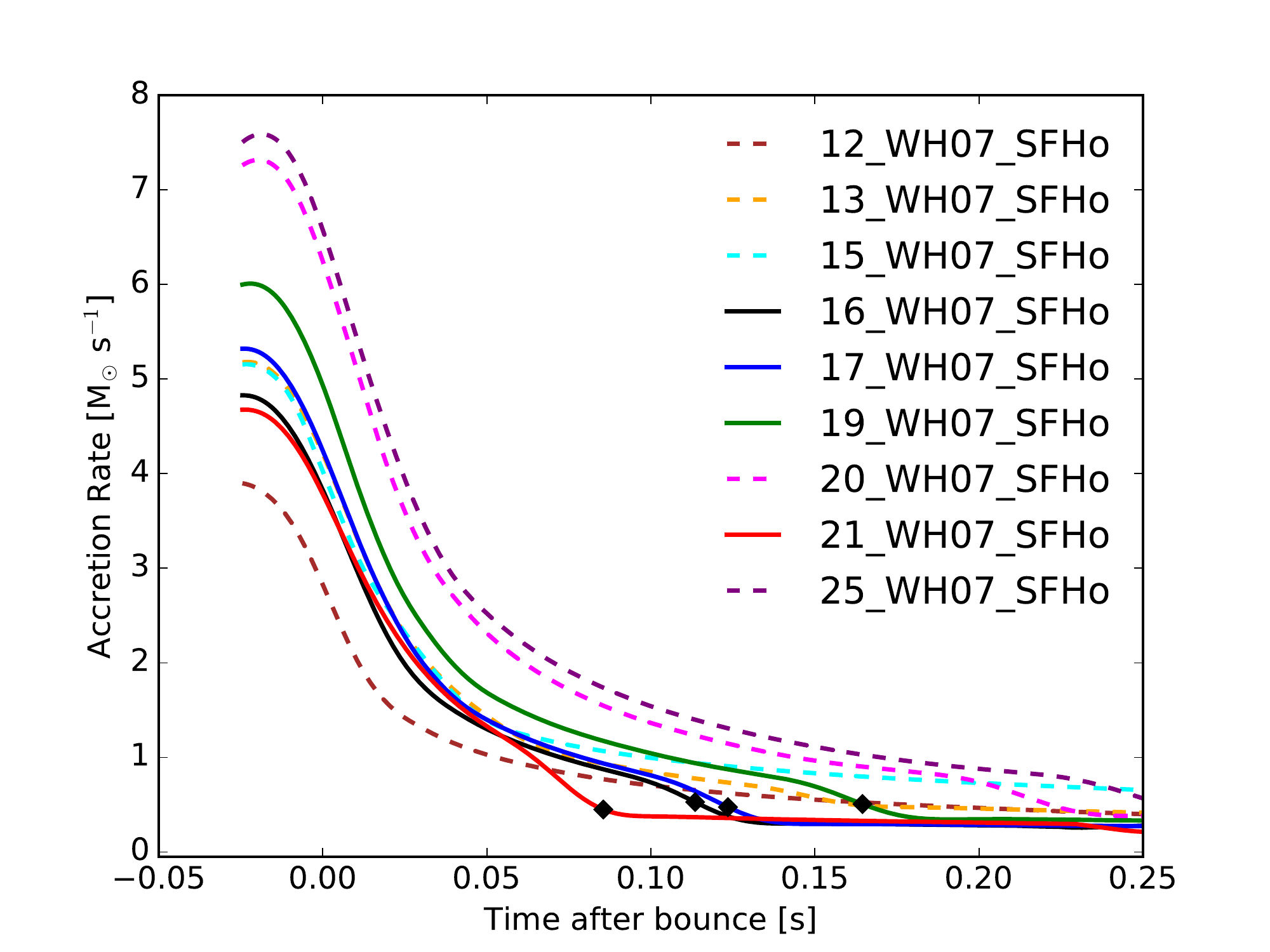}
\caption{\textbf{Left panel:} Initial density profiles (in g cm$^{-3}$) against enclosed mass (in M$_{\odot}$) for the nine progenitors (12-, 13-, 15-, 16-, 17-, 19-, 20-, 21-, and 25-M$_{\odot}$) taken from \protect\cite{2007PhR...442..269W}. We find that four of the nine benchmark models (including inelastic scattering off electrons and nucleons together with the many-body correction) explode, and their  density profiles are illustrated in thick solid lines. The remaining five density profiles (dashed) correspond to non-exploding benchmark models. The exploding models have pronounced Si-O interfaces interior to 1.7 M$_{\odot}$ as seen by the sharp density drop-off at several $\times$ 10$^6$ g cm$^{-3}$. These models have steeper density gradients interior to the interface and shallower profiles exterior, which we argue promotes explosion. \textbf{Right panel:} Accretion rates (in M$_{\odot}$ s$^{-1}$) at 500 km for the six WH07 progenitors plotted against time (in seconds) after bounce. The majority of the models sustain accretion rates of over 1 M$_{\odot}$ s$^{-1}$ for the first 300 ms. Note the approximate order of increasing accretion rate with progenitor mass for the non-exploding models (dashed). The accretion rates plummet over a spread of 150 ms for the various progenitors, similar to the spread in explosion times. All exploding models (solid, with explosion times marked with black diamonds) feature an early dip in accretion rate that corresponds to accretion of the Si-O interface and, subsequently, to explosion (see Fig.\,\ref{fig:3}); note that the exploding models with Si-O interfaces further interior in mass (\textbf{left}) are the first to dip in accretion (\textbf{right}).}
\label{fig:1}
\end{figure*}

In Sec.\,\ref{sec:methods}, we introduce the numerical methods and setup for our simulations. In Sec.\,\ref{sec:diagnostics}, we remark on basic diagnostics of our results and explore the role of the Si-O interface accretion in explosion outcome. We expand these diagnostics in Sec.\,\ref{sec:results}, where we look at explosion energies and probe properties in the gain region. We also focus on a study of microphysical and macrophysical inputs, as well as  progenitor dependence, illustrating that all models considered can explode with changes to opacities, moderate rotation and/or perturbations to infall velocities. We further explore the electron-fraction distribution of the ejecta mass for the exploding models, and look for evidence for the Lepton-number Emission Self-Sustained Asymmetry but find none. In Sec.\,\ref{sec:ns}, we comment on the properties of the resulting neutron stars. We compare our 2D and 1D simulations in Sec.\,\ref{sec:1D}. Finally, we summarize our results and  present our conclusions in Sec.\,\ref{sec:con}. 
 
\begin{figure*}
\includegraphics[width=\textwidth, angle = 0]{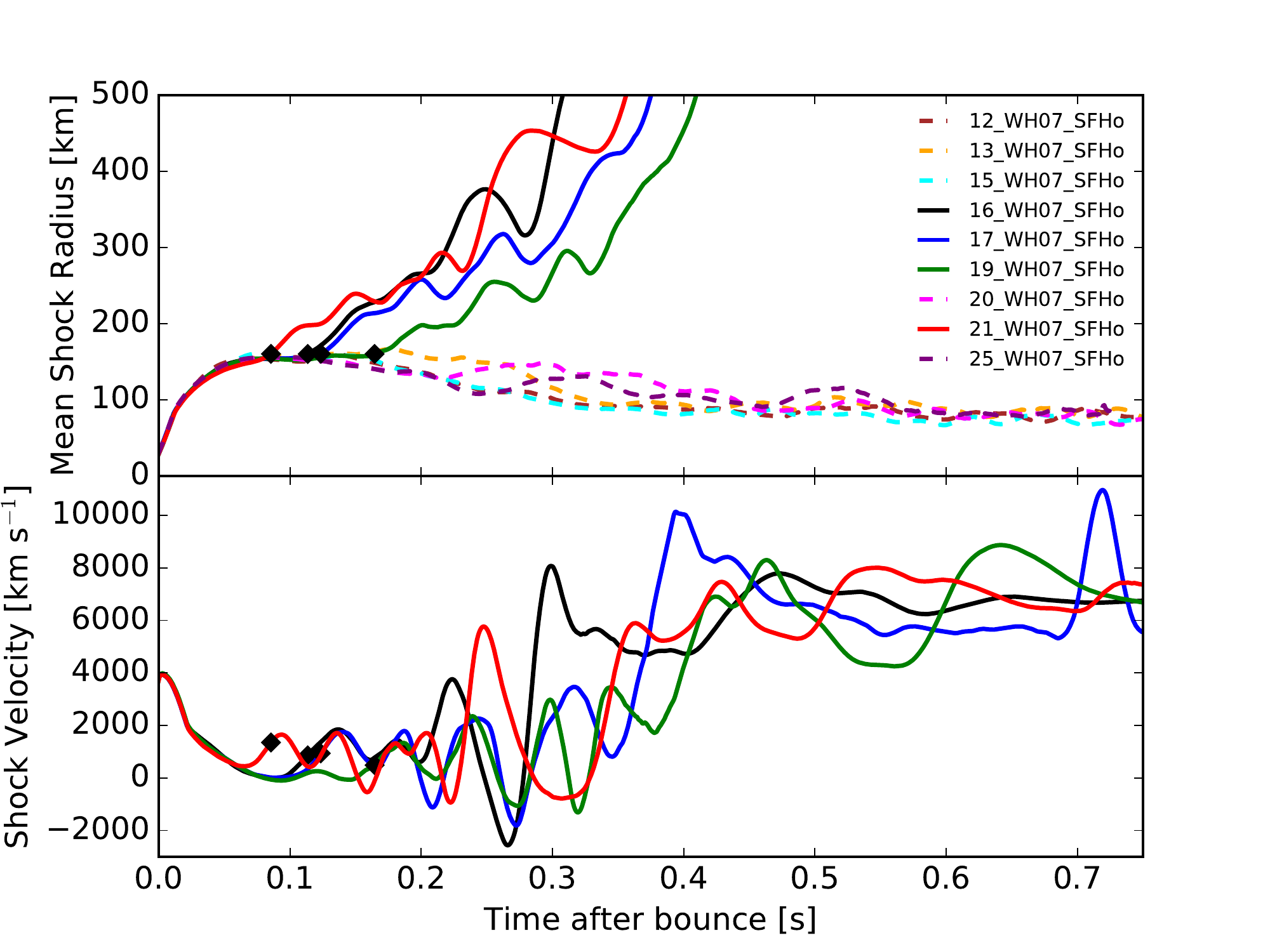}
\caption{\textbf{Top panel}: Plotted are the mean shock radii (in kilometers) for our 2D simulations of the 12-, 13-, 15-, 16-, 17-, 19-, 20-, 21-, and 25-M$_{\odot}$ progenitors (\citealt{2007PhR...442..269W}) as a function of time (in seconds) after bounce. Exploding models are illustrated in solid, and non-exploding models in dashed. The shock radii evolve almost identically until $\sim$100 ms after bounce. We see no monotonic correlation with progenitor mass and explosion. However, we note the correlation between the position of the Si-O interfaces in Fig.\,\protect\ref{fig:1} and explosion times. Interfaces located deeper in the progenitor correspond to earlier explosions, with explosion order of 21-, 16-, 17-, and 19-M$_{\odot}$, suggesting that earlier accretion of these interfaces prompts earlier explosion. \textbf{Bottom panel}: We plot average shock velocities as a function of time (in seconds) after bounce. Note that the shock is moving outwards at early times, accumulating mass and decreasing in velocity until 100 ms. The 25-M$_{\odot}$ progenitor has the largest shock velocity early on, peaking over 20,000 km s$^{-1}$, roughly 6$\%$ the speed of light. At late times, all shock velocities asymptote to approximately 10,000 km s$^{-1}$. This figure is smoothed using a running time average of 10 ms.}
\label{fig:2}
\end{figure*}

\section{Progenitors and Setup}\label{sec:methods}
We consider nine progenitors from \cite{2007PhR...442..269W} spanning 12 to 25 M$_{\odot}$. Their density profiles are illustrated in Fig.\,\ref{fig:1}. We evolve these models in two dimensions out to 20,000 km, until the maximum shock radius reaches the grid boundary, using F{\sc{ornax}}, a new multi-dimension, multi-group radiation/hydrodynamic code developed to study core-collapse supernovae (\citealt{2016ApJ...817..182W}; \citealt{2016ApJ...831...81S}, \citealt{Burrows2018}; Skinner et.\,al 2018, in prep.). F{\sc{ornax}} solves the comoving-frame velocity dependent transport equations to order O(v/c). The hydrodynamics uses a directionally-unsplit Godunov-type finite-volume scheme and computes fluxes at cell faces using an HLLC Riemann solver. It employs a dendritic grid that deresolves at small radii to overcome CFL limitations in evolution time while approximately preserving cell size and shapes to keep the timestep independent of resolution. Our default resolution is 608 radial cells by 256 angular cells. The radial grid extends out to 20,000 km and is spaced evenly with $\Delta$r $\sim$0.5 km for r $\leq$ 50 km and logarithmically for r $\geq$ 50 km, with a smooth transition between. The angular grid resolution varies smoothly from $\sim$0.95$^\circ$ at the poles to $\sim$0.65$^\circ$ at the equator. For this project, we use a monopole approximation for gravity. We include an approximate general relativistic term following \cite{2006A&A...445..273M} and employ the SFHo equation of state (\citealt{2013ApJ...774...17S}) which is consistent with all currently known nuclear constraints (\citealt{2017ApJ...848..105T}). 

\begin{figure*}
\subfigure{\includegraphics[width=0.5\textwidth]{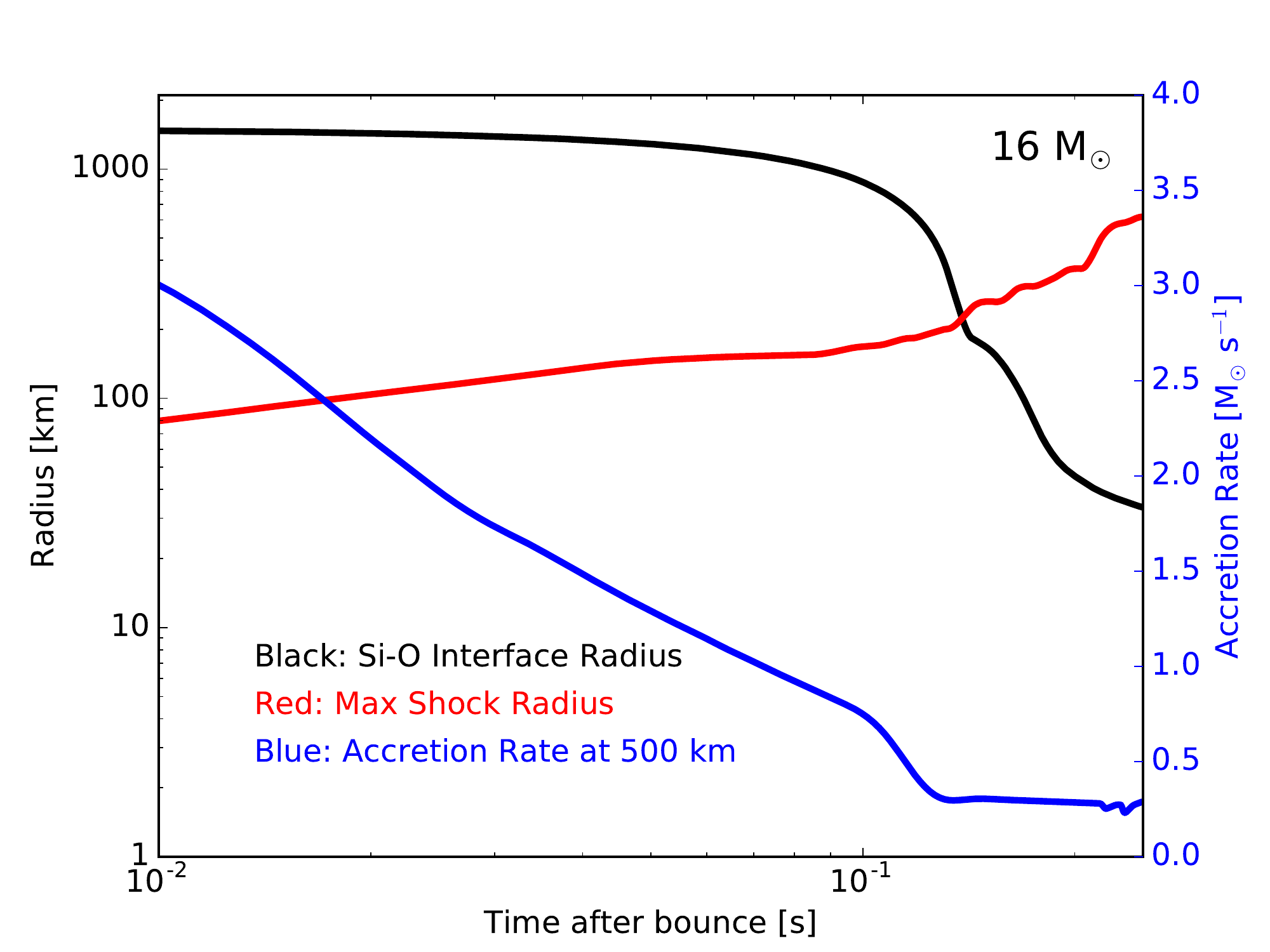}}\hfill
\subfigure{\includegraphics[width=0.5\textwidth]{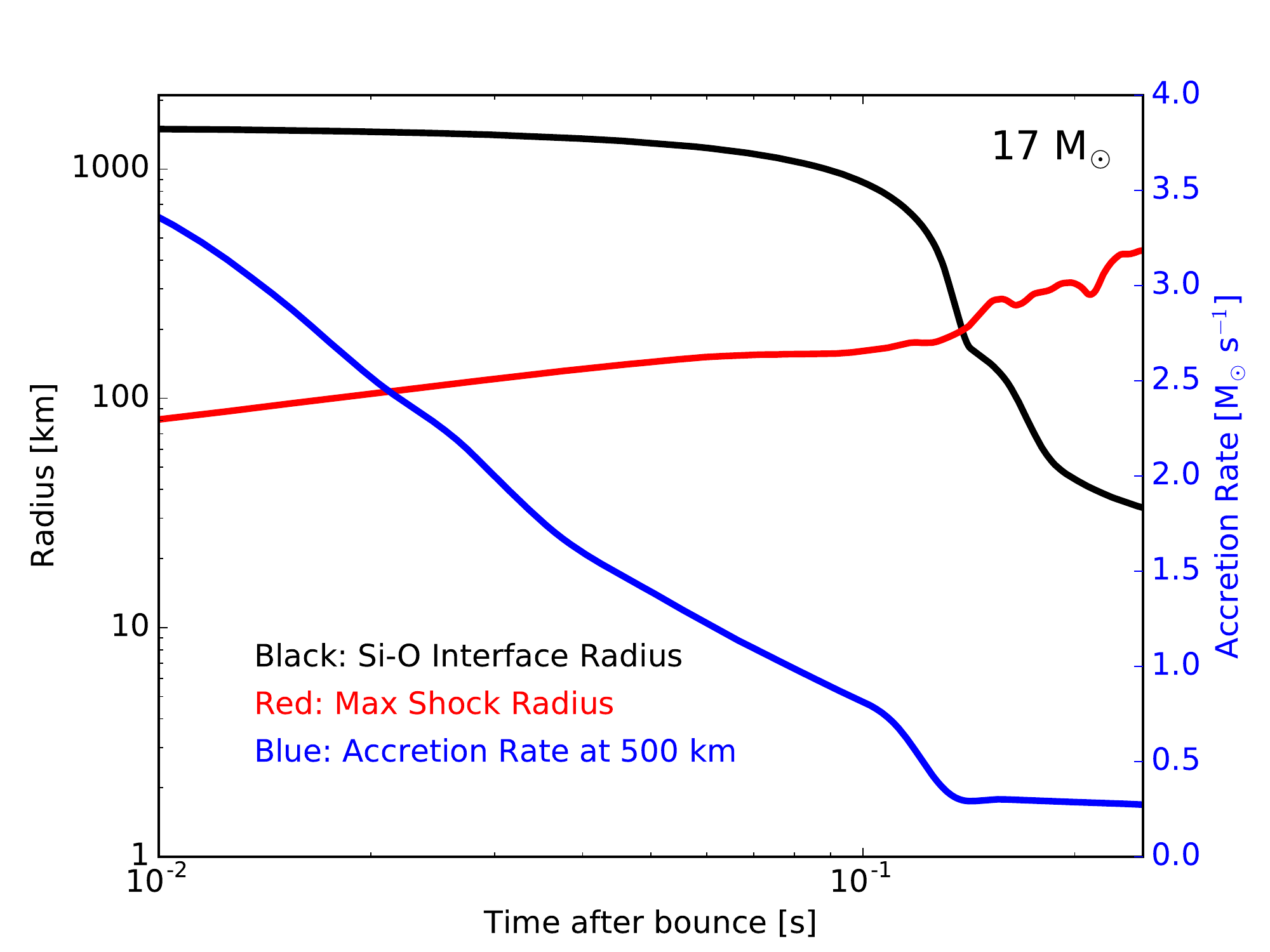}}
\subfigure{\includegraphics[width=0.5\textwidth]{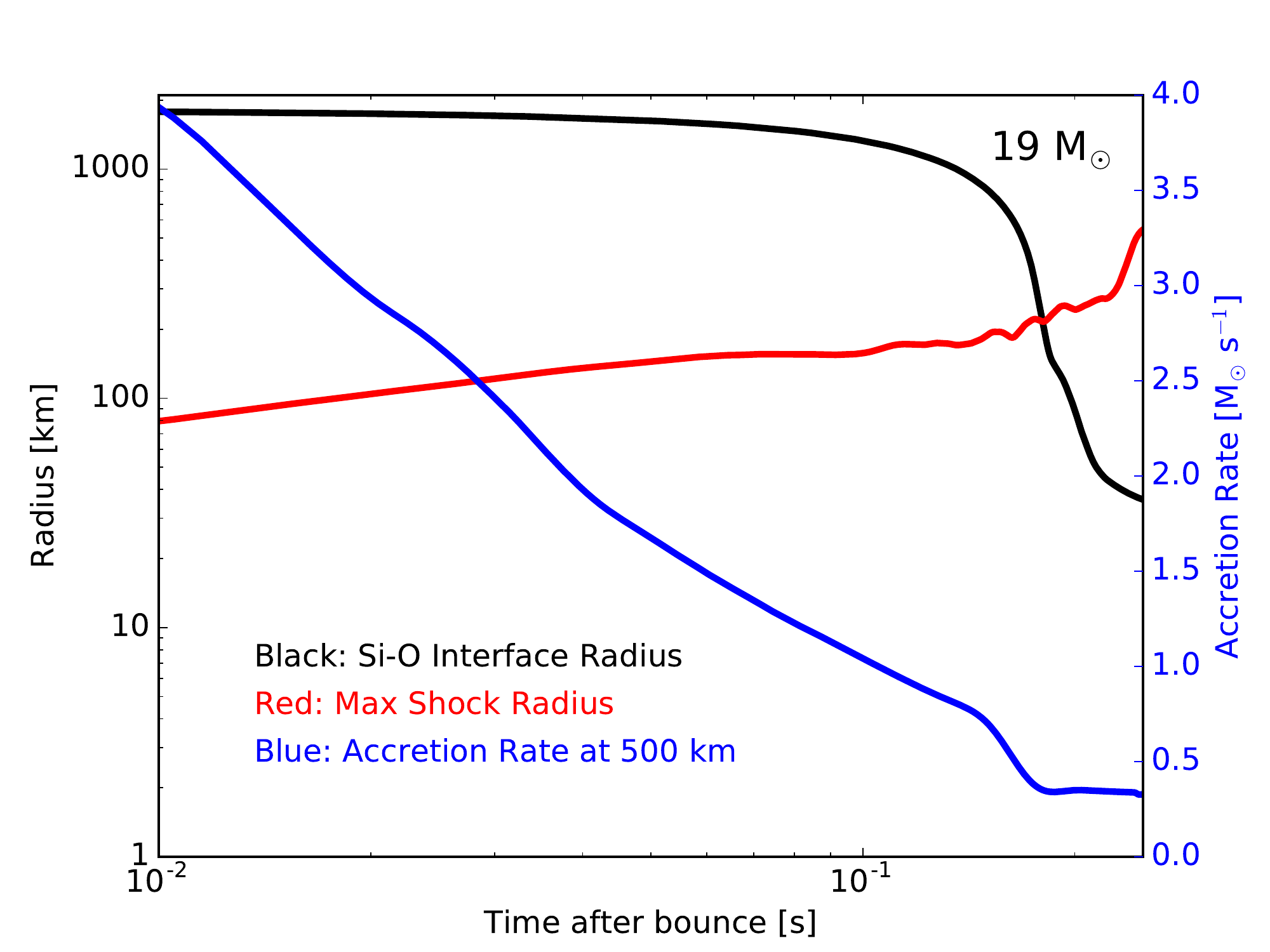}}\hfill
\subfigure{\includegraphics[width=0.5\textwidth]{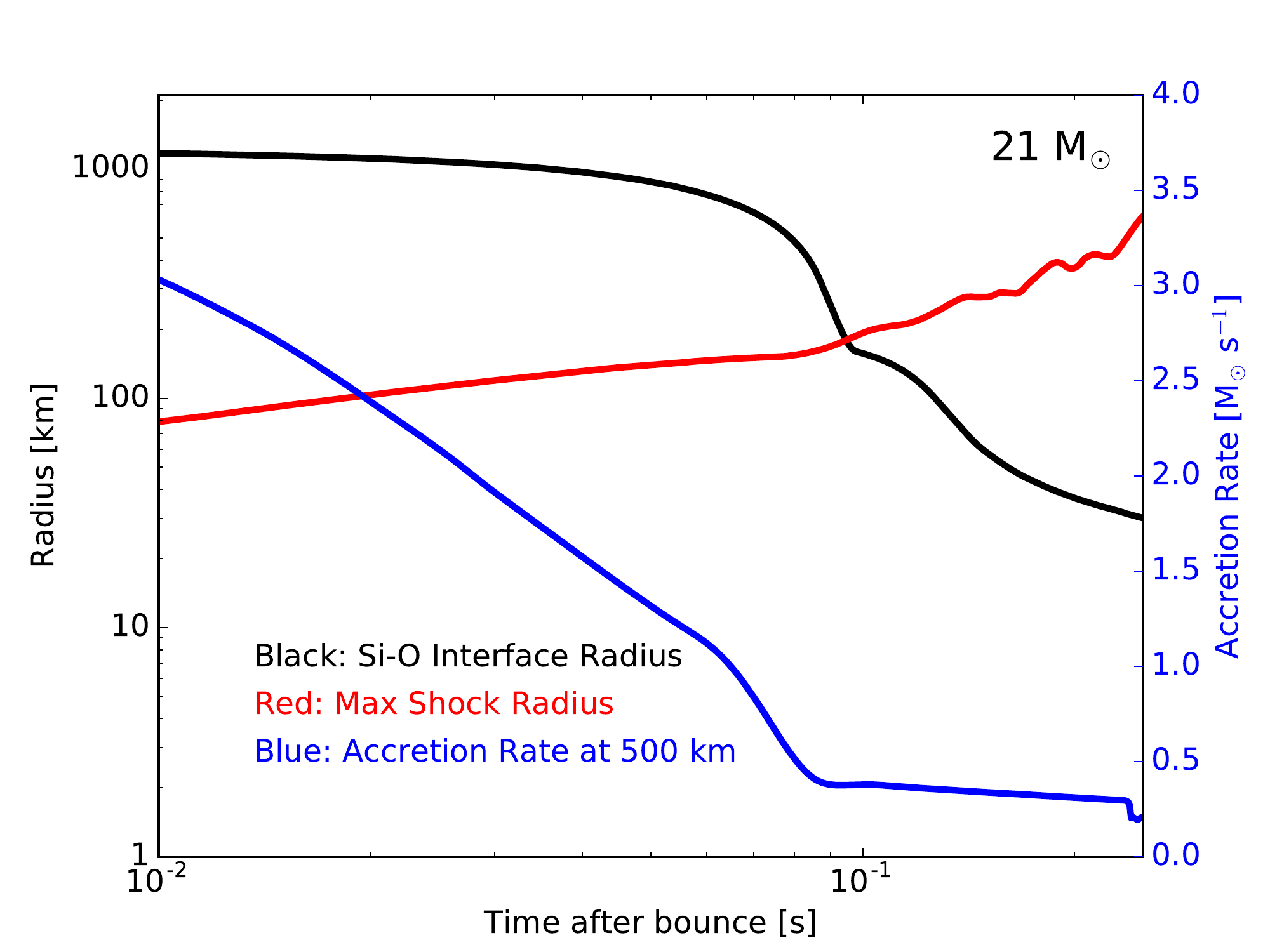}}
\caption{Maximum shock radius (in km, red), location of the Si-O interface (in km, black), and accretion rate at 500 km (in M$_{\odot}$ s$^{-1}$, blue) as a function of time after bounce (in seconds) for the exploding 16-, 17-, 19-, and 21-M$_{\odot}$ progenitors, left to right. The accretion rate radius is chosen to be close to, but outside, the maximum shock radius at early times post-bounce. We see a corresponding drop in the accretion rate just as the Si-O interface passes 500 km, which happens roughly 0.1 seconds after bounce for these models. Shortly afterwards, the maximum shock radius reaches the Si-O interface, and we witness the expansion of the maximum shock radius towards explosion. Note the simultaneous onset of variations in the maximum shock radius with the accretion of the Si-O interface.}
\label{fig:3}
\end{figure*}

We solve for radiation transfer using the M1 closure scheme for the second and third moments of the radiation fields (\citealt{2011JQSRT.112.1323V}). We follow three species of neutrinos: electron-type, anti-electron-type, and treat the heavy neutrinos as a single species, ``$\nu_\mu$." We use 20 energy groups spaced logarithmically between 1 and 300 MeV for electron neutrinos and to 100 MeV for anti-electron- and ``$\nu_\mu$"-neutrinos.

We follow the notation of \cite{Burrows2018} for our progenitors, with IES{\_}INS{\_}MB indicating inelastic scattering of neutrinos off electrons (IES) and nucleons (INS) and the many-body (MB) correction to the neutrino-nucleon opacities. The neutrino-matter interactions follow \cite{2006NuPhA.777..356B}, with inelastic neutrino-nucleon scattering incorporated using a modified version of \cite{2003ApJ...592..434T}.

For a more detailed discussion of the numerical methods employed by F{\sc{ornax}}, see \cite{2016ApJ...831...81S}; \cite{2017ApJ...850...43R}; \cite{Burrows2018}; Skinner et. al 2018 (in prep.).

\section{Explosion Dynamics}\label{sec:diagnostics}
We present results that the progenitor structure is crucial in determining explosion outcome for our model suite of simulations. In Fig.\,\ref{fig:1}, right panel, we illustrate the accretion rates (in M$_{\odot}$ s$^{-1}$) as a function of time after bounce (in seconds) at 500 km for the first several hundred milliseconds post-bounce. In the top panel of Fig.\,\ref{fig:2}, we illustrate the mean shock radii (in km) and the shock velocity (km s$^{-1}$) as a function of time after bounce (in seconds) in the top and bottom panels, respectively, for our suite of nine WH07 progenitors from 12 to 25 M$_{\odot}$. Explosion times, defined as when the mean shock radius reaches 160 km after passing an inflection point, are indicated subsequently in our figures as black diamonds. The explosion times are non-monotonic with progenitor mass and have a spread of approximately 100 ms for the various explosions. The mean shock radii evolve almost identically until 100 ms post-bounce, then continue to rise without stalling for all our exploding models. Models 16-, 17-, 19- and 21- M$_{\odot}$ all explode, with the heaviest, the 21 M$_{\odot}$ model exploding first. These four have not been the focus of previous studies in recent simulations. Both \cite{2016ApJ...818..123B} (see also their Paper 1, \citealt{2013ApJ...767L...6B}) and \cite{2016ApJ...825....6S} studied the 12-, 15-, 20- and 25-M$_{\odot}$ WH07 progenitors, finding that all of them explode, but using the ray-by-ray approximation to neutrino transport and the LS220 equation of state (\citealt{1991NuPhA.535..331L}). We find that none of these models explodes for our default setup, but we show later (Sec.\,\ref{sec:results}) that, with moderate macrophysical modifications, all progenitors models can be nudged into explosion. However, using M1 transport rather than ray-by-ray for neutrino transport, \cite{2018ApJ...854...63O} find delayed explosions for the 15-, 20-, and 25-M$_{\odot}$ progenitors. There are significant physical differences between our two approaches, including a different analytic closure (\citealt{1978JQSRT..20..541M} vs. \citealt{2011JQSRT.112.1323V}), a different equation of state (LS220 vs. SFHo), and different energy resolutions (12 vs. 20 energy groups). In particular, the LS220 equation of state has shown to be more explosive than the SFHo (\citealt{2017PhRvL.119x2702B}). Be that as it may, we show in Sec.\,\ref{sec:results} that all progenitors are close to criticality for explosion.
 
Models 16- and 17-M$_{\odot}$ explode in short succession of each other, with the 19-M$_{\odot}$ progenitor exploding last, roughly 50 ms later. As we will discuss in Sec.\,\ref{sec:results}, explosion order anti-correlates with explosion energy. In the bottom panel of Fig.\,\ref{fig:2}, we plot the mean shock velocity versus time after bounce for the four exploding models. At late times, the shock velocities undulate with no regard to progenitor mass, asymptoting to roughly 7000 km s$^{-1}$ with values as high as 10,000 km s$^{-1}$, around 0.03 times the velocity of light, $c$. However, the early rise in velocities at $\sim$0.1s after bounce follows the explosion ordering seen in the shock radii, with earlier explosions showing higher shock velocities at early times.

\begin{figure*}
\subfigure{\includegraphics[width=0.5\textwidth]{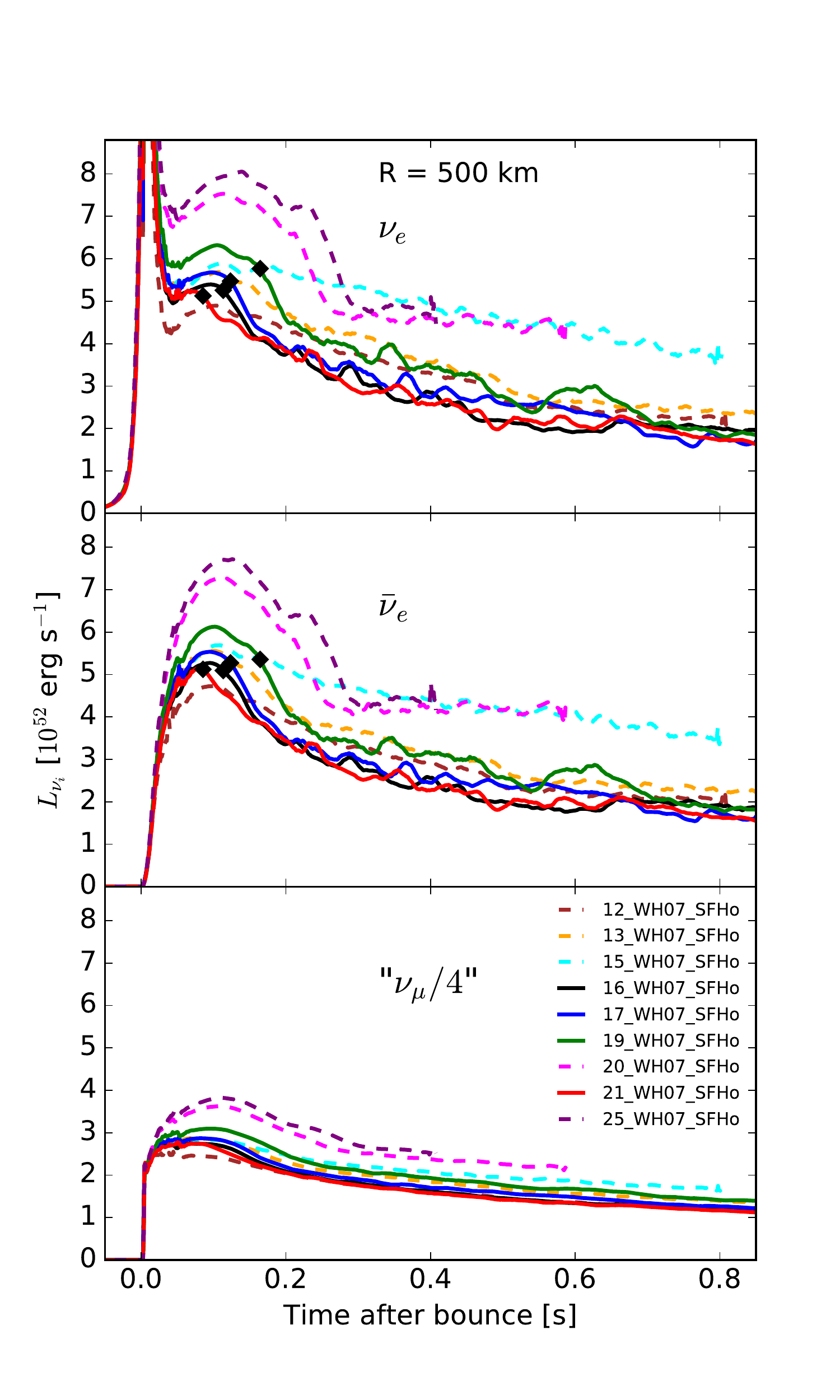}}\hfill
\subfigure{\includegraphics[width=0.5\textwidth]{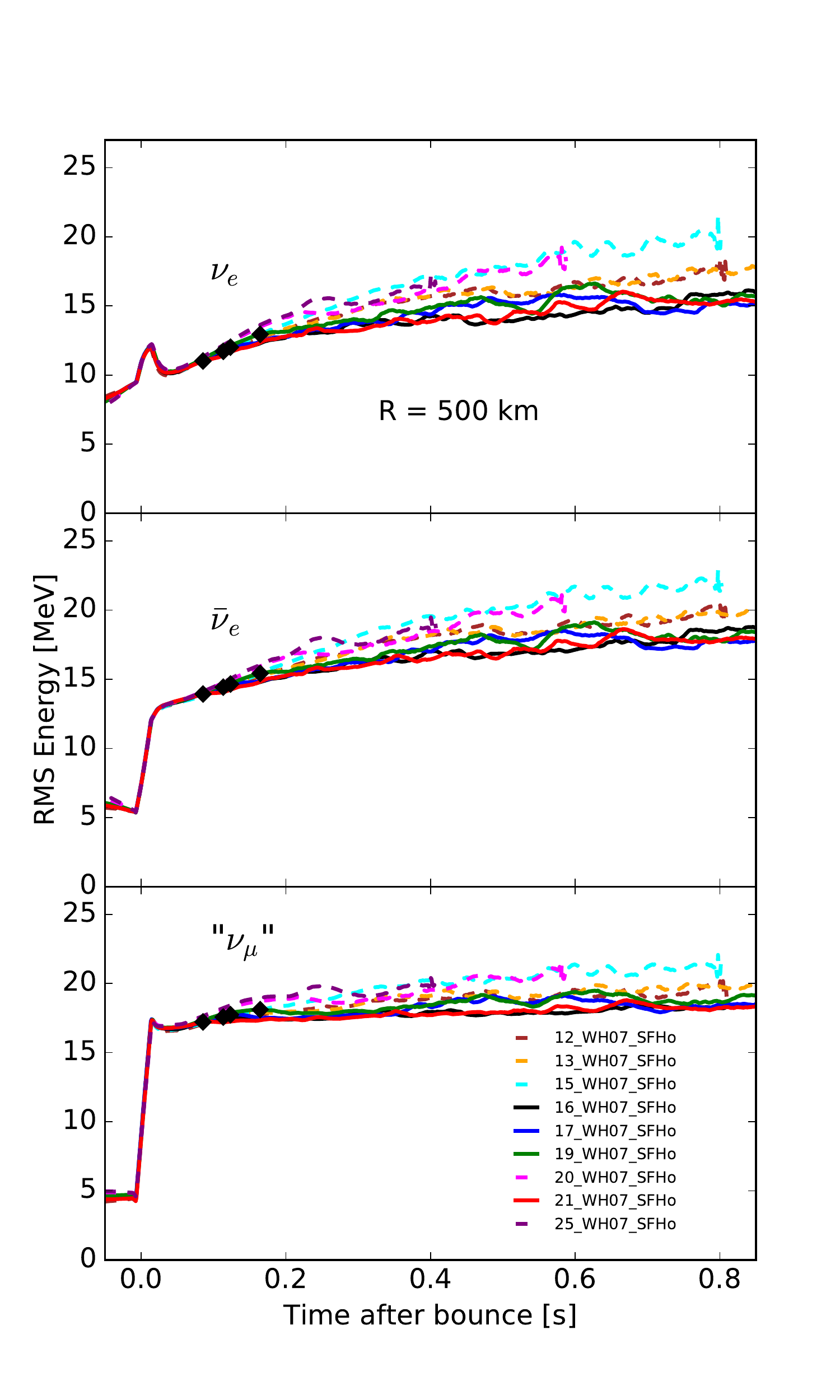}}
\caption{Lab-frame neutrino luminosities (ergs s$^{-1}$, \textbf{left panel}) and RMS neutrino energies (MeV, \textbf{right panel}) for our progenitor suite relative to time after bounce (in seconds) at 500 km. For the non-exploding models (dashed), the luminosities for all three neutrino species increases with progenitor mass. For the exploding models (solid), neutrino luminosity tracks explosion time, with earlier explosions having lower neutrino luminosities. Interestingly, we see no such feature in the RMS neutrino energies. Rather, all exploding models have similar RMS neutrino energies of under 15 MeV at late times for electron-type neutrinos. Non-exploding models have RMS energies several MeV higher for all neutrino species.}
\label{fig:4}
\end{figure*}

For the heaviest progenitors, the accretion rates remain over 1 M$_{\odot}$ s$^{-1}$ until as late as 200 ms post-bounce, over a hundred of milliseconds longer than found by \cite{2017ApJ...850...43R} for a suite of lower mass (8.1 - 11 M$_{\odot}$) progenitors. We naively expect lower mass progenitors to have systematically lower accretion rates, resulting in less ram pressure to overcome to achieve explosion. While the five non-exploding models do have accretion rates which increase with mass, the four exploding models (16-, 17-, 19-, 21-M$_{\odot}$) see their accretion rates dip earlier for models that explode earlier. Note that even for our conservatively early definition of explosion time (shock radius reaching only 160 km, marked by the black diamonds), the accretion rate dips in advance of explosion, suggesting that the reduced accretion rate prompts explosion and is not a result of it. The magnitude of the accretion rate itself is not the determinant of explosion, but rather the interplay between accretion rate and accretion luminosity (\citealt{1993ApJ...416L..75B}; \citealt{2015MNRAS.448.2141M}, \citealt{2016ApJ...816...43S}). For instance, the 12-M$_{\odot}$ progenitor has the lowest accretion rate, but also a low luminosity (Fig.\ref{fig:4}, top right panel) and does not explode. The sensitivity of explosion to the progenitor mass suggests that small differences in density profiles can be significant. 

We propose that the early accretion of the Si-O interface promotes explosion (see also, e.g. \citealt{2008ApJ...688.1159M}).  Looking at Fig.\,\ref{fig:1}, we see that those models that do explode (solid) have several characteristics in their density profiles that distinguish them from models that do not explode (dashed). Namely, the density is quite high in the interior, out to 1.2 M$_{\odot}$, but then drops sharply. Furthermore, the exploding models all have a Si-O interface located interior to 1.7 M$_{\odot}$, where the density drops by a factor of several over a thin mass region. Our default 20- and 25-M$_{\odot}$ models also have Si-O burning interfaces, but these are located further out. The 12-, 13-, and 15-M$_{\odot}$ models do not feature prominent interfaces. None of these five models explodes during the physical time they were followed here, and all five models feature mean shock radii stagnating at 100 km, 700 milliseconds post-bounce. The variation in outcome $-$ explosion or failure $-$ over progenitors differing only by a solar mass suggests that certain density profiles are most prone to explosion, and that early accretion of the Si-O interface can be one key to explosion. In fact, looking at Fig.\,\ref{fig:1}, we see that all four exploding models accrete Si-O interfaces early on, as indicated by the dip in accretion rates. Comparing with the shock radius evolution depicted in Fig.\,\ref{fig:2}, we find explosion occurring soon after interface accretion. 

In Fig.\,\ref{fig:3}, we simultaneously show accretion rates at 500 km (in M$_{\odot}$ s$^{-1}$, blue), the evolution of the Si-O interface (in km, red), and the maximum shock radius (in km, black) as a function of time after bounce (in seconds) for our four exploding models. Once the Si-O interface passes 500 km, at roughly 80 ms for the 21-M$_{\odot}$ progenitor, the corresponding accretion rate falls by a factor of five, from $\sim$10 to $\sim$2 M$_{\odot}$ s$^{-1}$. Simultaneously, the maximum shock radii begins to climb towards explosion once it intersects the Si-O interface. The high accretion rates prior to interface accretion enhance the accretion luminosity interior to the stalled shock. The subsequent drop in accretion rate reduces the ram pressure of the infalling material exterior to the stalled shock, while still allowing the stalled shock to benefit for a short time interval from the high luminosity due to earlier accretion. This paints a coherent picture of the critical role of the interface in explosion.

The role of the Si-O interface has been studied earlier in literature. Using BETHE-hydro, \cite{2008ApJ...688.1159M} explored the evolution of the 11.2- and 15-M$_{\odot}$ progenitors, and found that accretion rates plummet following accretion of the Si-O interface. \cite{2013ApJ...770...66H} evolve a 27-M$_{\odot}$ progenitor in both 2D and 3D and find strong shock expansion ensues after Si-O infall due to the resulting decrease in mass accretion. \cite{2018ApJ...852...28S} find that explosion for an artificially-rotating 15-M$_{\odot}$ progenitor follows shortly after accretion of the Si-O interface, but argue that a strong spiral SASI mode has set the grounds for explosion even earlier. Using Zeus-2D for 12-100 M$_{\odot}$ progenitors, \cite{2016ApJ...816...43S} similarly found shock expansion associated with the density, and hence ram pressure, jump  around the Si-O interface. However, shock expansion lasted briefly and the models did not explode. Recently, \cite{2017arXiv171201304O} performed 3D simulations using the GR multi-group radiation hydrodynamics code ZELMANI. Omitting inelastic scattering processes and velocity dependence and with 12 energy groups spaced logarithmically between 1 and 248 MeV, they find similar results regarding the role of Si-O interface accretion to prompting explosion, perhaps even more critical than the compactness parameter. Similarly, for a suite of 2D progenitors from 11-28 M$_{\odot}$, \cite{2017IAUS..329..449S} find a steep drop in density at the Si-O interface corresponds to a reduction of the accretion ram pressure and subsequent strong shock expansion.

\begin{figure*}
\subfigure{\includegraphics[width=0.5\textwidth]{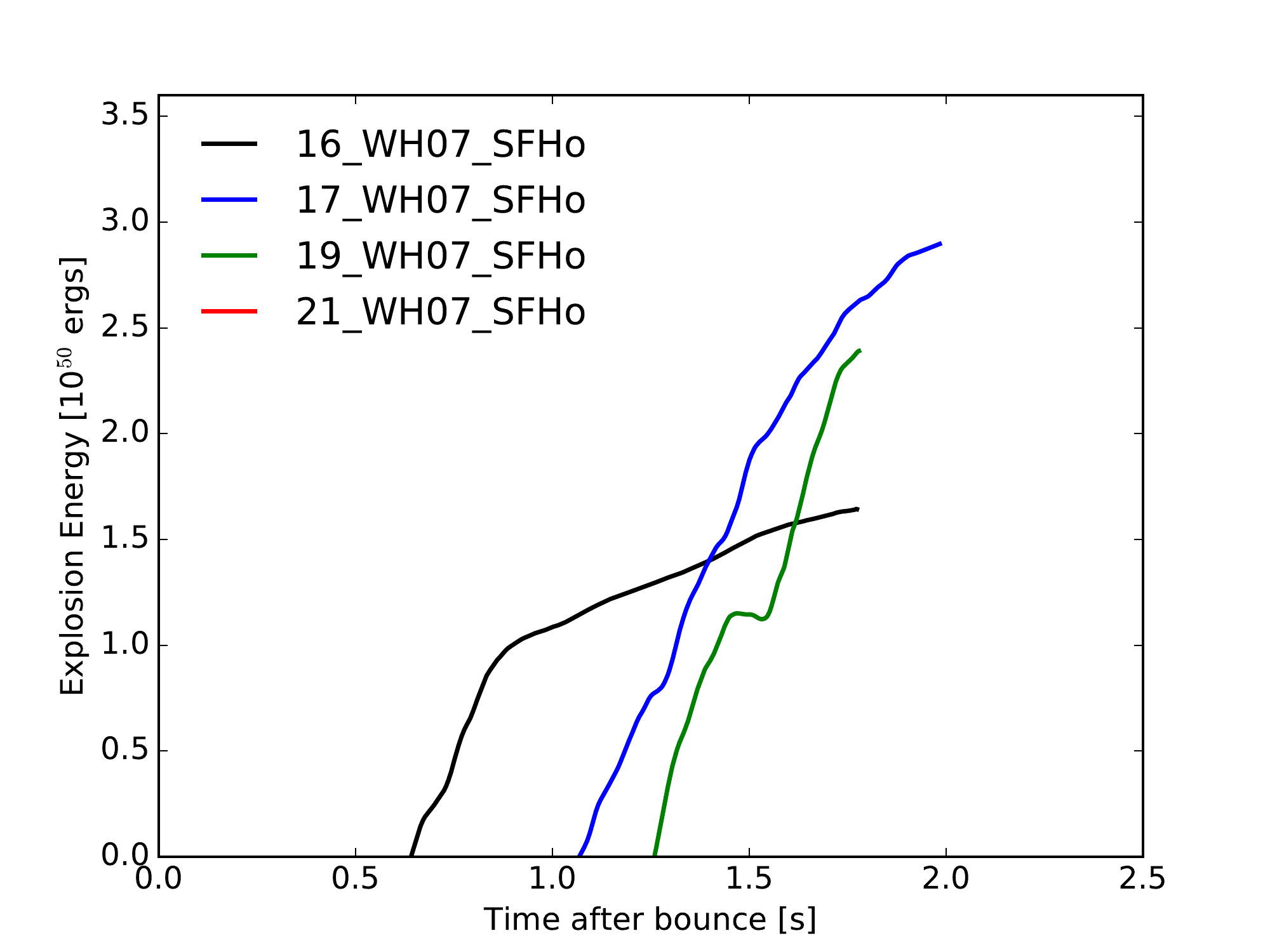}}\hfill
\subfigure{\includegraphics[width=0.5\textwidth]{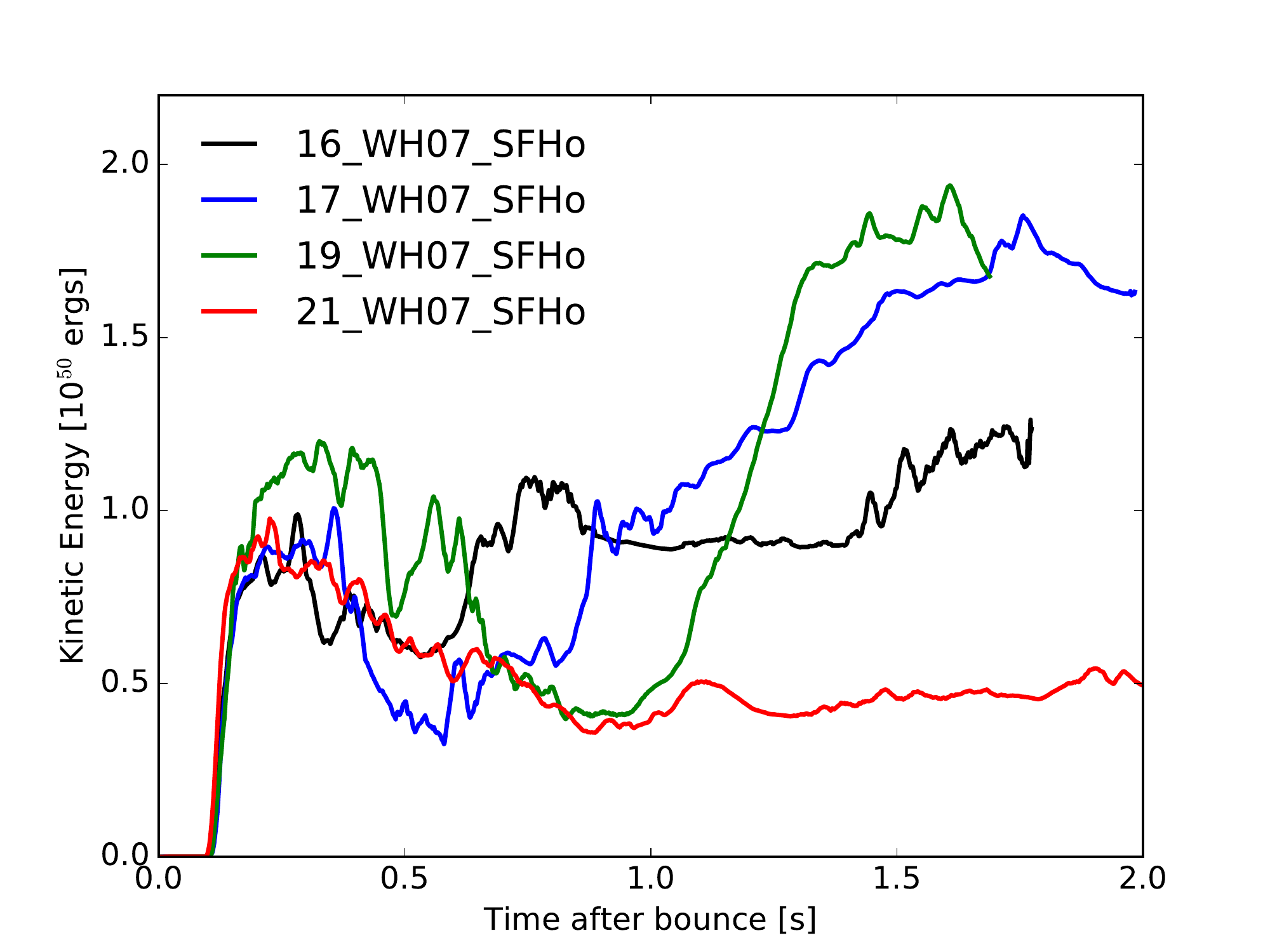}}
\subfigure{\includegraphics[width=0.5\textwidth]{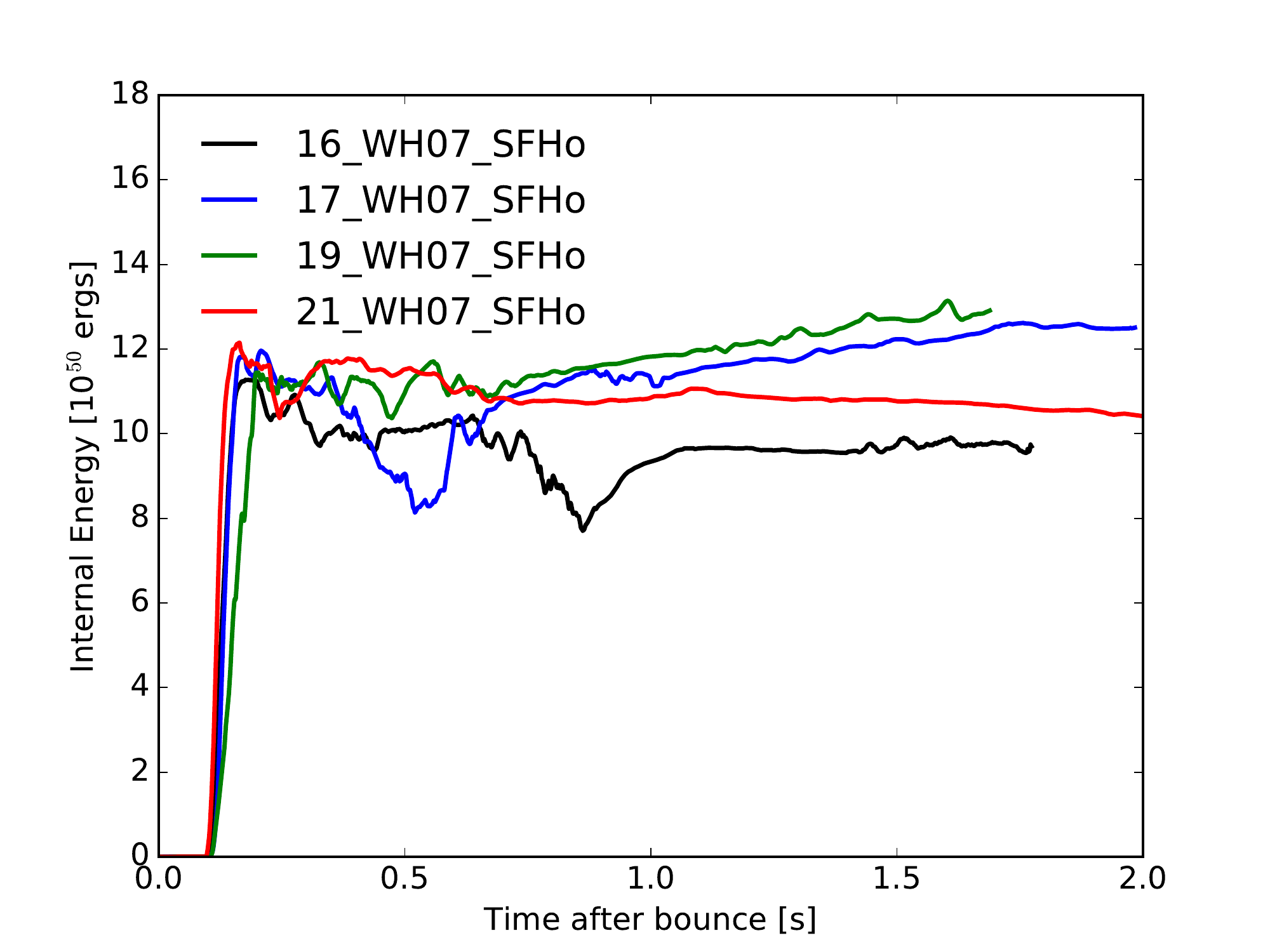}}\hfill
\subfigure{\includegraphics[width=0.5\textwidth]{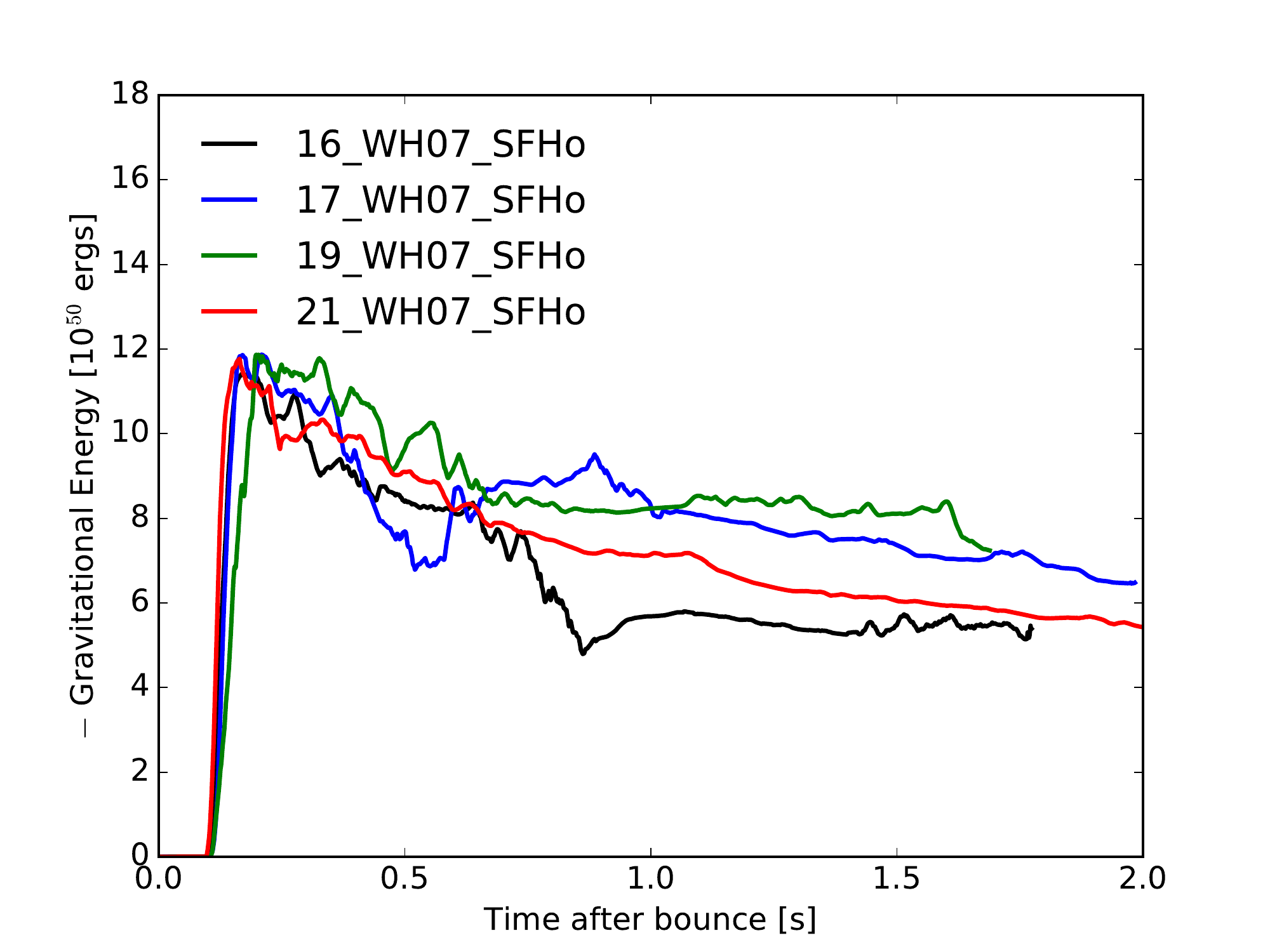}}
\caption{Explosion energies (in units of 10$^{50}$ ergs) against time after bounce (in seconds) of the ejecta (defined as energetically unbound material) for the four exploding models. We plot the total explosion energy (\textbf{top left}), kinetic energies (\textbf{top right}), internal energies (\textbf{bottom left}), and gravitational binding energies (\textbf{bottom right})  multiplied by negative one to be positive. The explosion energy is defined in the text and includes the gravitational binding energy of the envelope exterior to our grid at 20,000 km. Note that, during the simulation, the 21-M$_{\odot}$ progenitor never has a positive explosion energy, despite being the first to explode. All energies are still rising at the end of the simulation, with the 16-M$_{\odot}$ progenitor having the slowest rate of increase. The 17-M$_{\odot}$ progenitor reaches the highest explosion energy, of 3$\times$10$^{50}$ ergs at the end of the simulation. Both the 17- and 19-M$_{\odot}$ models rise rapidly in kinetic energy after 0.8 seconds post-bounce, corresponding to a sharp increase in total energies. By comparison, both the internal energies and gravitational energies are relatively flat with time. Because we truncated our simulations when the shock radius neared the grid edge, with energies still rapidly rising, we highlight the significance of performing simulations over larger radial domains.}
\label{fig:5}
\end{figure*}

Literature is replete with analytical parametrizations of explosions, spanning the compactness (\citealt{2011ApJ...730...70O}; \citealt{2013ApJ...762..126O}; \citealt{2015PASJ...67..107N}) and Ertl parameters (\citealt{2016ApJ...818..124E}), the antesonic condition (\citealt{2012ApJ...746..106P}; \citealt{2018arXiv180102626R}); critical luminosity curves (\citealt{1993ApJ...416L..75B}; \citealt{2016ApJ...825....6S}; \citealt{2018ApJ...852...28S}) to scaling relations (\citealt{2016MNRAS.460..742M}) and integral conditions (\citealt{2017ApJ...834..183M}). We propose early accretion of the Si-O interface as one possible empirical condition.

\section{Energetics and Diagnostics}\label{sec:results}

Here, we present an analysis of the energetics of our results, explode the failed explosions of the prior section, and comment on the Y$_e$-mass ejecta distribution and the absence of the LESA.

In Fig.\,\ref{fig:4}, we depict the lab-frame neutrino luminosities (in 10$^{52}$ erg s$^{-1}$, top panel) and RMS energies (in MeV, bottom panel) as a function of time after bounce (in seconds), evaluated at 500 km and redshifted out to infinity for our nine progenitors. We assume a forward-peaked radial neutrino distribution as in \cite{2017ApJ...850...43R}. At early times, RMS energies and luminosities are monotonic with progenitor mass for the non-exploding progenitors, with the 12-M$_{\odot}$ progenitor fielding neutrino luminosities 50\% smaller and RMS energies 30\% smaller than its more massive counterparts. Interestingly, the models that explode later reach higher post-breakout luminosities. Note that the exploding models show an expected dip in luminosities and RMS energies after explosion reverses accretion. Electron-type neutrino luminosities asymptote to $\sim$2$\times$10$^{52}$ ergs s$^{-1}$ by $\sim$1 second post-bounce for the exploding models.

\begin{figure*}
\centering
\includegraphics[width=\textwidth]{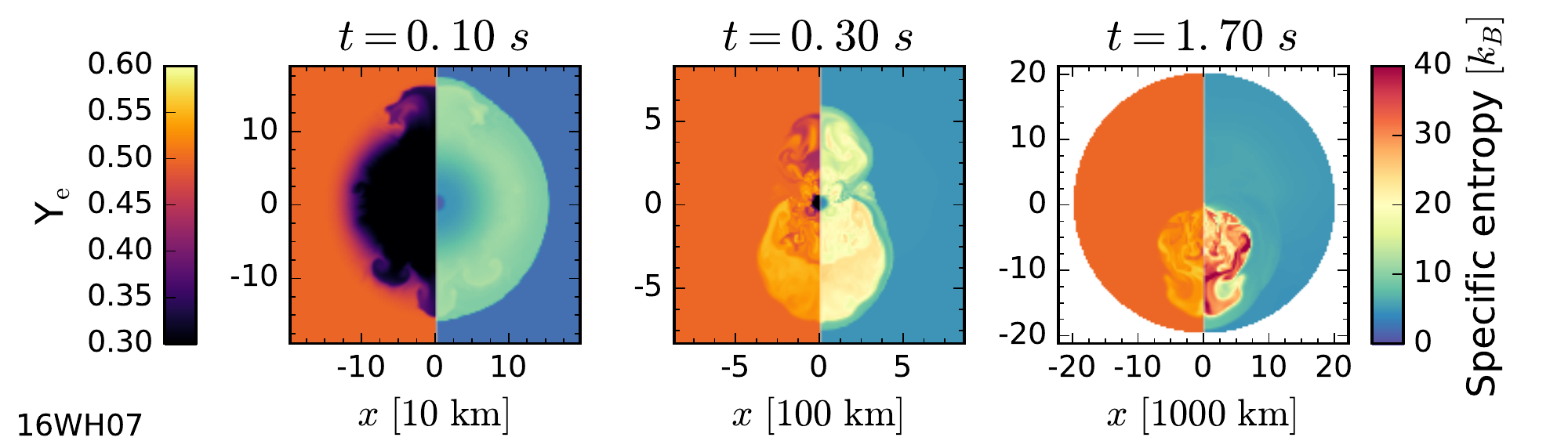}
\hfill
\includegraphics[width=\textwidth]{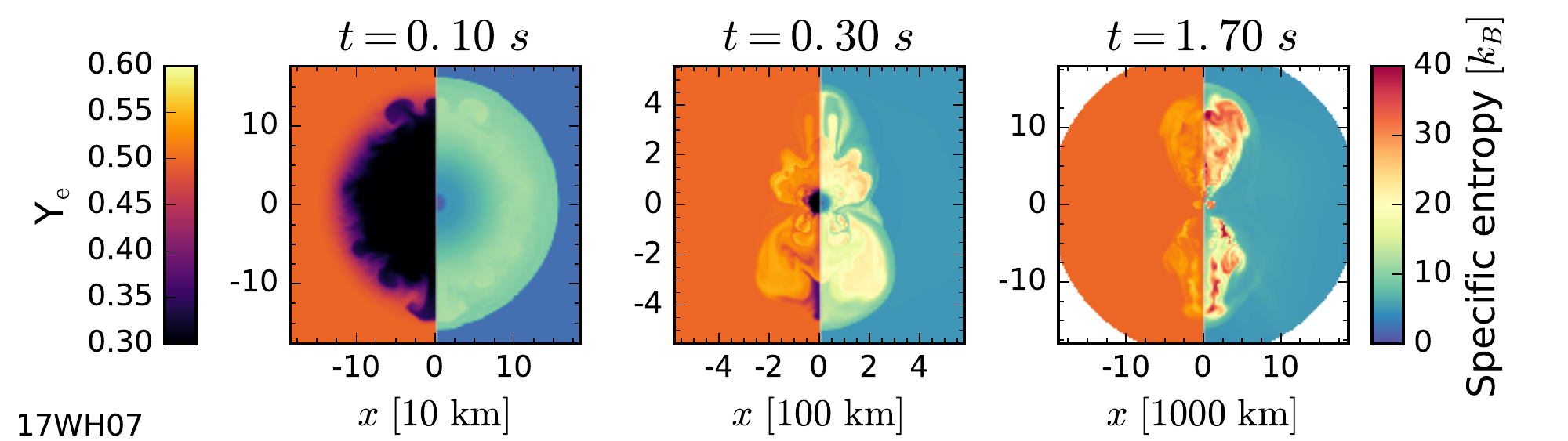}
\hfill
\includegraphics[width=\textwidth]{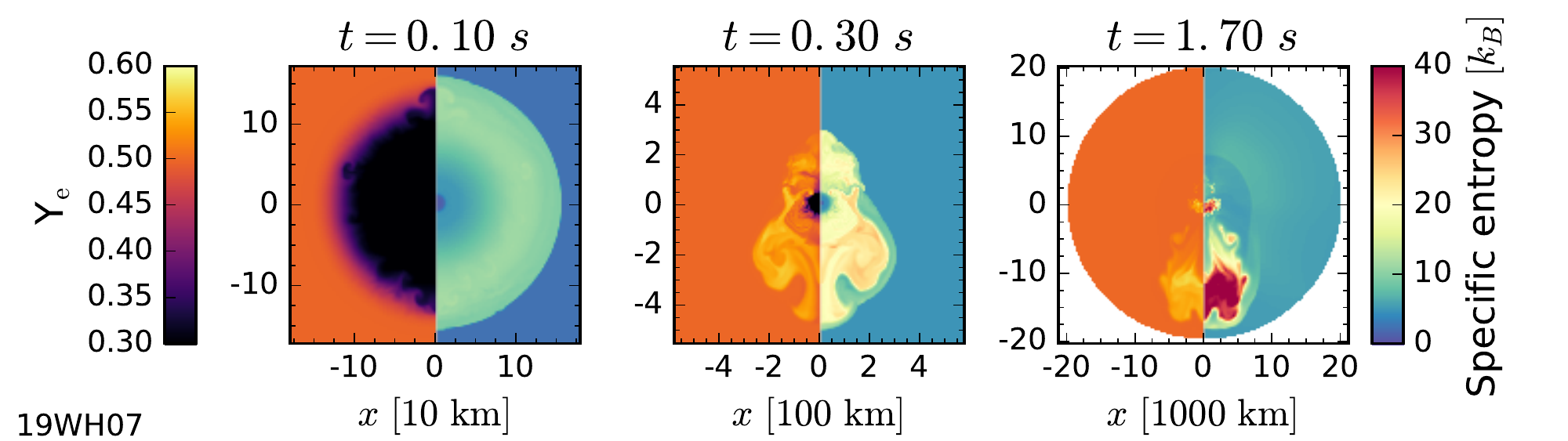}
\hfill
\includegraphics[width=\textwidth]{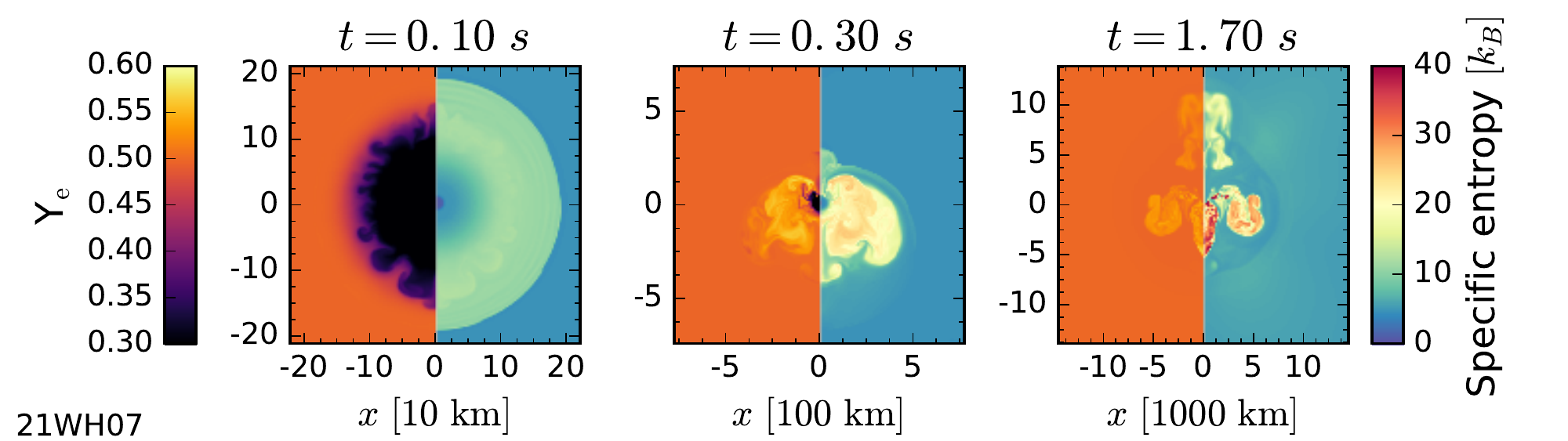}
\hfill
\caption{Ye (\textbf{left}) and specific entropy ($k_B$/nucleon, \textbf{right}) snapshots of the four exploding progenitors at 100, 300 and 1700 milliseconds post-bounce. As early as 100 ms post-bounce, we see nascent convection in the proto-neutron star. At late times, all explosions are very asymmetric. We find that models with multiple wide plumes have greater explosion energies than those localized in a single hemisphere. Because the two-dimensional nature of the simulation may artificially promote axial anisotropies, we presume that 3-D simulations will produce more isotropic explosions with consequently greater explosion energies.}
\label{fig:6}
\end{figure*}

\begin{figure}
\centering
\includegraphics[width=\columnwidth]{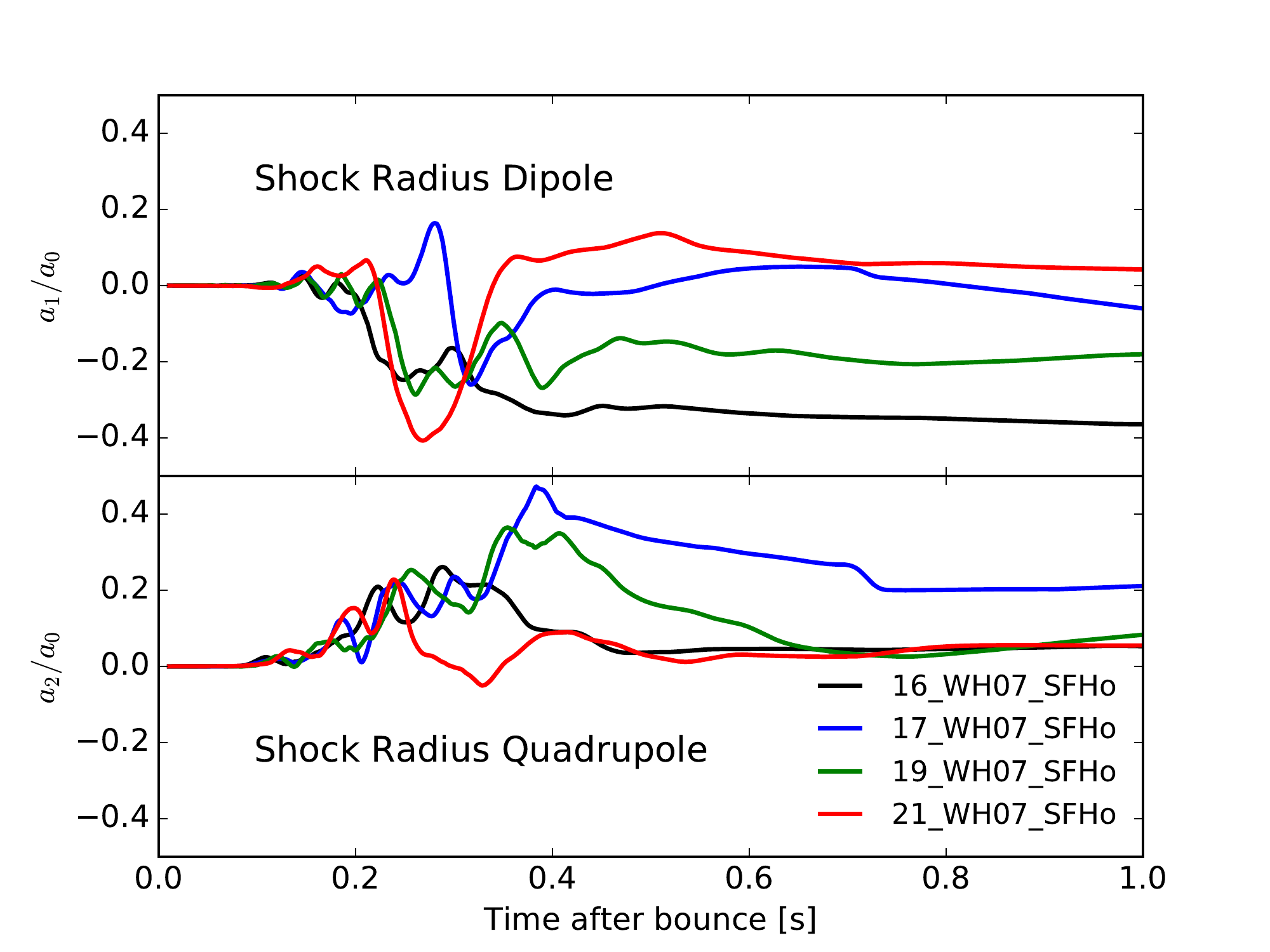}
\caption{Dipole (\textbf{top panel}) and quadrupole (\textbf{bottom panel}) moments of the shock radii, normalized to the monopole moments and plotted against time after bounce (in seconds) for our four default exploding models. Note the presence of a strong dipole moment for the different models. Furthermore, all the models $-$ with the possible exception of the 21-M$_{\odot}$ progenitor $-$ have positive quadrupole moments, corresponding to equatorial pinching, with the 21-M$_{\odot}$ model showing equatorial wings as seen in the entropy profiles in Fig.\,\ref{fig:6}.}
\label{fig:7}
\end{figure}

We calculate diagnostic energies (\citealt{2012ApJ...756...84M}) as the sum of the kinetic, internal, gravitational binding and nuclear binding energy interior to the grid and correct for the gravitational binding energies exterior to our 20,000 km grid. We list the binding and final energies in Table\,\ref{table:1}, with the latter summed over all zones where the final energy is positive.

The final explosion energies, with external binding energies subtracted, are illustrated in the top-left panel of 
Fig.\,\ref{fig:5} as a function of time after bounce for the four exploding progenitors. The 17-M$_{\odot}$ progenitor yields the highest diagnostic energy of $\sim3$ $\times$10$^{50}$ ergs at 2.0s after bounce. Its energy is still steeply rising at the end of the simulation, which ends once the outer shock radius reaches the grid edge at 20,000 km. This suggests the need to repeat such simulations over much larger radial domains and is consistent with the results found by \cite{2016ApJ...825....6S}, who performed simulations over a 10,000 km grid and found explosion energies still rising for their suite of 12-, 15-, 20- and 25-M$_{\odot}$ WH07 progenitors. Note, however, that they plot the diagnostic energy and do not correct for the gravitational binding burden off the grid, which would result in much smaller and even negative final energies for them. For the same progenitor suite, \cite{2016ApJ...818..123B} find final explosion energies corrected for the gravitational overburden $-$ nearly an order larger, around 1 Bethe (1B = 10$^{51}$ ergs). The 21-M$_{\odot}$ progenitor does not reach positive explosion energy by the end of our simulation since it has not yet overcome its gravitational overburden. Note also that the explosion energies anti-correlate with explosion times $-$ the models that explode later have higher energies. The same behavior appears for the luminosities in Fig.\,\ref{fig:6}, suggesting that a delayed explosion is more energetic. Interestingly, though the 16- and 17-M$_{\odot}$ progenitors explode almost simultaneously, their explosion energies are quite different, with the former asymptoting by the end of the simulation. 

In the top right panel, we show the kinetic energies which, early after bounce, are only a fraction of the internal energy. At late times, the kinetic energies of the 17- and 19-M$_{\odot}$ progenitors rise steeply, paralleling the larger explosion energy at late times for these models and accounting for roughly two-thirds of the final energies. Furthermore, from the entropy profiles in Fig.\,\ref{fig:6}, we see that exploding models with multiple plumes covering a wider spread of solid angle have higher explosion energies. The 16-M$_{\odot}$ progenitor has an asymmetric explosion concentrated around the southern pole at late times, and has a correspondingly smaller explosion energy. On the other hand, the 17- and 19-M$_{\odot}$ progenitors have multiple wide plumes in both hemispheres with correspondingly higher explosion energies. The rapid rise in kinetic energy together with multiple expanding plumes, which drive this kinetic outflow, suggest that the morphology of the unbound material is significant in producing robust explosion energies. Furthermore, simulations in 3D will not suffer from axial artifacts present in 2D; we thus expect more isotropic explosions in 3D with correspondingly higher explosion energies than in 2D. However, 3D simulations are required to draw consistent conclusions about explosion morphologies and energies. In the bottom right panel, we illustrate the (negative) gravitational binding energy interior to the grid, with magnitudes comparable to the internal energies. The 25-M$_{\odot}$ model does not have an exceptionally high interior binding energy; rather, a combination of low kinetic energy and high exterior binding overburden prevents its explosion energy from becoming positive. Its final energy at the end of our simulation is roughly $-7\times10^{49}$ ergs and rising.

\begin{figure}
\centering
\includegraphics[width=\columnwidth]{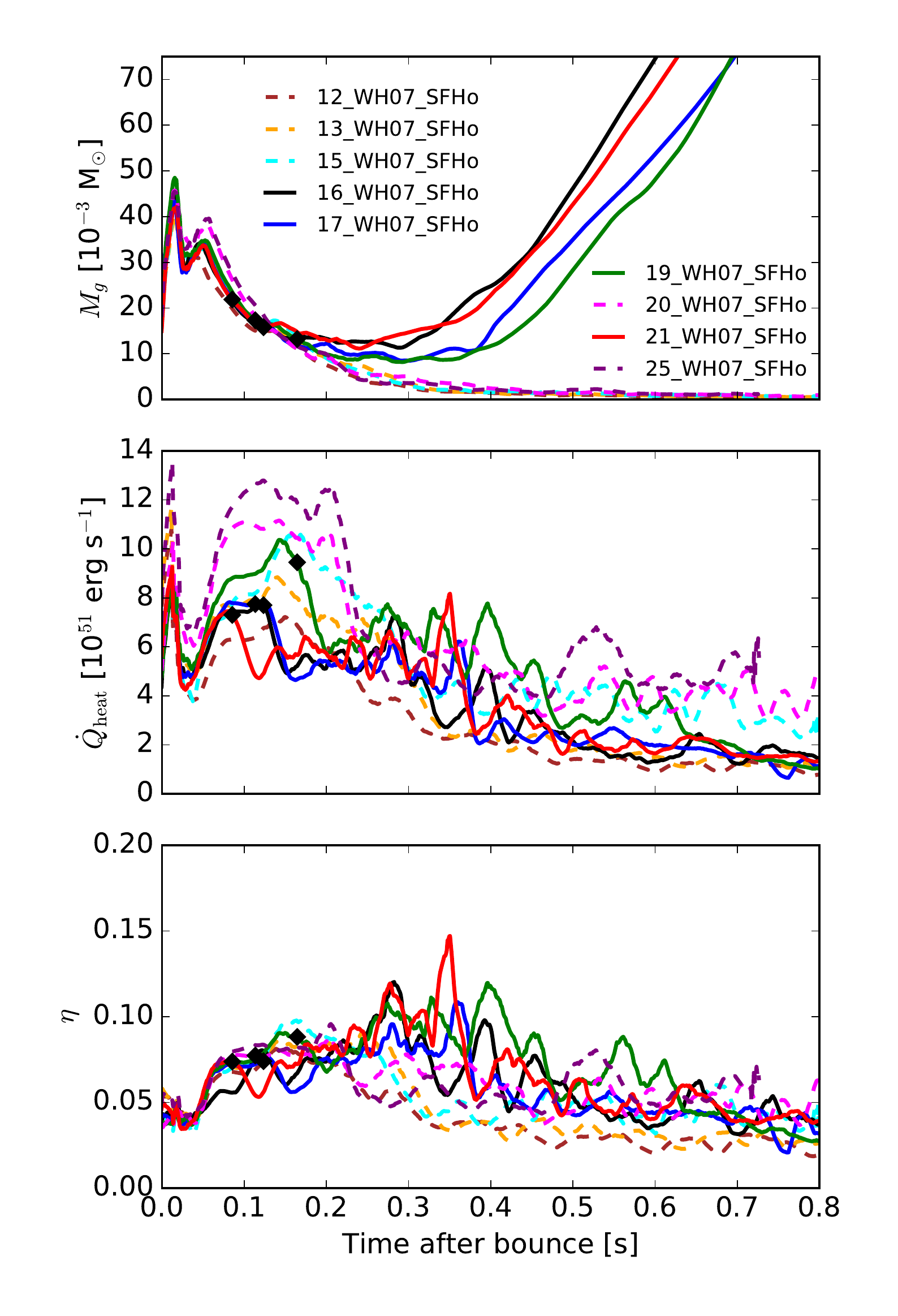}
\caption{\textbf{Top panel}: Gain mass (in 10$^{-3}$ M$_{\odot}$) as a function of time after bounce (in seconds). Through 100 ms post-bounce, the gain mass is similar for all the models. The gain mass then continues to rise for all the exploding models (solid lines), in order of explosion, and plummets for the non-exploding models (dashed). \textbf{Middle panel}: Heating rates (in 10$^{51}$ erg s$^{-1}$) plotted against time after bounce (in seconds) for the nine models. The heating rates of the non-exploding models increase with increasing mass. Though the 20- and 25-M$_{\odot}$ progenitors have the highest heating rates, they are unable to overcome the greater gravitational binding energy. However, among the exploding models, earlier explosion corresponds to a lower heating rate.  \textbf{Bottom panel}: Neutrino heating efficiency, $\frac{\dot{Q}_{heat}}{L_{\nu_e}+L_{\bar{\nu}_e}}$, of energy deposition. Note that the exploding models are distinguished by uniformly higher efficiency than the non-exploding models after explosion, as cooling fades.}
\label{fig:8}
\end{figure}

As a final point on morphology, in Fig.\,\ref{fig:7}, we plot the dipole (top panel) and quadrupole moments of the shock radii, normalized to the monopole moments and plotted against time after bounce (in seconds) for our four exploding models. All models feature a strong dipole moment, and with the possible exception of the 21-M$_{\odot}$ progenitor, have positive quadrupole moments, indicating equatorial pinching. Note that the 16-M$_{\odot}$ progenitor sustains a significant dipole moment even at late times and has a correspondingly smaller explosion energy. The 17-M$_{\odot}$ progenitor sustains a larger quadrupole moment at late times, and has a correspondingly larger explosion energy. These observations lend credence to our proposal that more isotropic explosions are more energetic, all else being equal.

\begin{figure*} 
\includegraphics[width=0.5\textwidth]{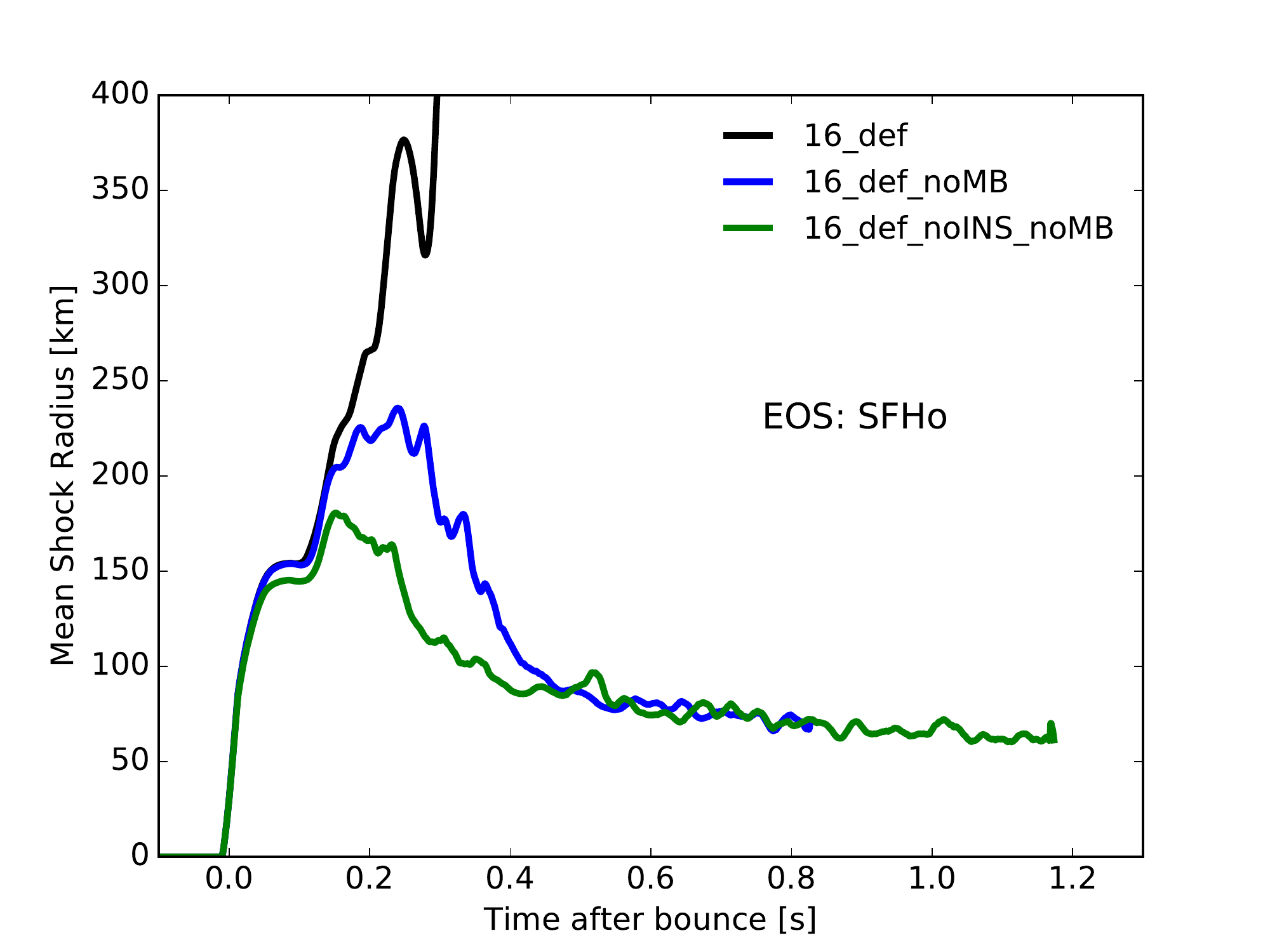}\hfill
\includegraphics[width=0.5\textwidth]{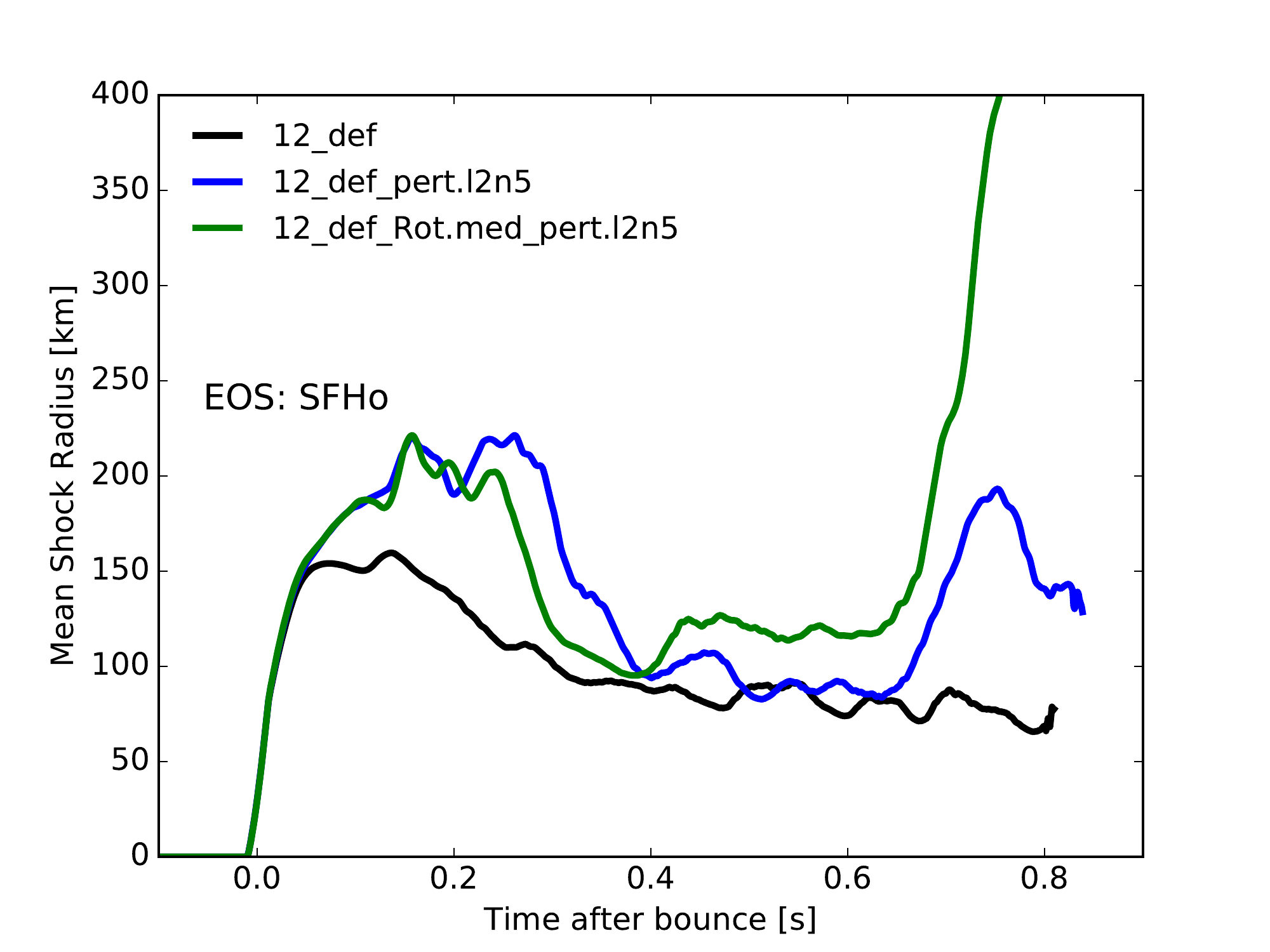}
\includegraphics[width=0.5\textwidth]{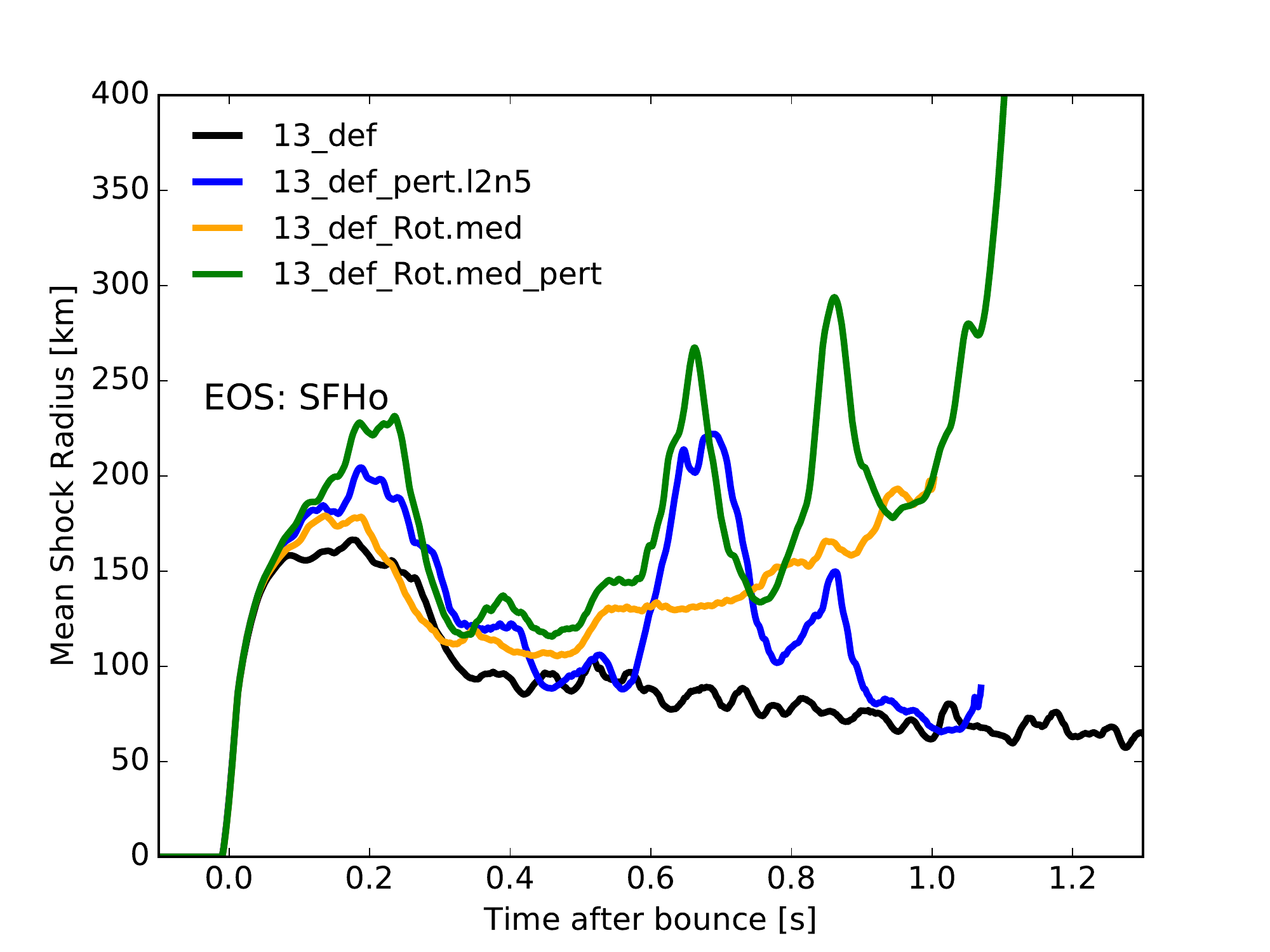}\hfill
\includegraphics[width=0.5\textwidth]{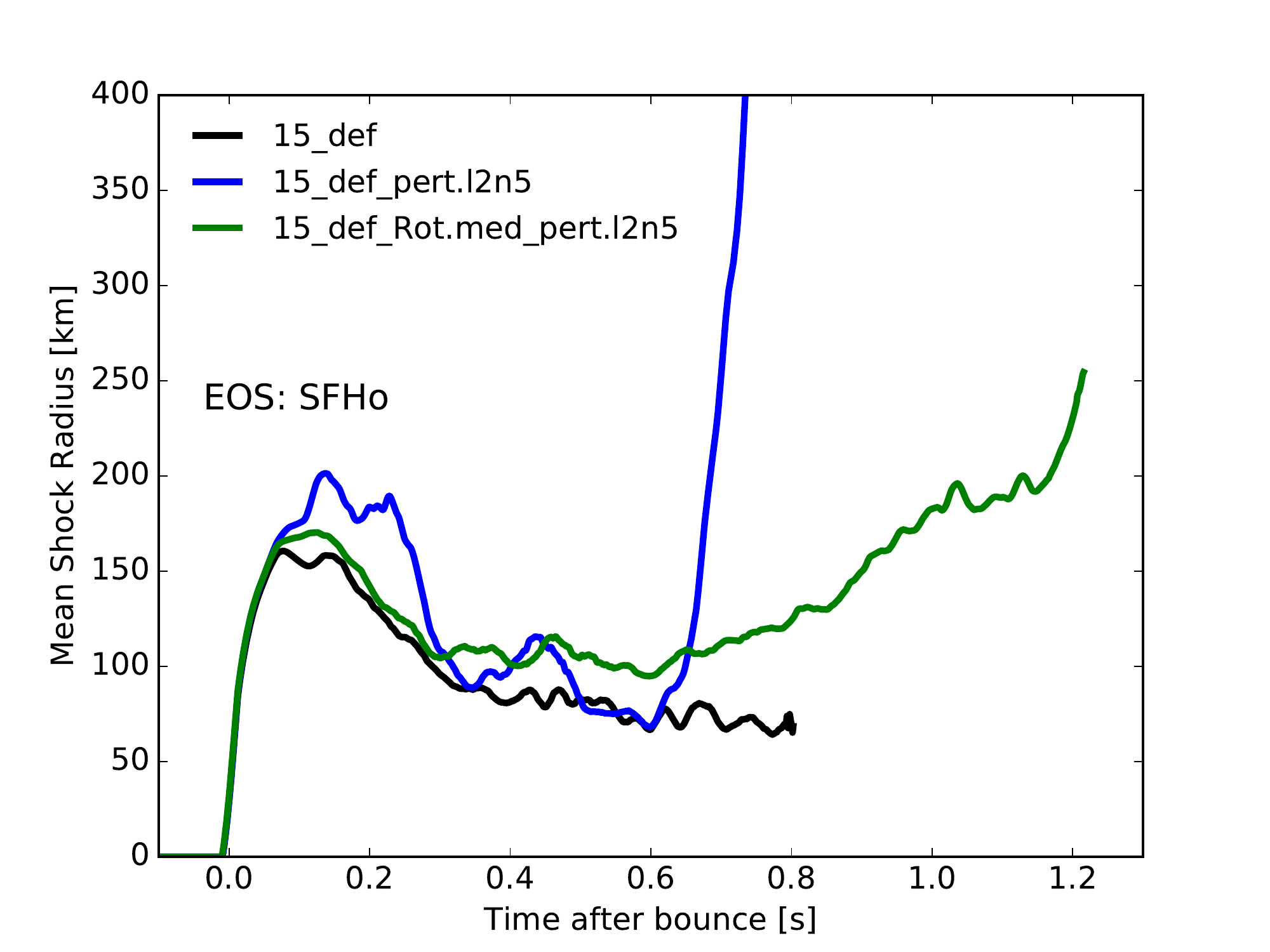}
\includegraphics[width=0.5\textwidth]{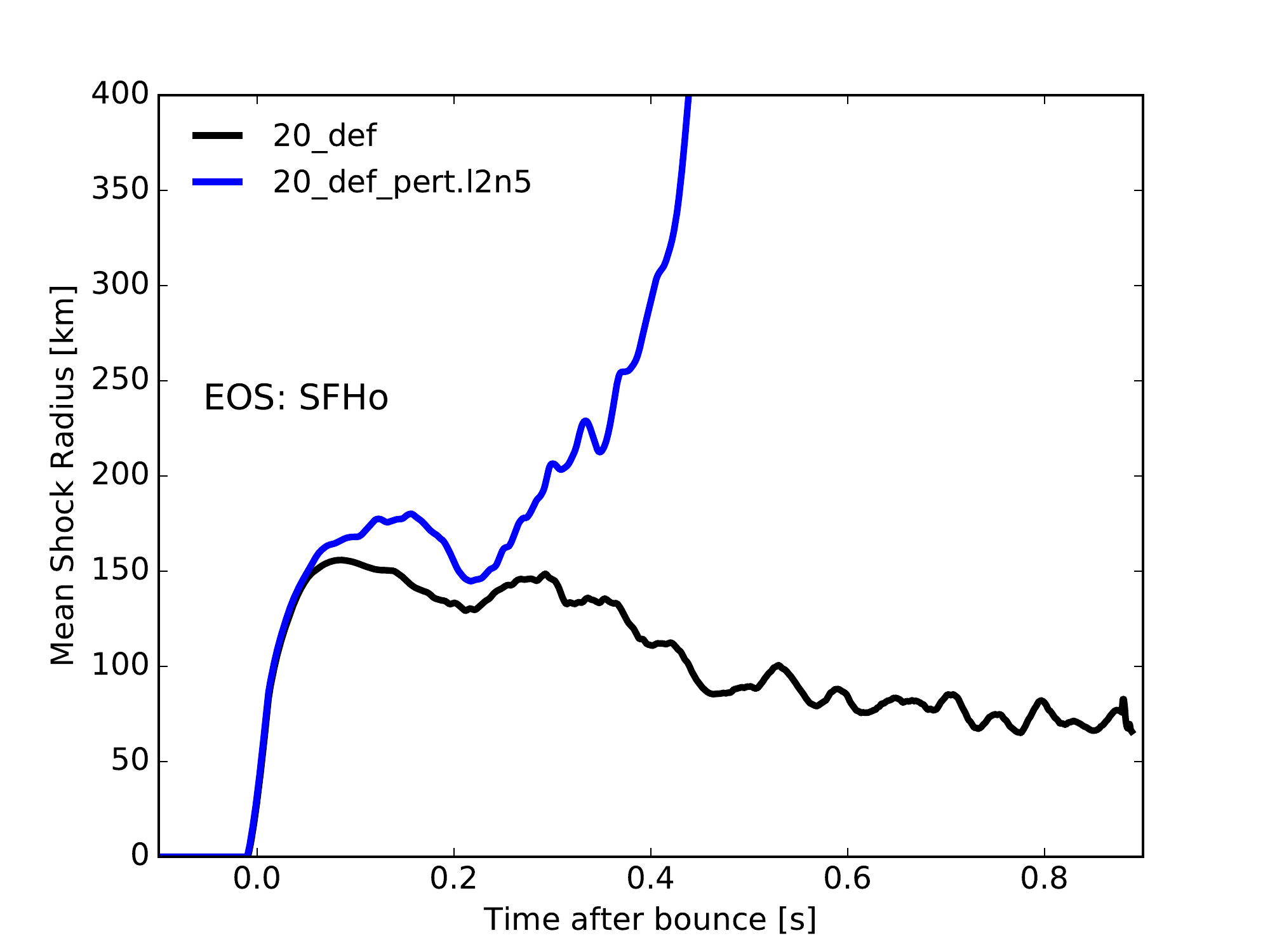}\hfill
\includegraphics[width=0.5\textwidth]{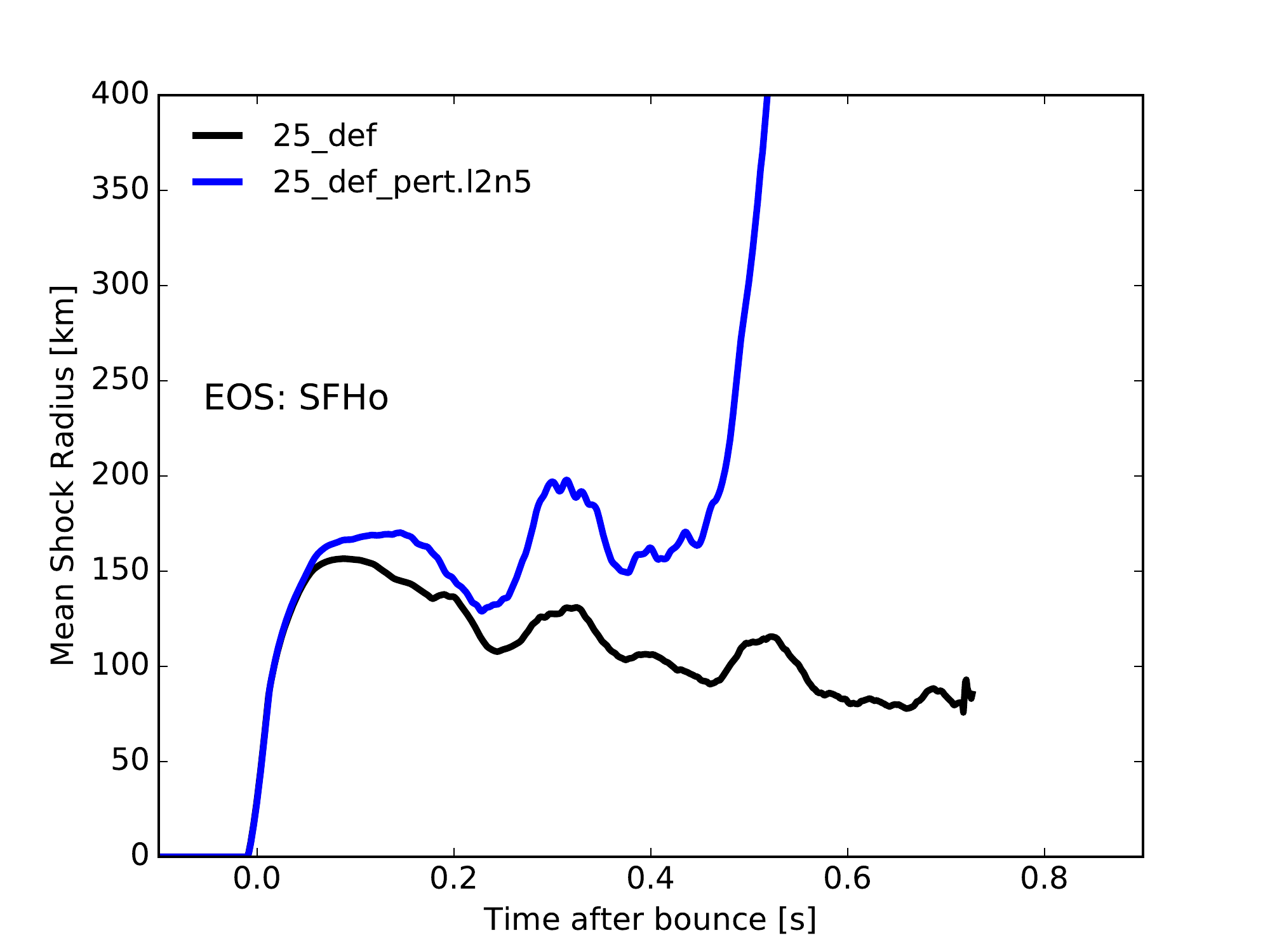}
\caption{Mean shock radius (in km) versus time after bounce (in seconds) illustrating the role of both microphysical (\textbf{top left}, 16-M$_{\odot}$) as well as macrophysical inputs (\textbf{remaining five panels}) in prompting explosion. For the latter, we show that the non-exploding models of Fig.\,\ref{fig:1} (12-, 13-, 15-, 20-, and 25-M$_{\odot}$) explode with perturbations to the infall velocities (indicated `pert.l2n5', blue) and moderate rotation (indicated with `Rot\_med', green). In the top left panel, we illustrate the role of inelastic scattering off nucleons (INS), as well as the neutral current many-body effect (MB), in driving the stalled shock radius further out and to explosion, respectively. However, these additions prove insufficient to explode the remaining five models shown. We follow \protect\cite{2015MNRAS.448.2141M} in implementing spherical harmonic perturbations to velocities on infall over three regions, extending to 6000 km.  All regions have $l=2, n=5$ and maximum velocity of 1000 km s$^{-1}$. We further implement modest cylindrical rotation (Rot\_med) where indicated (see 13-M$_{\odot}$ for a model with rotation alone, orange), with $\Omega_0 = 0.2$  radians s$^{-1}$ along the pole and a characteristic half-radius of 10,000 km. Note that, while rotation is essential for explosion (e.g., for the 12- and 13-M$_{\odot}$ progenitor), it delays explosion for the 15-M$_{\odot}$ progenitor.}
\label{fig:9}
\end{figure*}

\begin{figure*}
\includegraphics[width=0.5\textwidth]{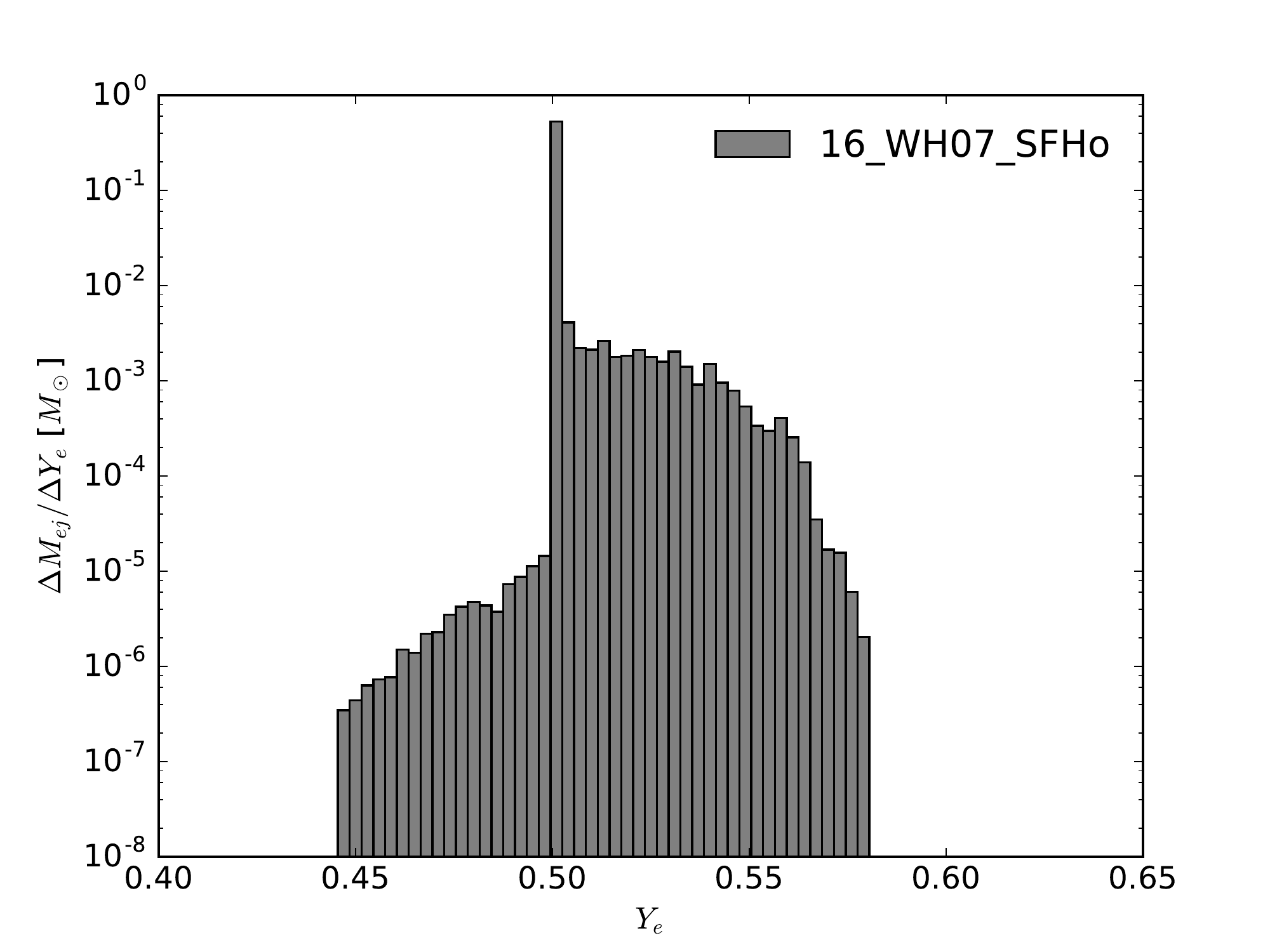}\hfill
\includegraphics[width=0.5\textwidth]{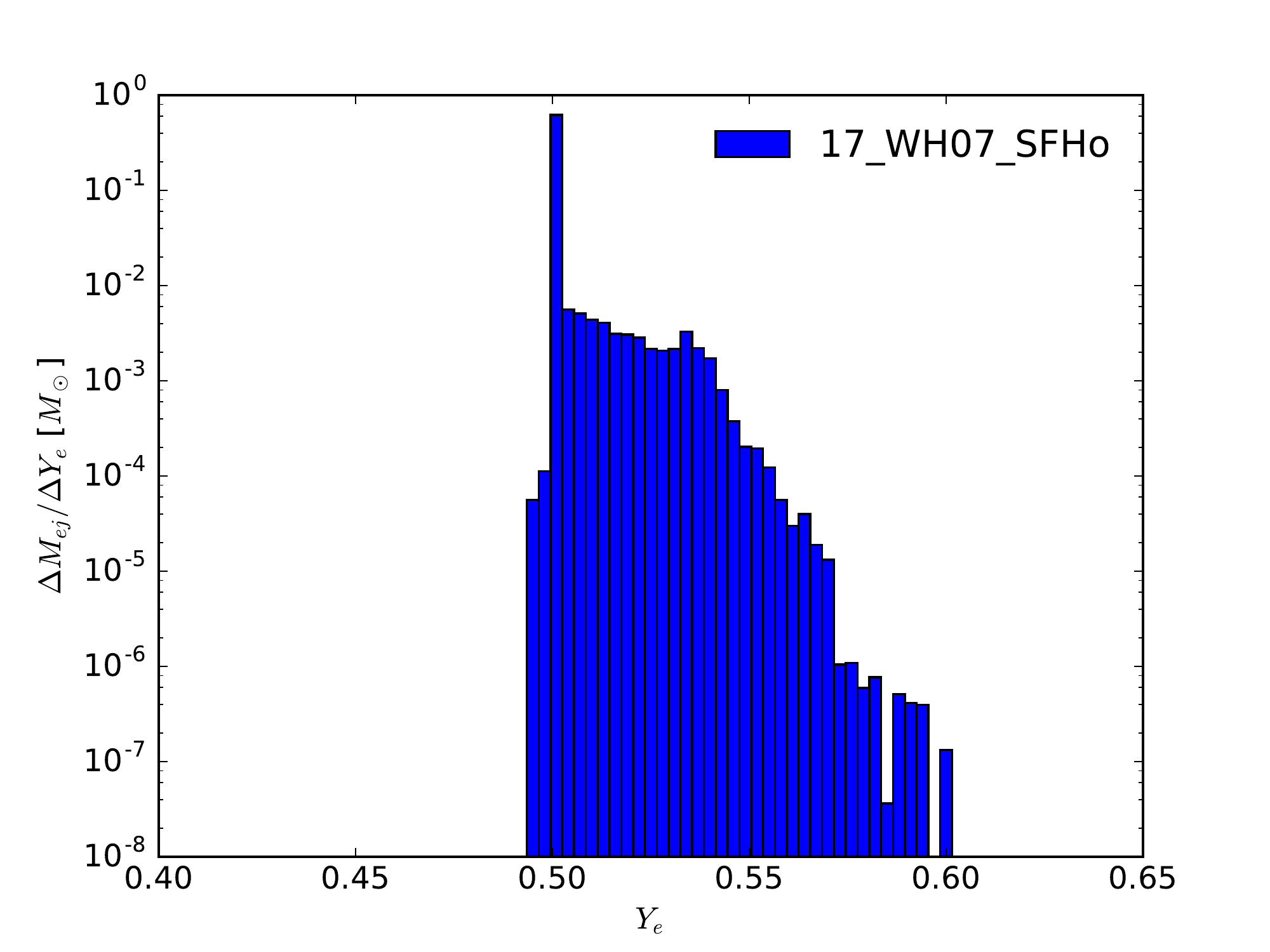}
\includegraphics[width=0.5\textwidth]{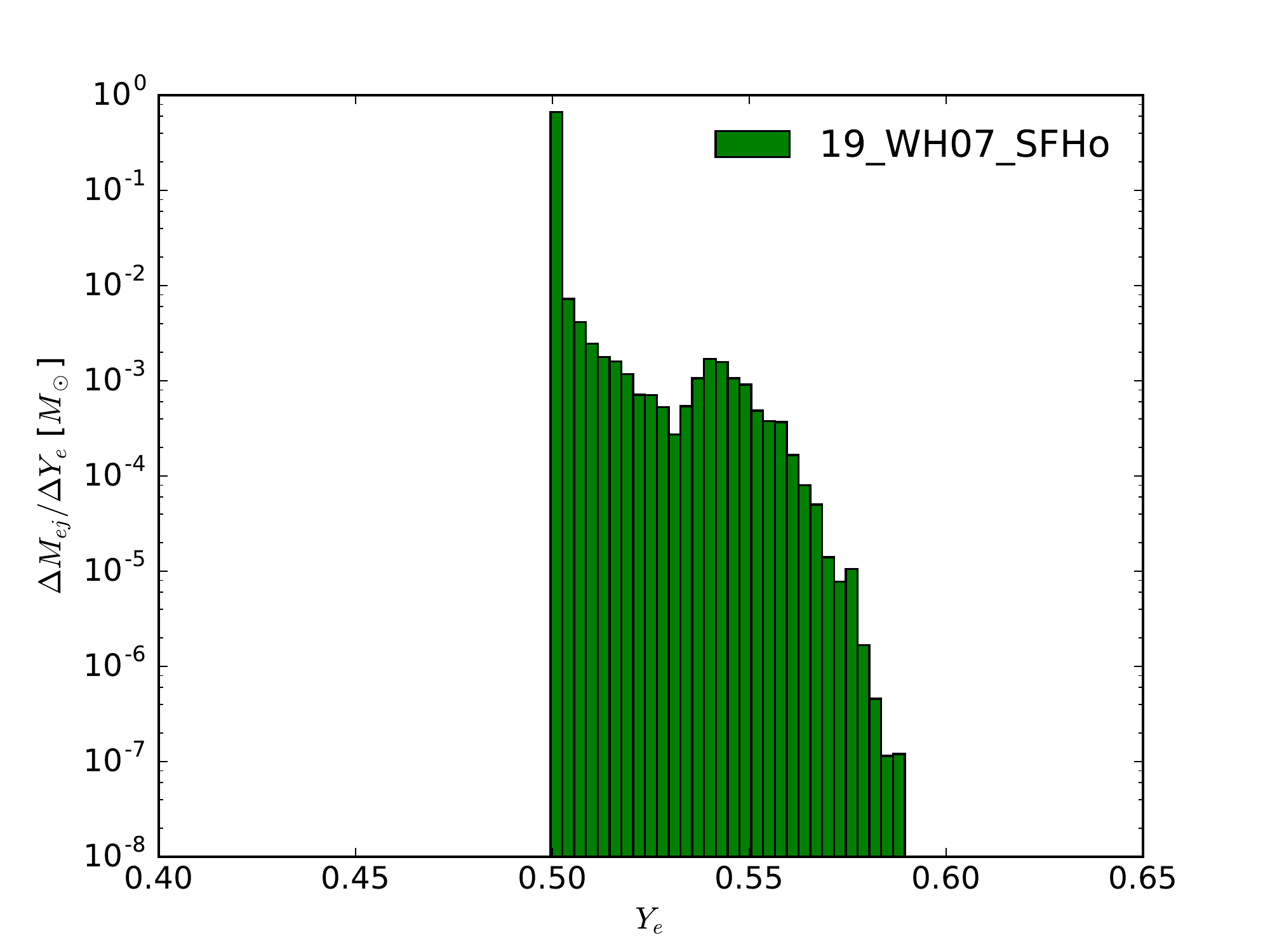}\hfill
\includegraphics[width=0.5\textwidth]{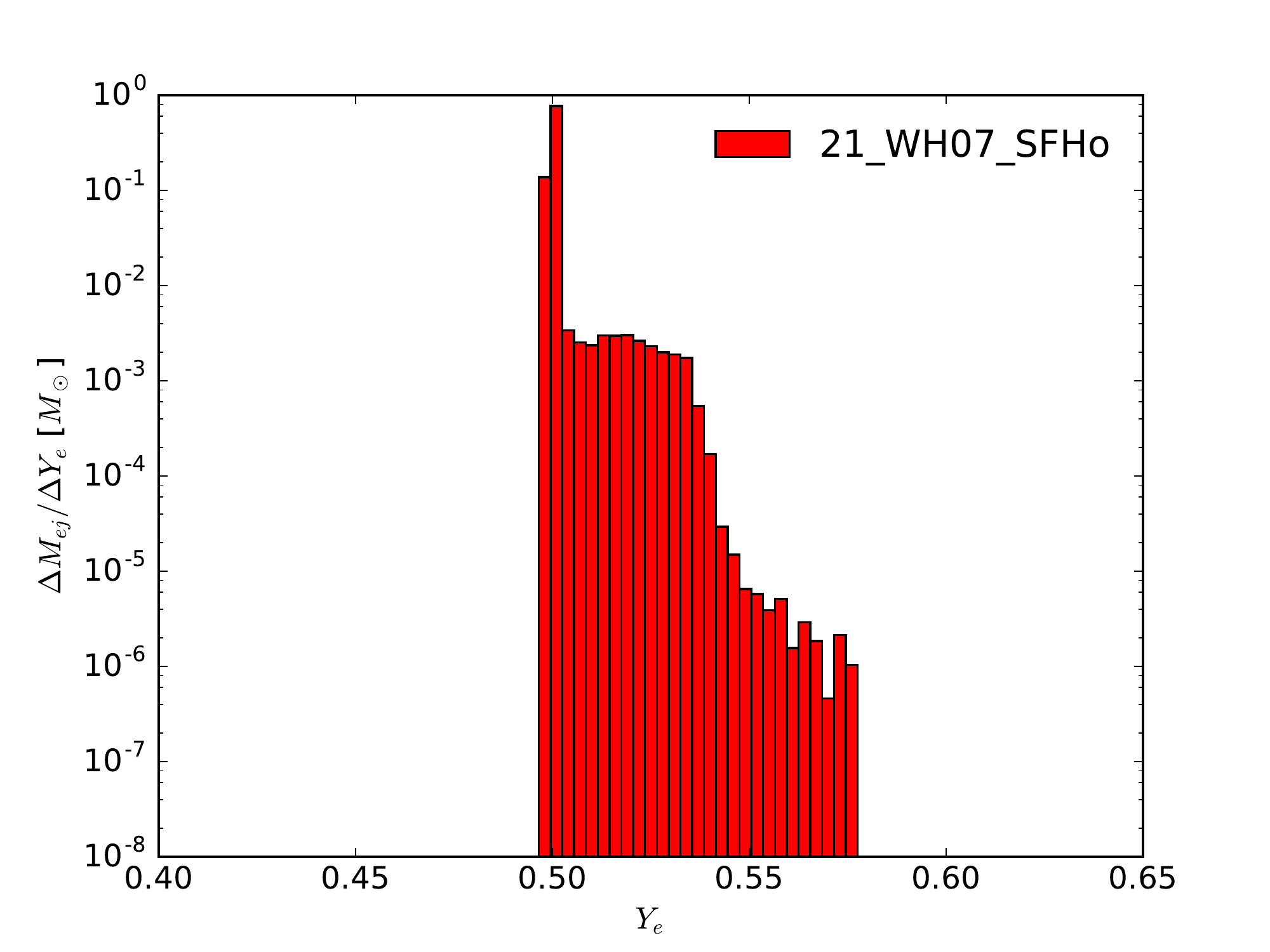}
\caption{Histograms illustrating the ejecta mass distribution function of Y$_e$ for our four default exploding models at the end of the evolution evaluated when the shock reaches the outer edge of the grid. The bins have width 0.003 Y$_e$. Surprisingly, all the models except the 16-M$_{\odot}$ show a similar distribution of Y$_e$, with a peak near 0.5 and distribution skewed towards larger values of Y$e$.}
\label{fig:10}
\end{figure*}

\begin{figure*}
\includegraphics[width=0.5\textwidth]{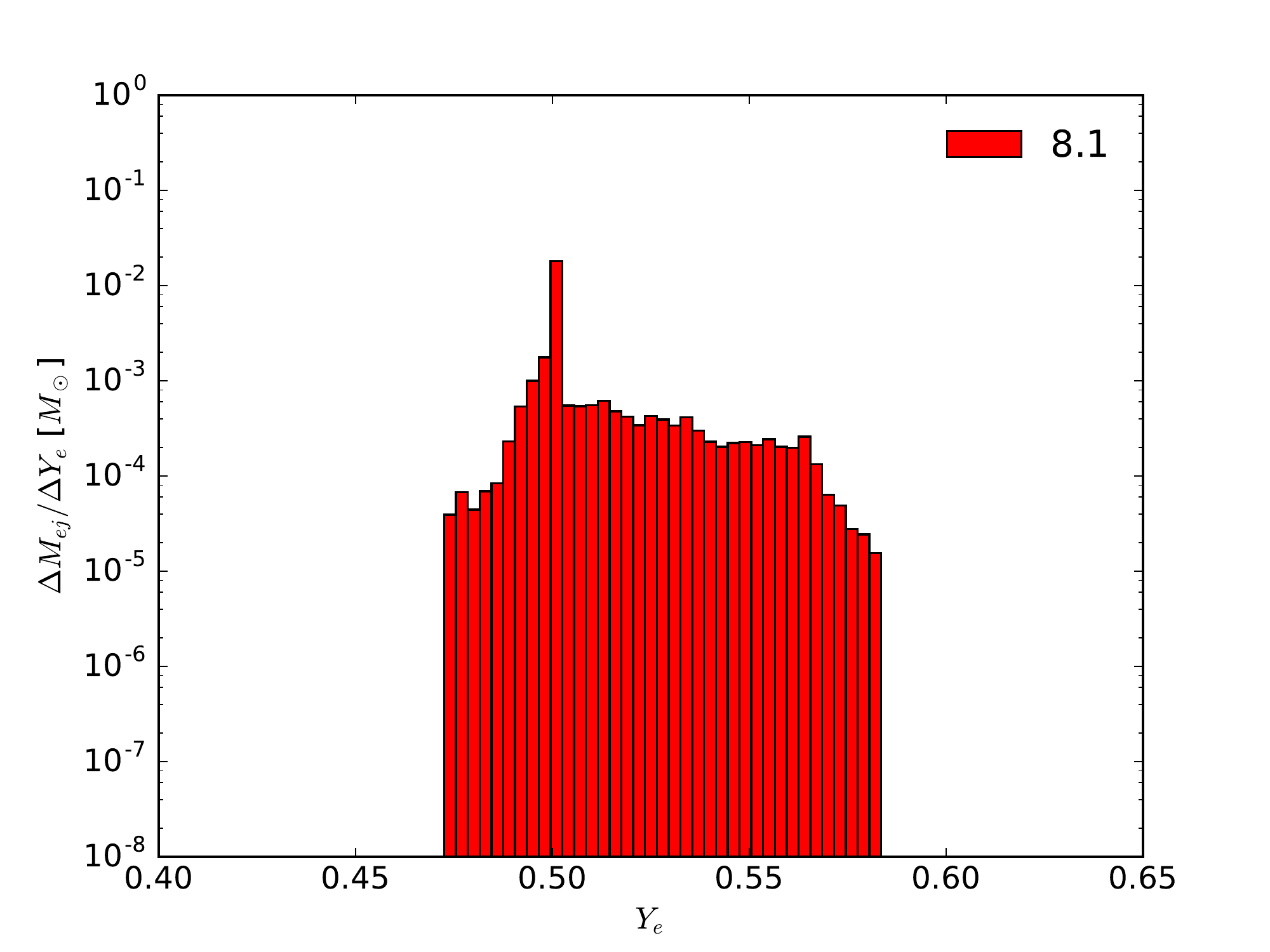}\hfill
\includegraphics[width=0.5\textwidth]{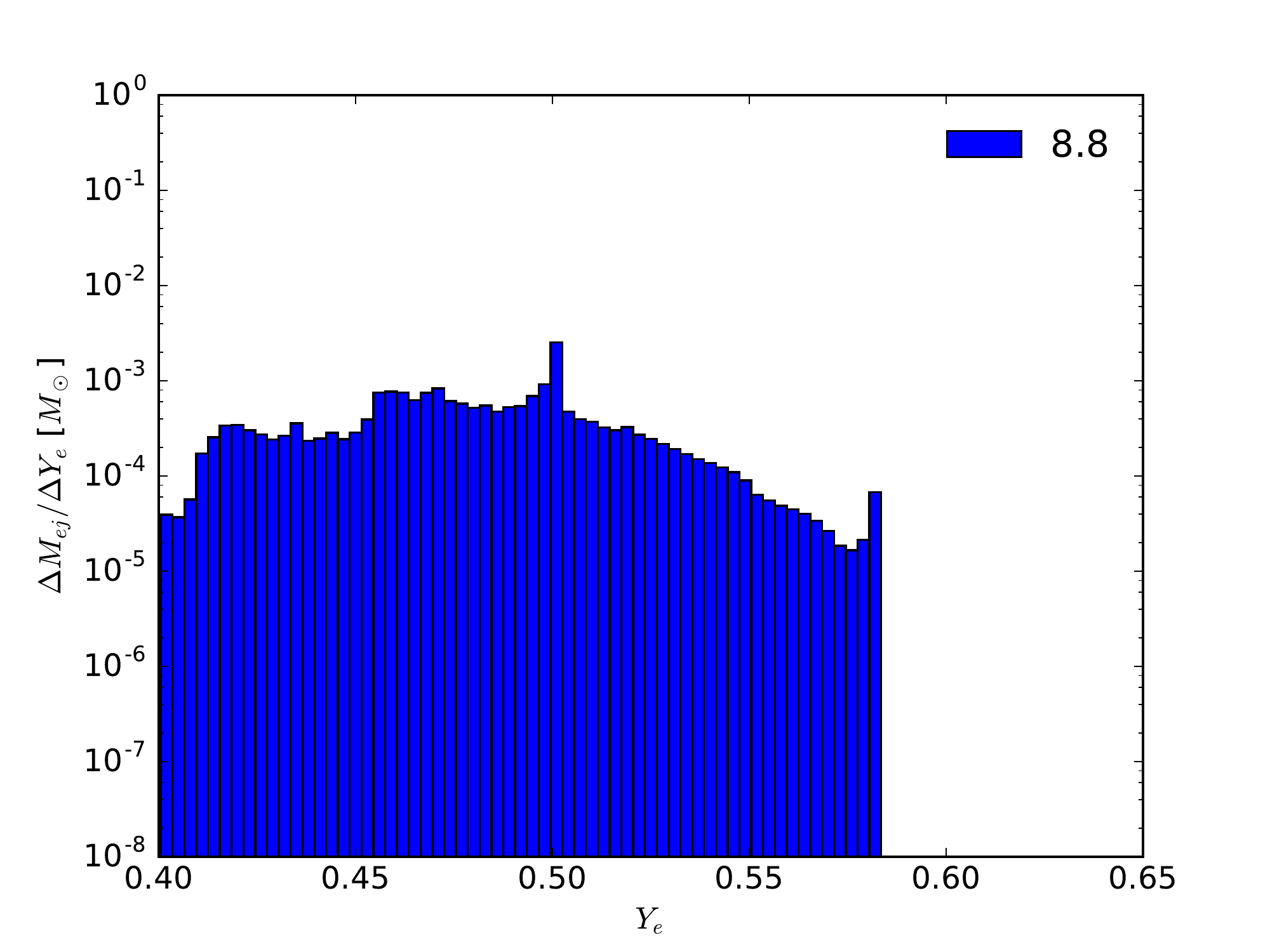}
\includegraphics[width=0.5\textwidth]{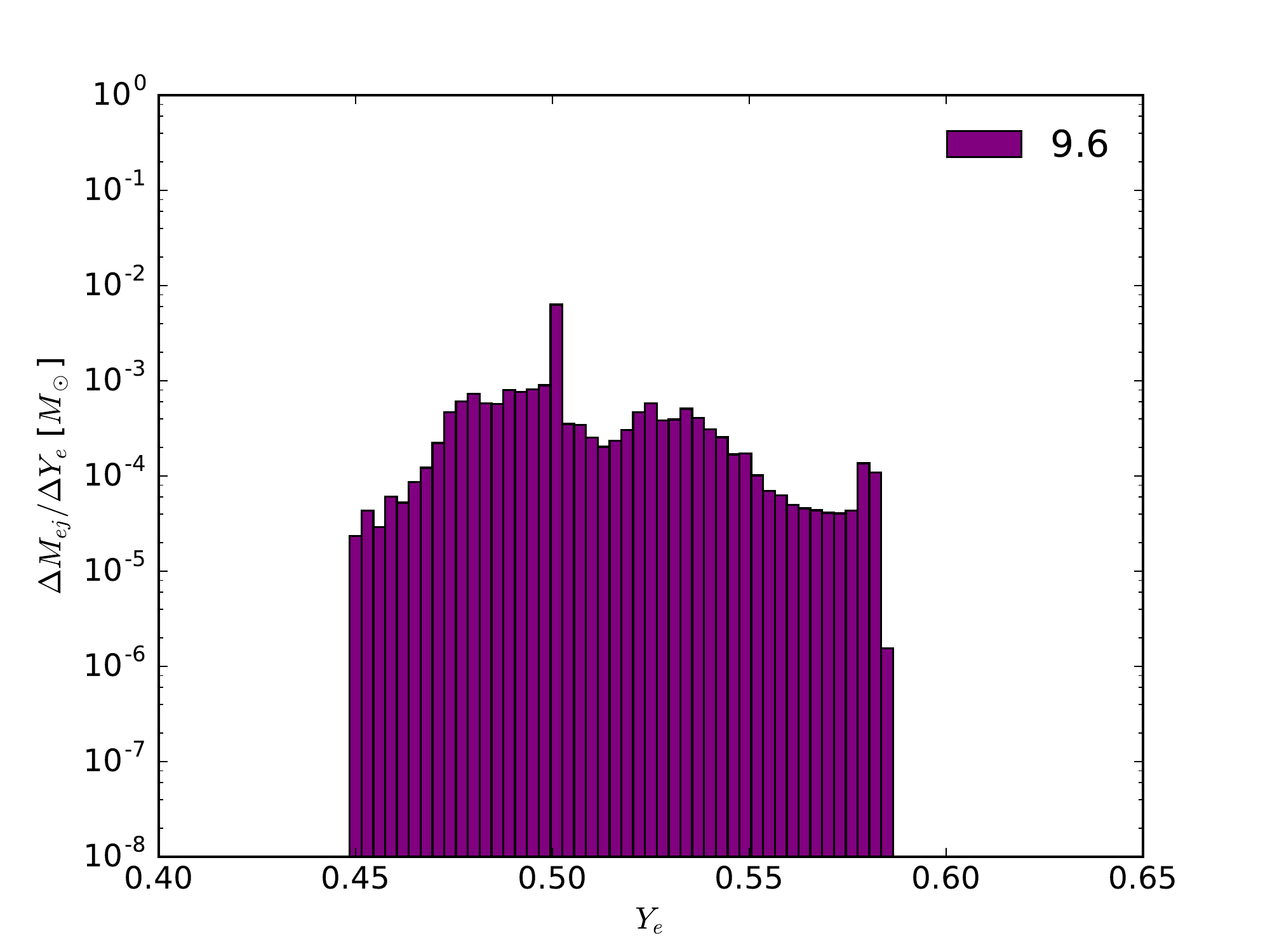}\hfill
\includegraphics[width=0.5\textwidth]{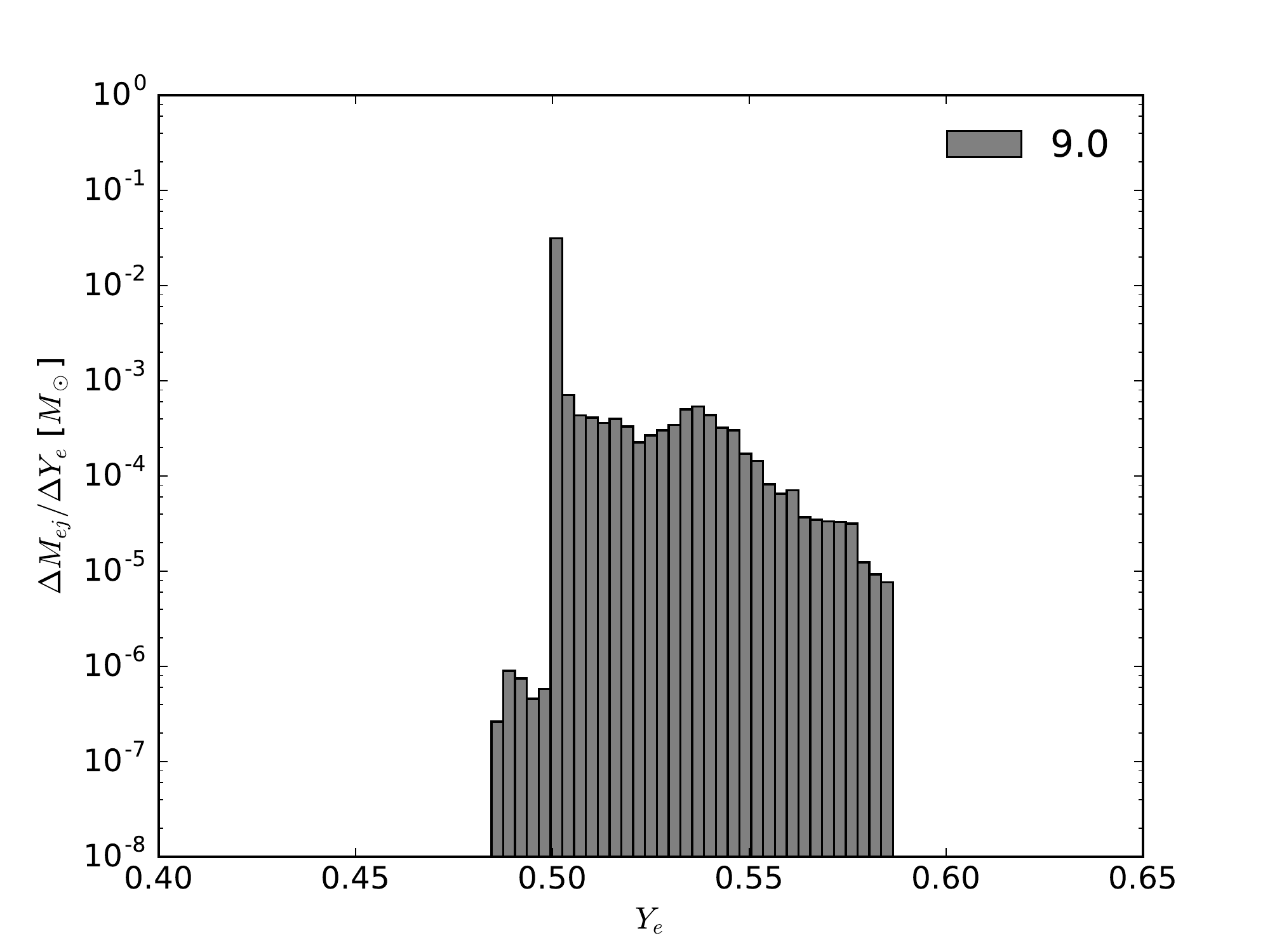}
\includegraphics[width=0.5\textwidth]{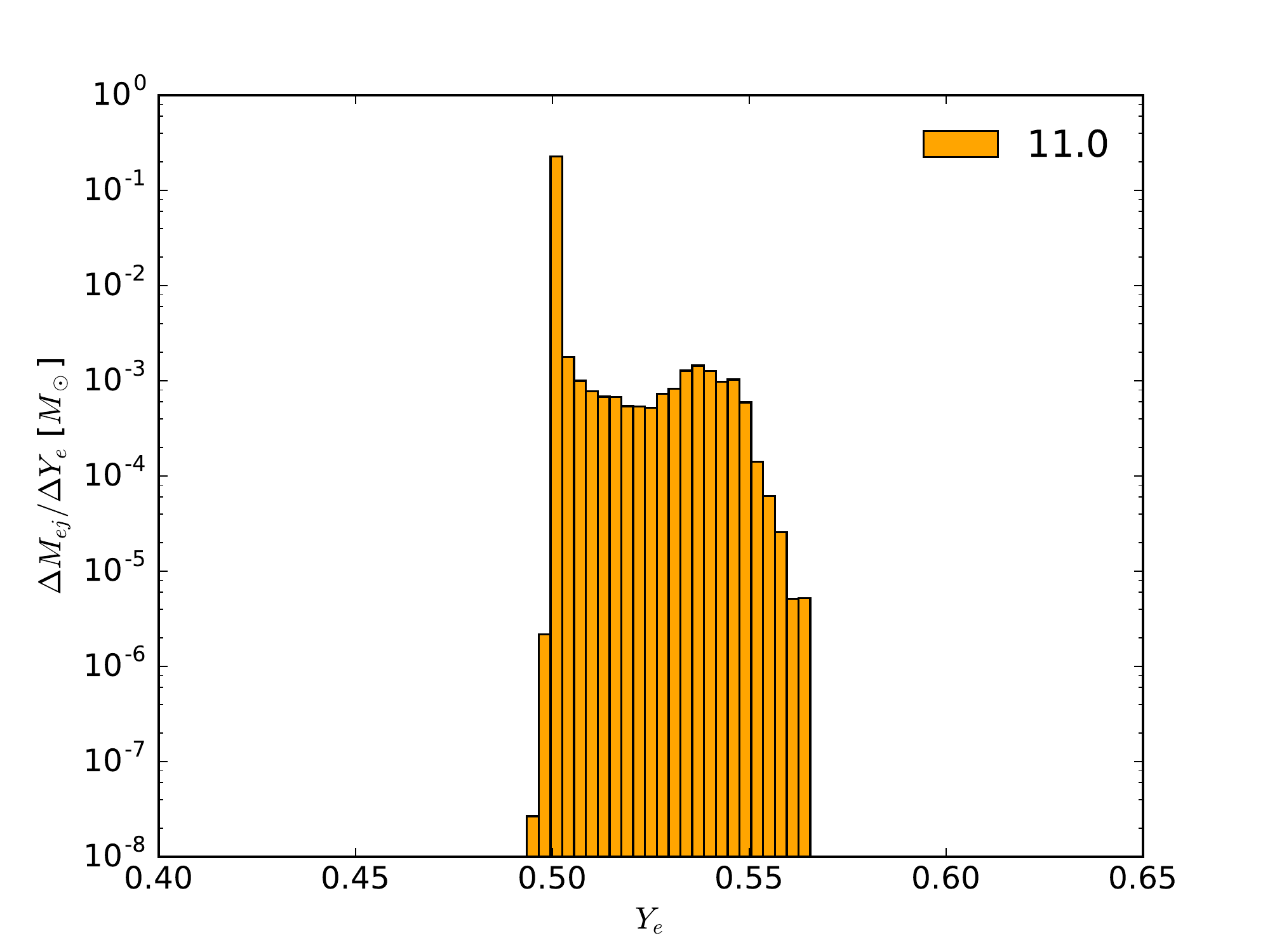}\hfill
\includegraphics[width=0.5\textwidth]{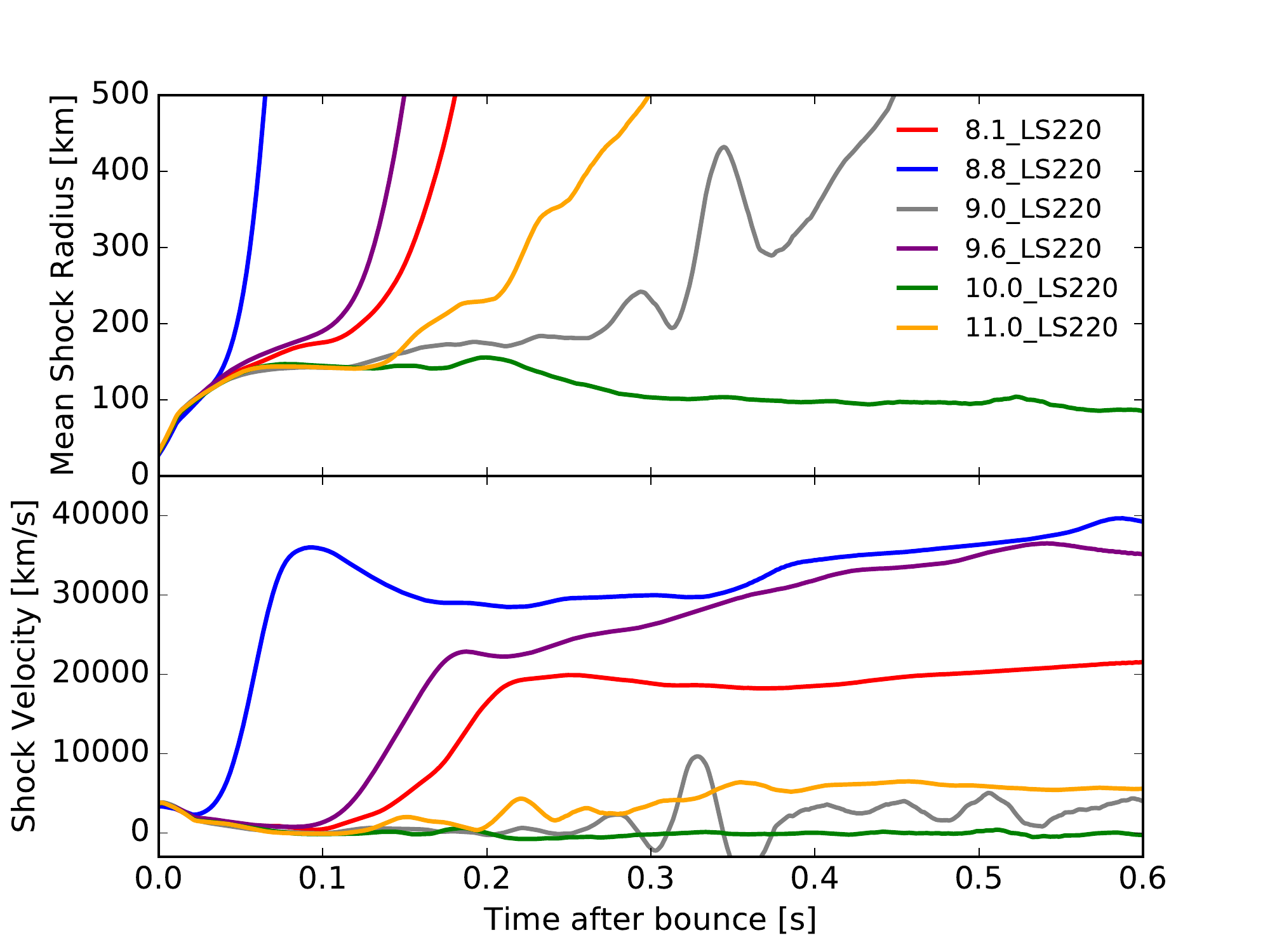}
\caption{Histogram (\textbf{first five panels}) illustrating the ejecta mass distribution function of Y$_e$ for the five exploding models models from \protect\cite{2017ApJ...850...43R} evaluated when the shock reaches the outer edge of the grid. The lower-mass progenitors with ejecta mass distributions extending to lower Y$_e$, namely the 8.1-, 8.8-, and zero-metallicity 9.6-M$_{\odot}$ progenitors, all have significantly higher mean shock velocities as seen in the \textbf{bottom right panel}, illustrating the mean shock radii (in km) and mean shock velocity (in km s$^{-1}$) as a function of time after bounce (in seconds). These are also the models that explode earlier. Note that the 10-M$_{\odot}$ progenitor does not explode, even with the many body correction. All models were evolved with the LS220 equation of state.}
\label{fig:11}
\end{figure*}

\subsection{Evolution of the Gain Region}

To probe the dependence of progenitor mass on explosion outcome, we study the properties of the models in the gain region, defined as where there is net neutrino heating. 

In Fig.\,\ref{fig:8}, we illustrate, from top to bottom, the mass of the gain region in units of 10$^{-3}$ M$_{\odot}$, the net heating rates $\dot{Q}$ in B s$^{-1}$, and the heating efficiency, defined as $\frac{\dot{Q}_{heat}}{L_{\nu_e}+L_{\bar{\nu}_e}}$, as a function of post-bounce time (in seconds) for the suite of nine progenitor masses. In the top panel, models that explode (solid) continue to grow in gain mass past the first 200 ms, while the remainder (dashed) do not. Furthermore, models with higher accretion rates (see Fig.\,\ref{fig:1}) have higher heating rates (middle panel) for the first 200 milliseconds of evolution, and hence earlier explosions feature lower heating rates. The heavier 20- and 25-M$_{\odot}$ models have the highest heating rates (not heating efficiency, bottom panel) early on, but these prove insufficient to overcome the greater ram pressure and the explosions are stifled. The exploding models have significantly higher heating efficiencies (bottom) after 200 milliseconds, following the explosion, than the non-exploding progenitors, with efficiency peaking at 0.15 for the 21-M$_{\odot}$ progenitor around 350 ms post-bounce. 

For comparison, \cite{2016ApJ...825....6S} study 12-, 15-, 20-, and 25-M$_{\odot}$ progenitors, finding maximum heating efficiencies in the gain region of $\sim$0.12, 600 milliseconds post-bounce, and \cite{2016ApJ...818..123B} find maximum efficiencies of $\sim$0.15, 200 milliseconds post-bounce. Changes in gain properties may be endemic of explosion and not necessarily its precursors.

\subsection{Microphysical Dependence}

We briefly explore the effects of inclusion of various physical processes to the explosion outcome. We use the 16-M$_{\odot}$ progenitor as a case study. In Fig.\,\ref{fig:9} (top left panel), we plot the mean shock radii (top panel) for three runs $-$ the default (`def') with IES$\_$INS$\_$MB, one with IES$\_$INS, and one with IES only.

We note a steady march towards explosion as we include additional physical processes. Adding inelastic scattering off nucleons leads to a shock radius stalling further out, and the inclusion of the many-body effect converts the failure to explosion.  Without all three effects, our 16-M$_{\odot}$ progenitor does not explode.

\cite{2006A&A...447.1049B} performed an early comparison of different neutrino interaction rates, finding reduced $\nu_e$, $\bar{\nu}_{e}$ opacities when including inelastic scattering of neutrinos off nucleons via \cite{1998PhRvC..58..554B}, rather than the elastic approach of \cite{1985ApJS...58..771B}. \cite{2012ApJ...756...84M} finds higher $\nu_e$, $\bar{\nu}_{e}$ luminosities when including inelastic scattering off nucleons. Furthermore, \cite{2017IAUS..331..107O} find increased neutrino heating due to the many-body effect (\citealt{PhysRevC.95.025801}). Finally, \cite{2018ApJ...854...63O} found that including inelastic scattering produced earlier explosions for the 12-, 15-, and 25-M$_{\odot}$ progenitors, though not early enough to overcome the discrepancy in explosion time between their work and that of \cite{2016ApJ...825....6S}.

As illustrated in \cite{Burrows2018}, however, one can prompt the model without the many-body effect to explode by including either perturbations to the infall velocities or modifying the opacity table to include the \cite{2016A&A...593A.103F} correction to the nucleon-nucleon bremsstrahlung (``bf'') and cutting the electron capture rate \cite{2010NuPhA.848..454J} on heavy nuclei to only 20\%. Non-exploding models can be made to explode with moderate changes to physical inputs.

\subsection{Exploding the ``Non-Exploding" Models}
Though the many-body effect was crucial in exploding the 16-M$_{\odot}$ progenitor, this default microphysical setup proved insufficient in exploding five of our other WH07 models.
Five (the 12-, 13-, 15-, 20-, and 25-M$_{\odot}$ models) of our nine WH07 models did not explode with the default setup, and we identify the absence of a sharp Si-O interface in the progenitor interior (see Fig.\,\ref{fig:1}) as one key difference. We find that with the inclusion of additional inputs, such as perturbations and/or moderate rotation, all these models explode.

We perturb the infalling velocities to 1000 km s$^{-1}$ over 3 regions using the prescription of \cite{2015MNRAS.448.2141M} (see also \citealt{2017ApJ...850...43R}). We use $n=5$ radial nodes and $l=2$ angular modes. Our perturbed regions span 1000-2000 km, 2100-4000 km, and 4100-6000 km. The inner region was chosen to be just outside our core at the start of the simulation and the outer region was approximated by the radial extent of matter that would be accreted during the first half second after bounce, by which time the default models have exploded. We find that the outcome is crucially sensitive to when these perturbed regions are accreted.

For our rotation prescription, we assume a cylindrical rotational profile following \cite{1985A&A...146..260E}. Our rotational angular frequency along the pole is a moderate 0.2 radians s$^{-1}$, corresponding to a period of just over 30 seconds. The characteristic radius, over which the frequency drops to half this value, is 10,000 km, much larger than normally assumed. We find that moderately rotation near the center that remains high at large radii is most promising for explosion (see Vartanyan et al. 2018b, in prep.). 

We plot our results in Fig.\,\ref{fig:9}. The 15-, 20- and 25-M$_{\odot}$ progenitors explode with only the addition of perturbations to infall velocities. However, the 12- and 13- M$_{\odot}$ models require the further inclusion of moderate rotation to explode. Note, however, that the 13-M$_{\odot}$ progenitor explodes with rotation alone (orange curve in third panel of Fig.\,\ref{fig:9}. For comparison, we also add rotation to the 15-M$_{\odot}$ progenitor and find, quaintly enough, that rotation delays explosion here by roughly 400 ms. The non-monotonic affect of rotation on explosion outcome will be further explored in Vartanyan et al. (2018b). 

\subsection{Electron Fraction Distribution}
We study the ejecta mass distribution (with the ejecta defined as the gravitationally unbound mass) with $Y_e$ at the end of our default simulations. Figure \ref{fig:10} illustrates a histogram of these results with Y$_e$ bin resolution of 0.003. Independent of progenitor mass, all models show a peak near Y$_e$ = 0.5, with a tail extending to $Y_e = 0.6$. Only the lowest mass progenitor, 16 M$_{\odot}$, shows a tail extending to lower $Y_e$ values.

Recently, \cite{2018ApJ...852...40W} found that, for their sample of four low-mass progenitor supernovae, lower-mass progenitors had more neutron-rich ejecta due to faster shock growth and, hence, less dwell time of the neutron-rich ejecta for neutrino processing. This holds true for the lower-mass progenitors, which are relatively isotropic in explosion. However, we find little correlation between the shock velocities of our more massive models (see Fig.\,\ref{fig:2}) and the ejecta distribution, Fig.\,\ref{fig:10}, where perhaps multidimensionality and ejecta anisotropies play a bigger role. For instance, our 16-M$_{\odot}$ progenitor is the only model with outflow concentrated in the southern hemisphere (see Fig.\,\ref{fig:6}), which we suggested earlier could lead to a correspondingly smaller explosion energy. Such an anisotropic explosion would also leave neutron-rich material in the northern hemisphere relatively untouched, possibly explaining the low-$Y_e$ tail for the 16-M$_{\odot}$ model seen in Fig.\,\ref{fig:10}.

To explore this claim, we add in Fig.\,\ref{fig:11} the $Y_e$ distribution of the ejecta mass for a set of low-mass progenitors from \citealt{2017ApJ...850...43R}, which we note uses the LS220 and not the SFHo EoS, as per our calculations (all else equal). We look at an 8.8-M$_{\odot}$ model (\citealt{1984ApJ...277..791N}; \citealt{1987ApJ...322..206N}); an 8.1-M$_{\odot}$ model (\citealt{2012ApJ...761...72M}); an initially metal-free 9.6-M$_{\odot}$ model (\citealt{2013ApJ...766...43M}); and 9-, 10-, 11-M$_{\odot}$ models (\citealt{2016ApJ...821...38S}), using the Lattimer-Swesty (LS220) equation of state with nuclear incompressibility of 220 MeV (\citealt{1991NuPhA.535..331L}). \cite{2017ApJ...850...43R} find that all models except for the 10-M$_{\odot}$ progenitor explode with the inclusion of inelastic scattering off electrons and nucleons as well as the \cite{PhysRevC.95.025801} many-body correction (the 10-M$_{\odot}$ model explodes with the further addition of perturbations to infall velocities). We plot in the final panel of Fig.\,\ref{fig:11} the mean shock radii (km) and shock velocities (km s$^{-1}$) as a function of time post bounce (in seconds) for these six low-mass progenitors. Note that the shock velocities show a bimodal clumping: 1) those weakly explosive models (9- and 11-M$_{\odot}$) with shock velocities less than 10,000 km s$^{-1}$ (together with the non-exploding 10-M$_{\odot}$ progenitor), and 2) the three more robust explosions with shock velocities spanning 20,000 to 40,000 km s$^{-1}$ (the 8.1-, 8.8-, and 9.6- M$_{\odot}$ progenitors). In our Y$_e$ histograms in Fig.\,\ref{fig:10}, we see that these three lower mass ECSN progenitors have more low-Y$_e$ ejecta together with greater shock velocities, in agreement with \cite{2018ApJ...852...40W}. For comparison, all four of our more massive exploding models have smaller shock velocities, asymptoting at roughly 7000 km s$^{-1}$ (see Fig.\,\ref{fig:2}), and the association between shock velocity and ejecta mass is less clear for these more massive progenitors, where we argue explosion anisotropies play a more decisive role in $Y_e$-ejecta mass distribution.

\begin{figure}
\centering
\includegraphics[width=\columnwidth]{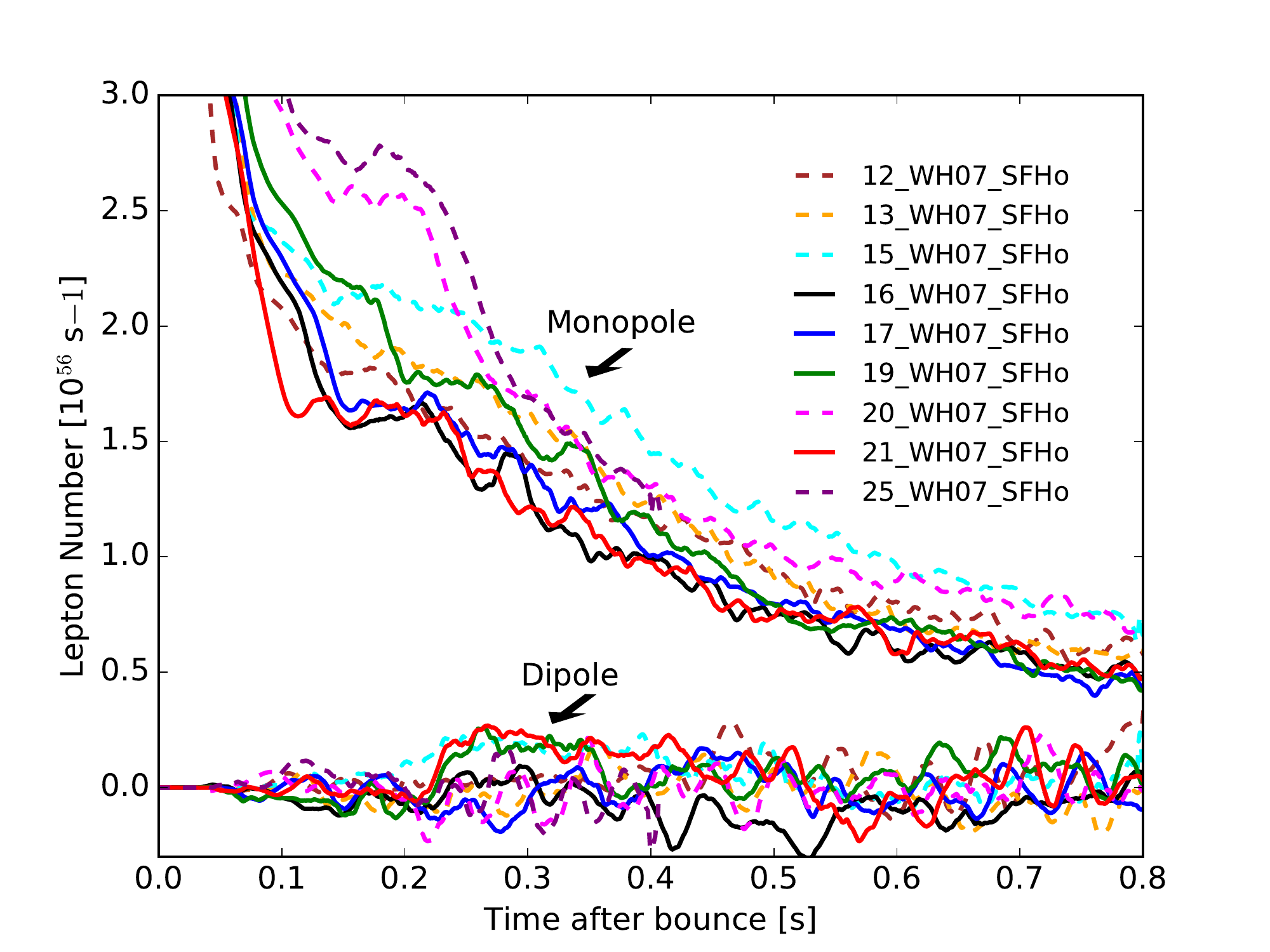}
\vfill
\includegraphics[width=\columnwidth]{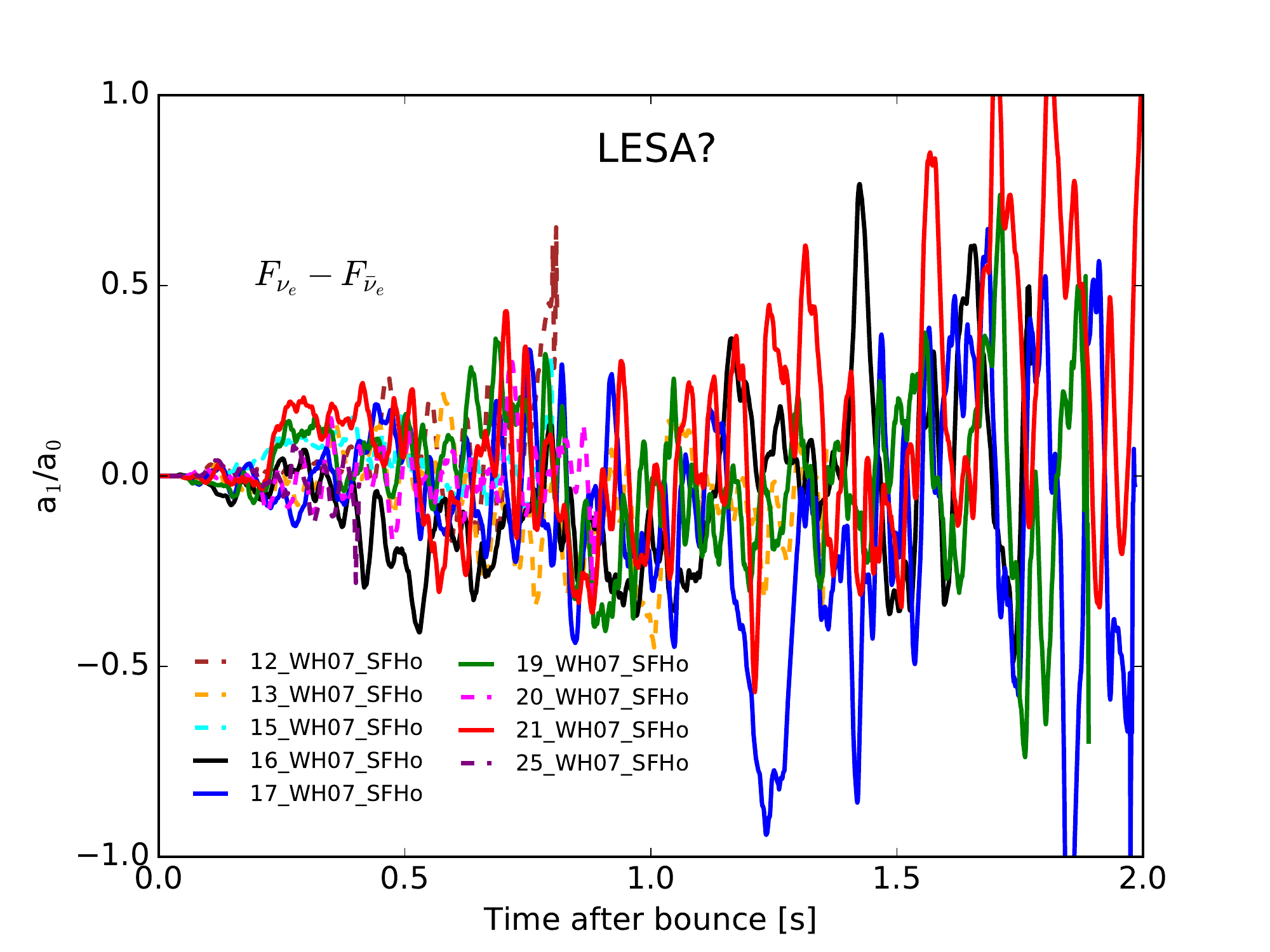}
\caption{Following \protect\cite{2014ApJ...792...96T}, we plot the ratio of the dipole and monopole (\textbf{top panel}) of the differences in lepton number fluxes at 500 km and their ratios (\textbf{bottom panel}) as a function of time after bounce (in seconds). Note the different x-axis scales. We find our dipole term to be an order of magnitude smaller, at least for the first several hundred milliseconds than \protect\cite{2014ApJ...792...96T} (who perform the simulations in 3D with the ray-by-ray plus approximation but find no explosions). This is in agreement with \protect\cite{2015ApJ...800...10D}, though we find our dipole component to be slightly larger. Furthermore, even when the dipole term is of the same order as the monopole term (around one second post-bounce, when the latter has decayed sufficiently), the amplitudes are oscillating and not sustained. Only for the 21-M$_{\odot}$ model (red) do we find a sustained dipole term from 200 to 500 ms post-bounce, but even this is smaller by an order of magnitude than the results found by \protect\cite{2014ApJ...792...96T} for their progenitor suite. Thus, we conclude that we do not find evidence for lepton-emission self-sustained asymmetry (LESA).}
\label{fig:12}
\end{figure}

\subsection{LESA}\label{sec:LESA}
Following \cite{2014ApJ...792...96T}, we look for evidence for the Lepton-number Emission Self-sustained Asymmetry (LESA), a neutrino-hydrodynamical instability that may set in shortly before explosion. In  Fig.\,\ref{fig:12}, we plot (top panel) the dipole and monopole moments of the neutrino number asymmetry (defined as the number flux of electron-type neutrinos minus anti-electron type neutrinos, $F_{\nu_e}-F_{\bar{\nu}_e}$,) and the ratio of the two (bottom panel). 
Relative to \cite{2014ApJ...792...96T} (who perform the simulations in 3D with the ray-by-ray plus approximation but find no explosions), we find our dipole term to be an order of magnitude smaller, at least for the first several hundred milliseconds. This is in agreement with \cite{2015ApJ...800...10D}, though we find our dipole component to be slightly larger. Furthermore, even when the dipole term is of the same order as the monopole term (around one second post-bounce, when the latter has decayed sufficiently), the amplitudes are oscillating and not sustained. Only for the 21-M$_{\odot}$ model (red) do we find a sustained dipole term from 200 to 500 ms post-bounce, but even this is smaller by an order of magnitude than the results found by \cite{2014ApJ...792...96T} for their progenitor suite. Thus, we conclude that we do not find evidence for lepton-emission self-sustained asymmetry (LESA), at least in 2D. However, we emphasize that thorough analysis requires performing the simulation in 3D with correct neutrino transport.

\begin{figure}
\centering
\includegraphics[width=\columnwidth]{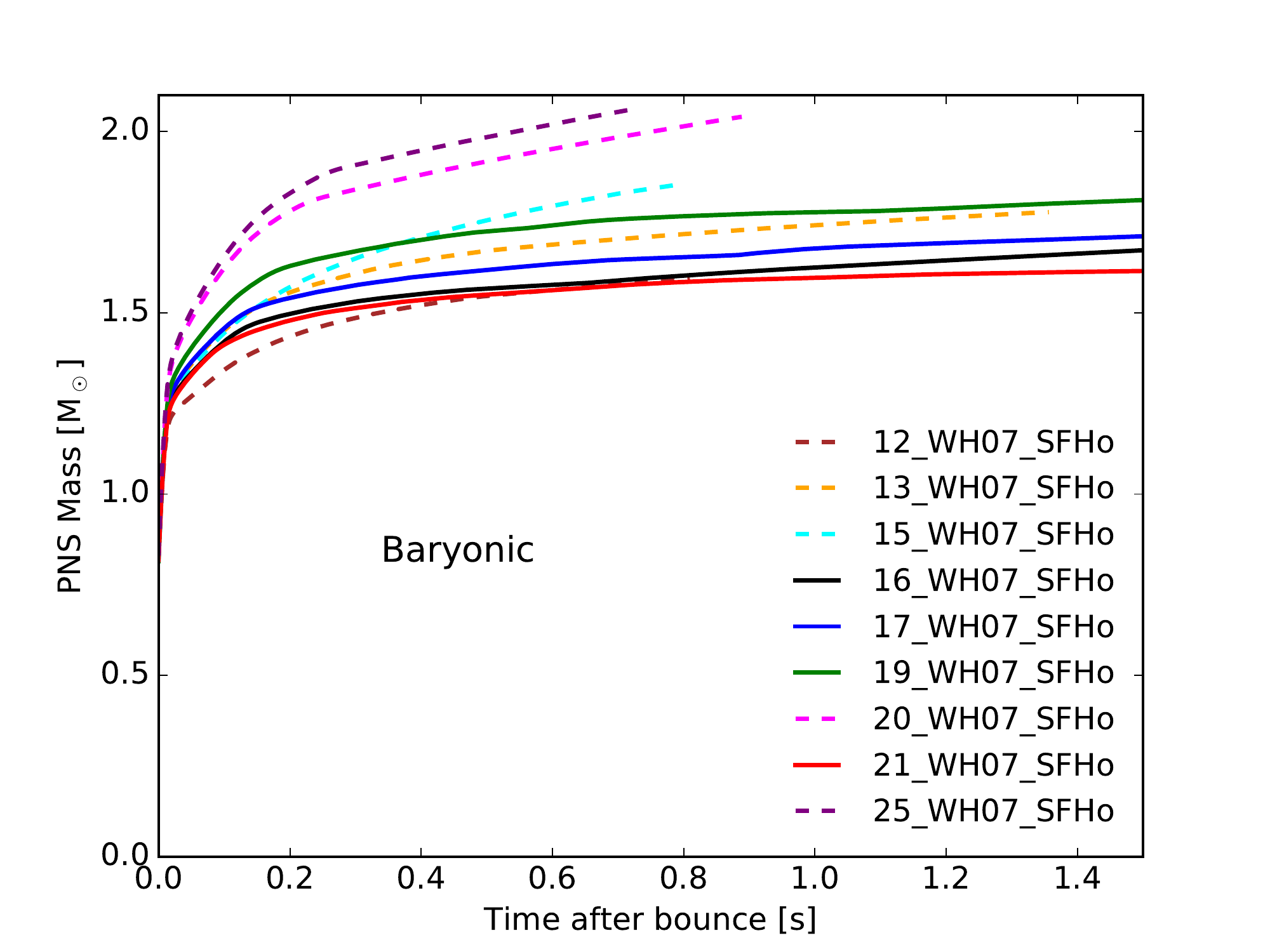}
\includegraphics[width=\columnwidth]{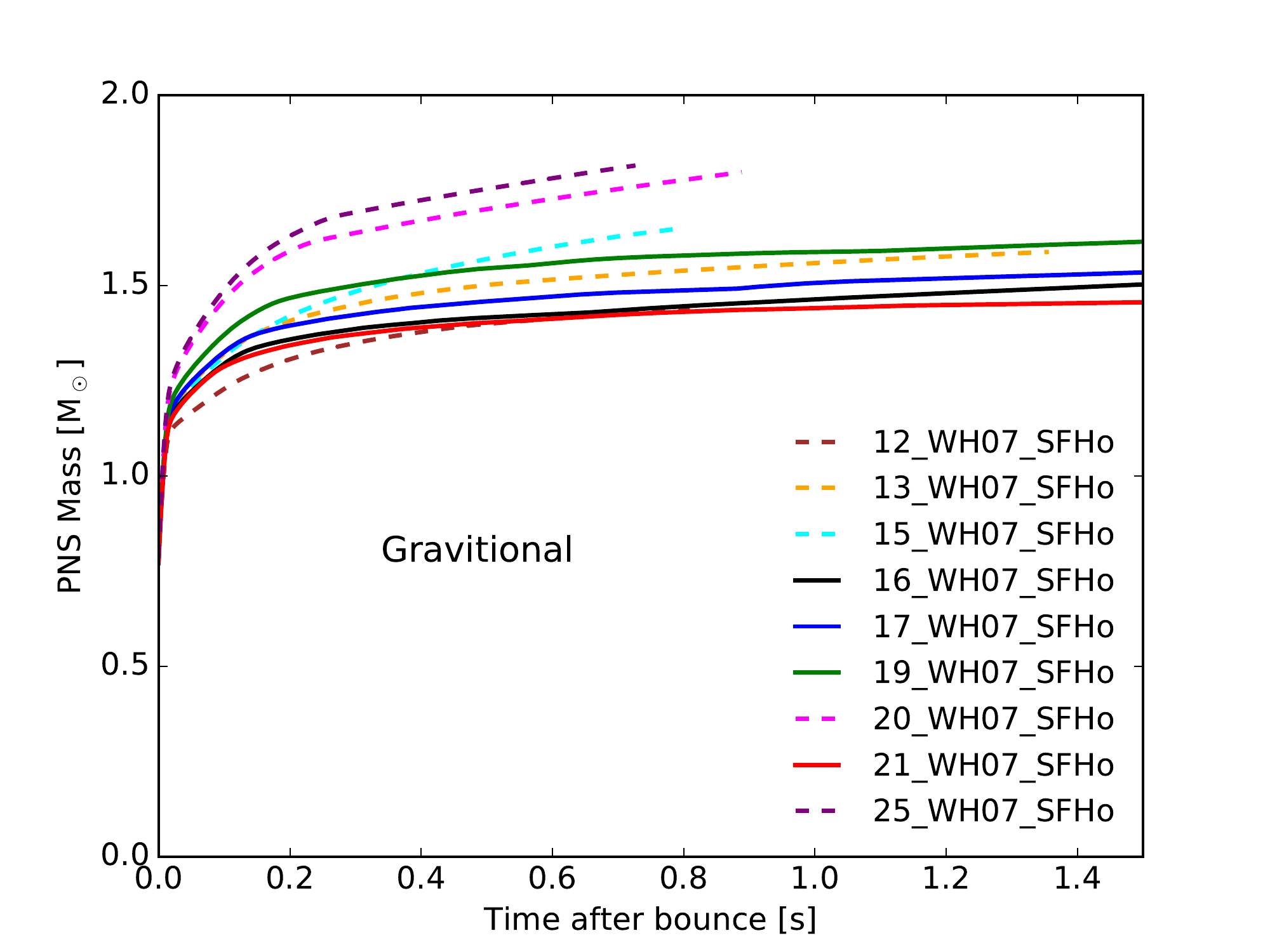}
\includegraphics[width=\columnwidth]{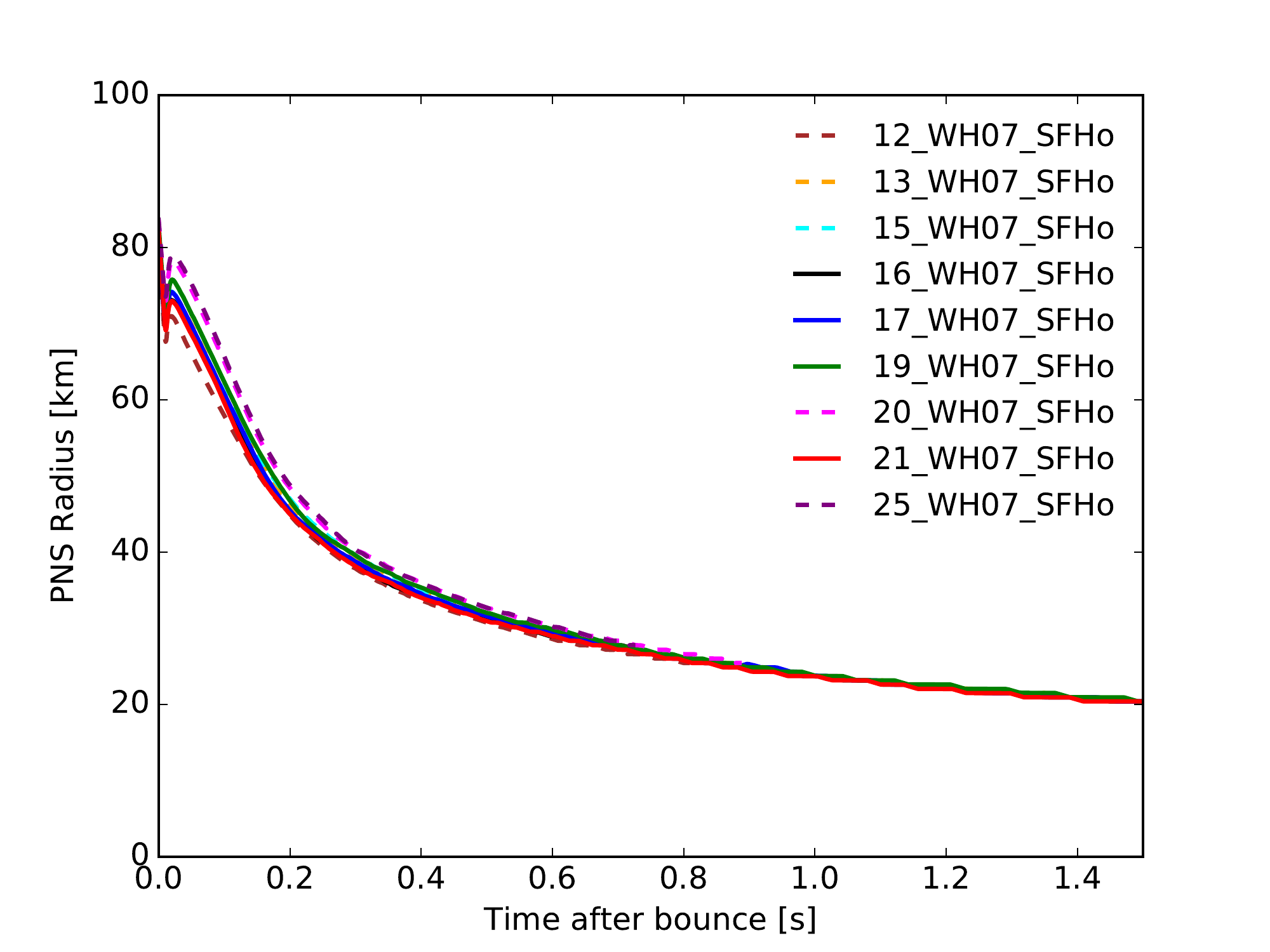}
\caption{PNS baryonic (\textbf{top}) and gravitational masses (\textbf{middle}) in M$_{\odot}$, and PNS radii (\textbf{bottom}) in km for our nine progenitors plotted against time after bounce (in seconds), with the PNS surface defined where the density is above 10$^{11}$ g cm$^{-3}$. The two most massive non-exploding progenitors, 20- and 25-M$_{\odot}$, reach almost 2 M$_{\odot}$ in baryonic PNS mass only half a second after bounce, whereas the others are clustered between 1.6 and 1.8 M$_{\odot}$. Note that the PNS masses for the non-exploding models (dashed) increase monotonically with progenitor mass, whereas they increase in order of explosion time (and thus accretion history) for the four exploding models (solid). The PNS radii evolve to be independent of progenitor mass within one second of bounce.}
\label{fig:13}
\end{figure}

\begin{figure}
\centering
\includegraphics[width=\columnwidth]{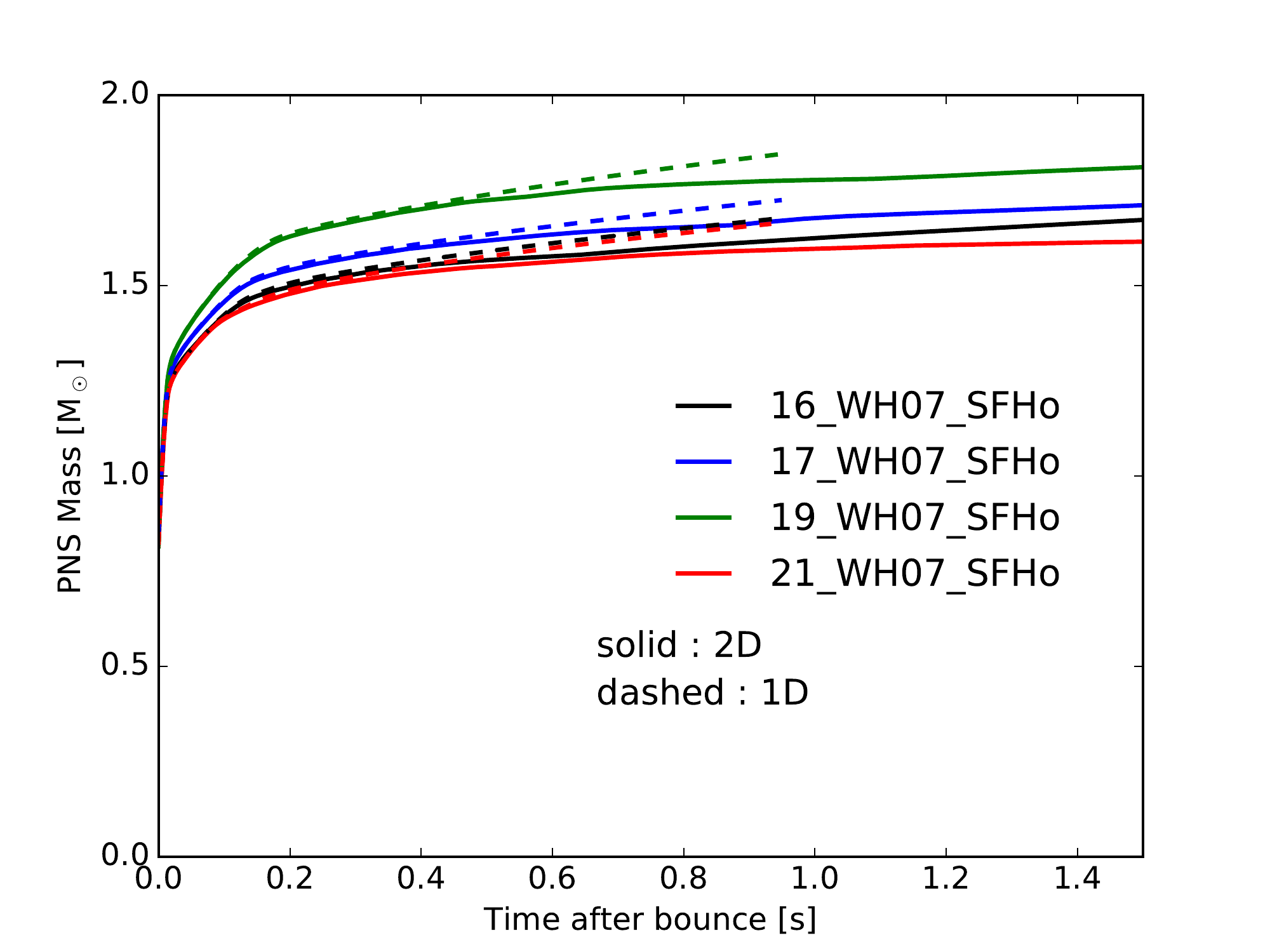}
\includegraphics[width=\columnwidth]{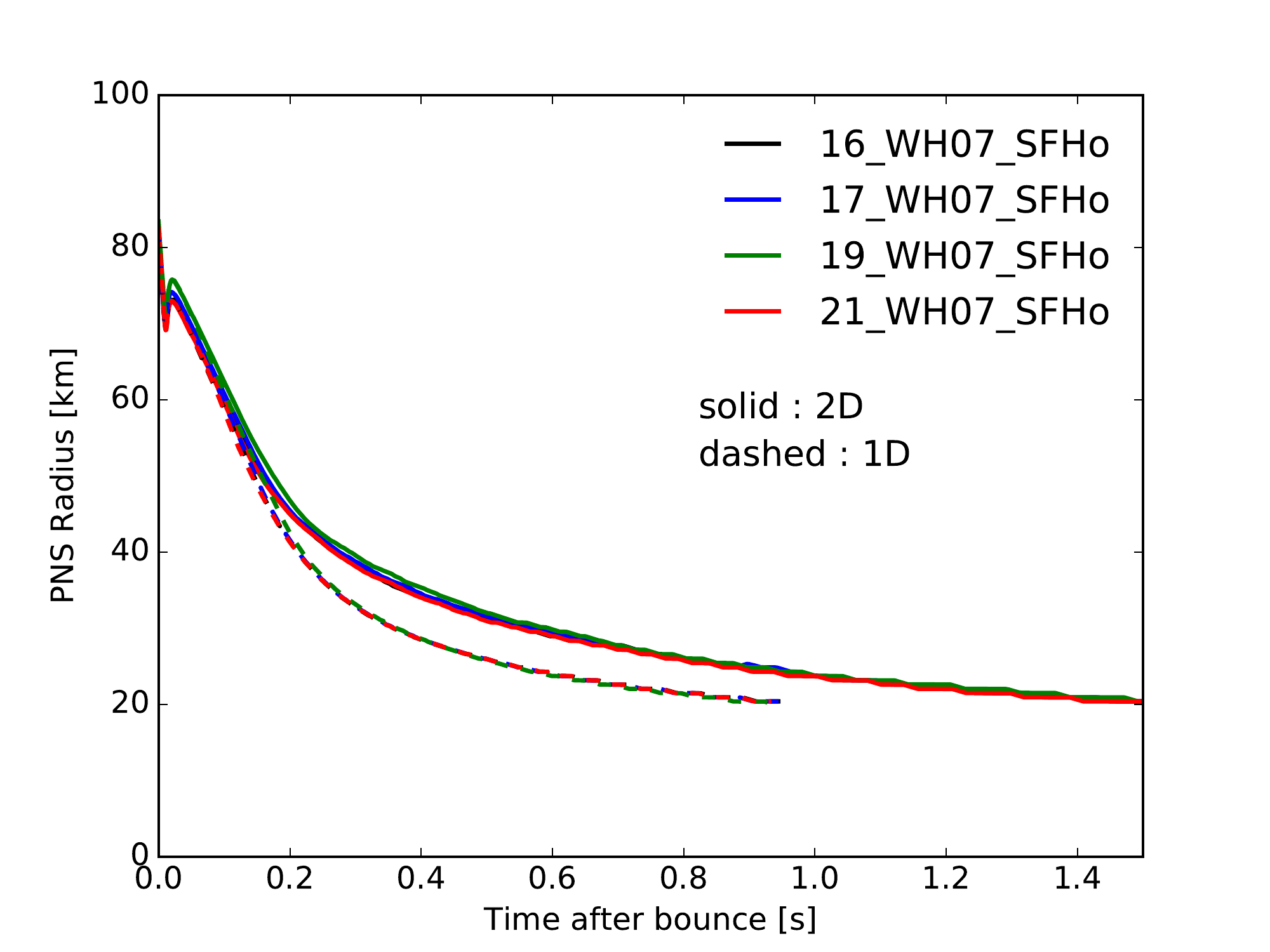}
\caption{PNS masses (\textbf{top}) in M$_{\odot}$ and radii (\textbf{bottom}) in km, defined as where the density is above 10$^{11}$ g cm$^{-3}$, for our four default exploding  progenitors in 2D (solid) and corresponding 1D (dashed) as a function of time (in seconds) after bounce. The PNS masses are roughly correlated with progenitor masses and indicate their respective accretion history, hence the higher PNS mass for the non-exploding 1D models. Note that the PNS radii all converge to a similar mass regardless of progenitor mass, as determined by the equation of state. The larger PNS radii in 2D is attributed to inner PNS convection.}
\label{fig:14}
\end{figure}

\section{Neutron Star Properties}\label{sec:ns}

Here, we provide a few of the properties of the remnant proto-neutron star in our model suite. We identify the proto-neutron star (PNS) radius where the density first drops below 10$^{11}$ g\,cm$^{-3}$, though the radius is quite insensitive to the precise density cutoff near this value.

In Fig.\,\ref{fig:13} and Fig.\,\ref{fig:14}, we study both the dependence upon progenitor mass and the detailed microphysics on PNS mass (top panel) and radius (bottom panel) for two sets of models: all nine progenitors in 2D, and the four exploding progenitors in 2D and their 1D counterparts. For the former, we also plot the gravitational mass following the approximate fit of \cite{1996ApJ...457..834T}. Since the density drops sharply at the PNS surface, the PNS radii are insensitive to the ambient pressure external to the core and, hence, to the progenitor mass (see also \citealt{2017ApJ...850...43R}), and we find that all PNS radii converge to the same mass by 1.5 seconds post-bounce for all the progenitors in 2D. However, we find that the PNS radii are sensitive to the dimensionality, with the 1D models' PNS radii roughly 20\% smaller than the 2D counterparts. \cite{2017ApJ...850...43R} find a similar result for their set of low-mass progenitors, attributing the larger radii in 2D to convection in the PNS.

\begin{figure}
\centering
\includegraphics[width=\columnwidth]{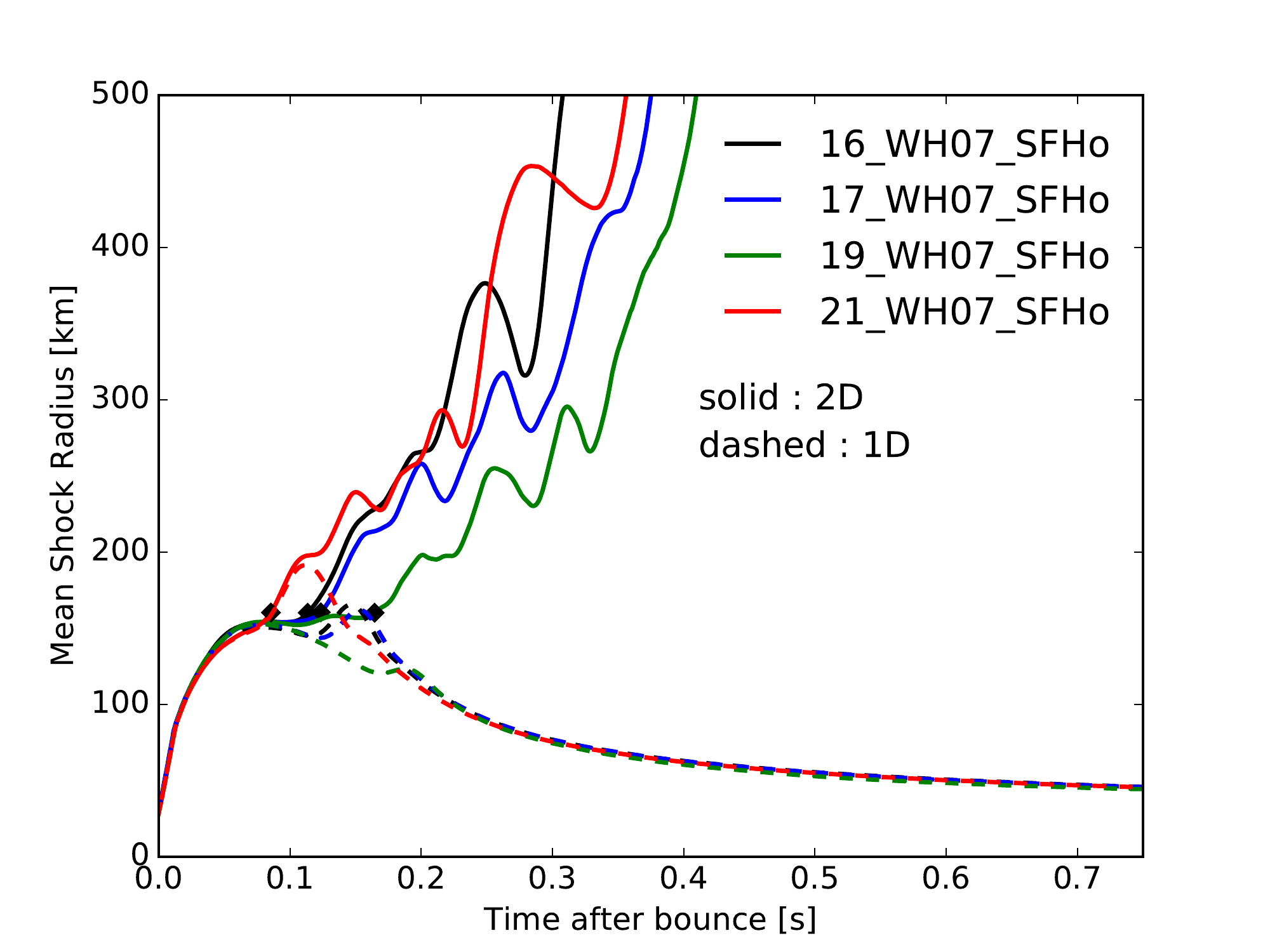}
\caption{Mean shock radii (in km) against time after bounce (in seconds) of the exploding models in 2D (solid) and their non-exploding 1D counterparts (dashed). The 1D counterpart evolve similarly until roughly 100 ms post-bounce, even featuring analogous bumps in shock radii with the same time ordering as the 2D models. None of the 1D models explodes but all asymptote to roughly 40 km at late times.}
\label{fig:15}
\end{figure}

The PNS baryon mass evolution on the other hand, simply tracks accretion history. For the non-exploding (dashed) models in Fig.\,\ref{fig:13}, this is monotonic with progenitor mass, spanning from $\sim$1.6 to $\sim$2.0 M$_{\odot}$ for the 12 to 25-M${\odot}$ progenitors, respectively, and correlates roughly monotonically with progenitor mass. For the 20- and 25-M$_{\odot}$ progenitors, the PNS exceeds 2.0 M$_{\odot}$ as early as 0.6 seconds post-bounce. Unlike the four exploding models, the non-exploding models have not yet asymptoted by the end of the calculation. Figure \ref{fig:17} illustrates the mass evolution comparing 1D and 2D. The latter explodes, reverting accretion, and hence, leaves behind a smaller PNS mass. 

\subsection{Effect of Microphysics on PNS Masses}
Due to the reduced neutrino opacities, we find that the many-body effect leads to a faster PNS contraction rate and a smaller PNS radii by $\sim$5\%, as was found to be the case for low-mass progenitors by \cite{2017ApJ...850...43R}. Furthermore, because the model with the many-body correction prompts an earlier explosion, it leads to smaller PNS masses because of the shorter accretion history. 

\begin{figure}
\centering
\includegraphics[width=\columnwidth]{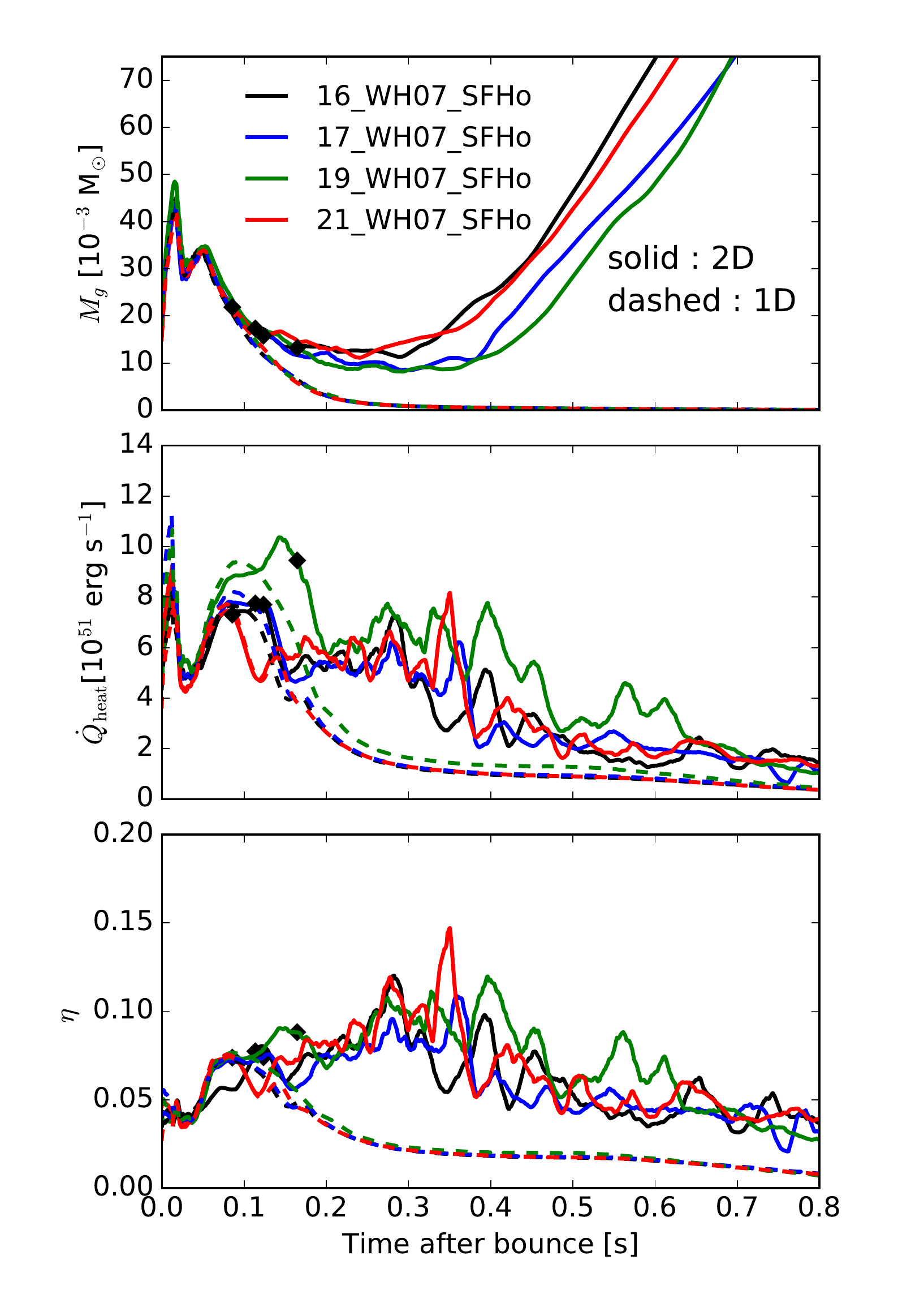}
\caption{Same as Fig.\,\ref{fig:8}, except comparing exploding models (solid) in 2D to their counterparts in 1D (dashed), none of which explodes. Up to 100 ms post-bounce, these diagnostics of the gain region $-$ mass, heating rate, and heating efficiency $-$ are quite similar for the two cases. After explosion, the gain mass, heating rates, and heating efficiency of the 2D models rise to values much higher than their 1D counterparts. Dimensionality of the simulation is directly reflected in the energetics of the exploding models.}
\label{fig:16}
\end{figure}

\section{1D Comparison}\label{sec:1D}

Here, we compare one-dimensional counterparts to the four models that explode in two dimensions. The shock radii (in km) for these eight models are shown in Fig.\,\ref{fig:15} as a function of time after bounce (in seconds). None of these models explodes by the end of our simulations, spanning at least 700 milliseconds post-bounce.

\begin{figure*}
\includegraphics[width=0.5\textwidth]{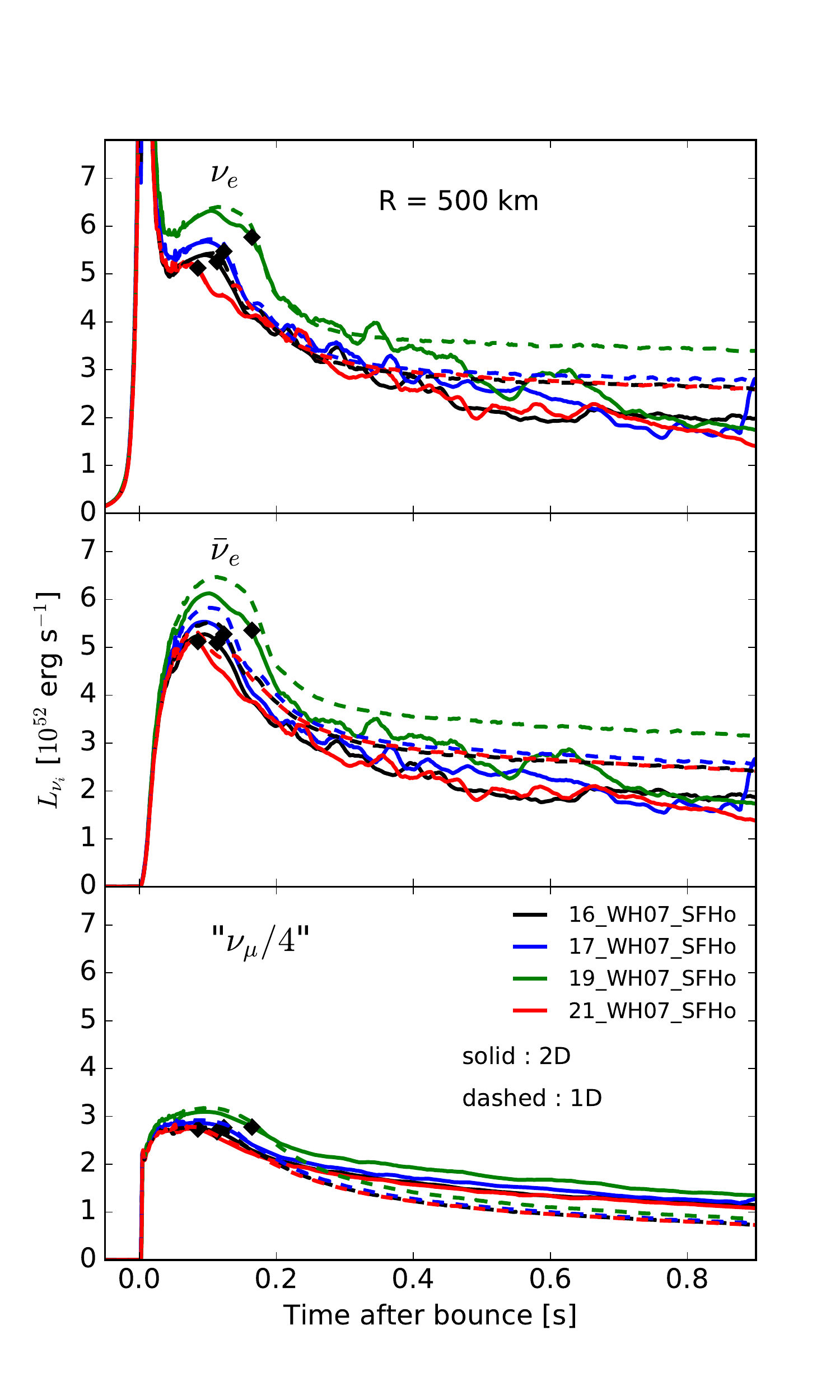}\hfill
\includegraphics[width=0.5\textwidth]{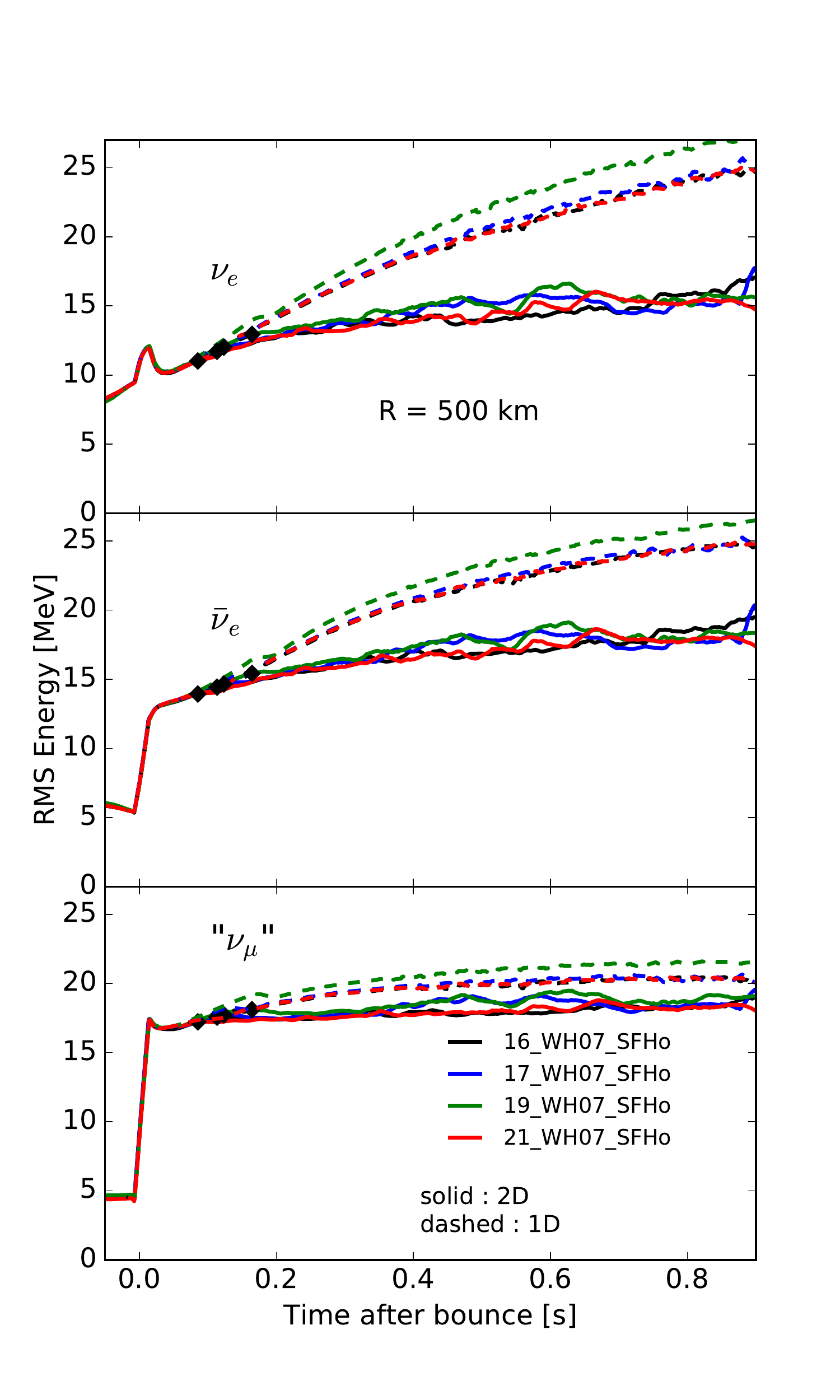}
\caption{Same as Fig.\,\ref{fig:5}, but now comparing the exploding models in 2D (solid) to their non-exploding 1D counterparts (dashed). The 2D models have lower electron- and anti-electron type neutrino luminosities, but higher heavy-type neutrino luminosities associated with PNS convection. Furthermore, peak neutrino luminosities are monotonically increasing with progenitor mass. All 1D RMS energies are consistently higher than their 2D counterparts.}
\label{fig:17}
\end{figure*}

Figure \ref{fig:17} depicts the luminosities (in 10$^{52}$ erg s$^{-1}$) and RMS energies (MeV) as a function of time after bounce (in seconds) for the 2D models and their 1D counterparts. The electron and anti-electron luminosities dip after explosion for the former as accretion is reversed. However, the 2D models have consistently higher $``\nu_\mu"$ luminosities by $\sim50$\%, which is in agreement with \cite{2017ApJ...850...43R} and \cite{2018ApJ...854...63O}, who cite increases in heavy-neutrino luminosities due to PNS convection in multi-dimensional simulations (see also \citealt{1993ApJ...418L..33B}; \citealt{1996ApJ...473L.111K}; \citealt{2006ApJ...645..534D}). In Fig.\,\ref{fig:6},  we see an inner convective region for the 2D models driven by a negative $Y_e$ gradient developing in the PNS as early as 100 ms post-bounce. Indeed, since our 1D models do not explode and accretion continues for longer, we may be underestimating the effects of convection in our comparison. 
    
In Fig.\,\ref{fig:16}, we compare gain region properties for the four exploding models (solid) in 2D to their counterparts in 1D (dashed). Up to 100 ms post-bounce, the gain region mass, heating rate, and heating efficiency are quite similar for the two cases. The similarities are short-lived at later times, the gain mass, heating rates, and heating efficiency of the 1D models plummeting to values much smaller than their 2D counterparts. Hence, dimensionality of the simulation is directly reflected in the energetics of the exploding models.

\section{Conclusions}\label{sec:con}

We have presented a series of 2D radiation-thydrodynamic simulations for nine progenitors with inelastic scattering processes off electrons and nucleons, as well as the many-body correction to neutrino-nucleon scattering opacities over a grid extending out to 20,000 km. We find that four of these models (16-, 17-, 19-, and 21-M$_{\odot}$ from \citealt{2007PhR...442..269W}) explode in this default configuration. These four models have Si-O interfaces featuring a significant density drop which reduces the accretion rate near the stalled shock and prompts early explosion. All four models explode within 200 milliseconds of bounce. The remaining five models do not have a prominent Si-O interface (12-, 13-, 15-M$_{\odot}$), or have one further out (20-, 25-M$_{\odot}$), suggesting that the timing of the accretion and dip at the Si-O interface could be critical to explosion. However, with the addition of moderate rotation and perturbations to infall velocities, even these five non-exploding models explode, suggesting that all progenitors are close to criticality for explosion. We also explore the microphysical dependence for the 16-M$_{\odot}$ progenitor, finding that it does not explode if either IES, INS, or MB is not included. Even removing the many-body correction prevents explosion. However, in \cite{Burrows2018}, we show that inclusion of the \cite{2016A&A...593A.103F} correction to the nucleon-bremsstrahlung and reducing the electron capture rate on heavy nuclei (\citealt{2010NuPhA.848..454J}) leads to explosion, corroborating our proposal that all models are near criticality and that modest changes to inputs can lead to explosion.

We calculate explosion energies for the four exploding models, summing kinetic, internal, and gravitational energies over our grid and substracting the absolute value of nuclear binding energy. We correct for the binding energy of the exterior overburden. All but the 21-M$_{\odot}$ progenitor have positive explosion energies at the end of our simulation of order a few $\times$10$^{50}$ ergs, and rising. The 17- and 19-M$_{\odot}$ progenitors are far from asymptoting and feature a corresponding rise in kinetic energy, suggesting the need to carry these calculations out on larger grids and for longer times to estimate final explosion energies. Furthermore, we see that the more energetically explosive models have multiple convective plumes with larger solid angles. Together with the rise in kinetic energy, this suggests that more isotropic morphology of outflow is significant in producing larger explosion energies. The gain region properties of exploding models further distinguish them from the non-exploding models, with the former growing in gain mass following explosion with correspondingly higher heating efficiencies.

Together with the low-mass progenitors models from \cite{2017ApJ...850...43R}, we show that lower-mass progenitors tend to have higher shock velocities and consequently, less dwell time of the neutron-rich ejecta for neutrino processing. This produces $Y_e$-ejecta mass histograms skewed towards lower $Y_e$ for the lower mass progenitors. We also find no evidence for Lepton-number Emission Self-sustained Asymmetry (LESA), finding rather that the dipole moment of the net neutrino number is an order of magnitude smaller than found in \cite{2014ApJ...792...96T}.

We find that PNS masses track accretion history and are systemically larger for non-exploding or later-exploding models. PNS radii, however, are largely insensitive to input physics, but are sensitive to dimensionality, with 1D models asymptoting to a smaller PNS radii than their 2D counterparts. \cite{2017ApJ...850...43R} found similar behavior for their set of low-mass progenitors, citing convection in the PNS in 2D for the larger PNS radii. Including the many-body effect, however, does lead to a faster PNS contraction rate.

We concluded by exploring 1D comparisons to our four exploding models in 2D. None of the models explodes in 1D. The electron- and anti-electron-type neutrino luminosities dip in 2D post-explosion, as accretion is reversed. The ``$\nu_\mu$"-type neutrino luminosities, however, are consistently higher, attributed to inner convection in the PNS. Mass, heating rate, and heating efficiency rise post-explosion for the exploding 2D models, but not for their 1D counterparts.

In the near future, we will explore these progenitor models in 3D using F{\sc{ornax}}. Early multi-group 3D simulations either did not explode, or exploded later; more recent simulations illustrate that 3D progenitors are only slightly less explosive (see review by \citealt{2016PASA...33...48M}). The inclusion of detailed microphysics, including the many-body effects, together with multi-dimensional neutrino transport, may bridge this gap. Moreover, we will explore whether 3D simulations produce more isotropic explosions and larger explosion energies that closely reproduce what we see in Nature. 

\section{Acknowledgements}
The authors thank Sydney Andrews and Viktoriya Morozova for helpful discussions and feedback. DR acknowledges support from a Frank and Peggy Taplin Membership at the Institute for Advanced Study and the Max-Planck/Princeton Center (MPPC) for Plasma Physics (NSF PHY-1523261). JD acknowledges support from the Laboratory Directed Research and Development program of Los Alamos National Laboratory. The authors acknowledge support under U.S. NSF Grant AST-1714267, the 
Max-Planck/Princeton Center (MPPC) for Plasma Physics (NSF PHY-1144374), 
and the DOE SciDAC4 Grant DE-SC0018297 (subaward 00009650). They employed 
computational resources of the Princeton Institute for Computational 
Science and Engineering (PICSciE) and of the National Energy Research 
Scientific Computing Center (NERSC), which is supported by the Office of 
Science of the US Department of Energy (DOE) under contract 
DE-AC03-76SF00098.

\bibliographystyle{mnras}
\bibliography{References}

\begin{table*}
\centering
\begin{tabular}{| c | c | c | c | c | c | c |}\hline
& & & & & \\ 
Model  & M$_{\mathrm{Bar}}$  & M$_{\mathrm{Grav}}$ & -E$_{\mathrm{Env}}$  & E$_{\mathrm{Tot}}$ & $\dot{\mathrm{E}}_{\mathrm{Tot}}$ \\
  & [M$_{\odot}$] & [M$_{\odot}$] &  [10$^{50}$ ergs] & [10$^{50}$ ergs] & [10$^{50}$ ergs $s^{-1}$] \\
\hline
 16    & 1.70 & 1.52 & 1.57 &  1.64 & 0.4 \\
 17    & 1.74 & 1.56 & 2.02 &  2.89 & 0.8 \\
 19    & 1.84 & 1.64 & 2.70 &  2.40 & 0.8 \\
 21    & 1.63 & 1.47 & 3.56 & -0.70 & 0.7 \\
 \hline

\end{tabular}
\caption{Table 1: Explosion diagnostics and PNS properties for the four exploding models at the end of our simulations. We list the baryonic (M$_{\mathrm{Bar}}$) and gravitational masses (M$_{\mathrm{Grav}}$) in solar masses, the  envelope energies (E$_{\mathrm{Env}}$) and total explosion energies (E$_{\mathrm{Tot}}$) in 10$^{50}$ ergs, the rate of increase in explosion energy ($\dot{\mathrm{E}}_{\mathrm{Tot}}$) in 10$^{50}$ erg s$^{-1}$, and the ejecta mass M$_{\mathrm{Ej}}$ in M$_{\odot}$ (defined as neutrino-processed, neutron-rich gravitationally unbound material on our computational grid). All values are calculated at the end of the simulation.}
\label{table:1}
\end{table*}

\bsp	
\label{lastpage}
\end{document}